\RequirePackage[switch]{lineno_babar_revtex}
\documentclass[twocolumn,showpacs,aps,floatfix,prd,nofootinbib]{revtex4}
\catcode`\@=11\relax
\def\@makecol{%
 \setbox\@outputbox\vbox{%
  \boxmaxdepth\@maxdepth
 \protected@write\@auxout{}{% 
 \string\@LN@col{\@ifnum{\pagegrid@cur=\@ne}{1}{2}}
      }%
  \@tempdima\dp\@cclv
  \unvbox\@cclv
  \vskip-\@tempdima
 }%
 \xdef\@freelist{\@freelist\@midlist}\global\let\@midlist\@empty
 \@combinefloats
 \@combineinserts\@outputbox\footins
  \set@adj@colht\dimen@
  \count@\vbadness
  \vbadness\@M
  \setbox\@outputbox\vbox to\dimen@{%
   \@texttop
   \dimen@\dp\@outputbox
   \unvbox\@outputbox
   \vskip-\dimen@
   \@textbottom
  }%
  \vbadness\count@
 \global\maxdepth\@maxdepth
}%
\def\balance@two#1#2{%
\outputdebug@sw{{\tracingall\scrollmode\showbox#1\showbox#2}}{}%
 \setbox\@ne\vbox{%
  \@ifvoid#1{}{%
   \unvcopy#1\recover@footins
   \@ifvoid#2{}{\marry@baselines}%
  }%
  \@ifvoid#2{}{%
   \unvcopy#2\recover@footins
  }%
 }%
 \dimen@\ht\@ne\divide\dimen@\tw@
 \dimen@i\dimen@
 \vbadness\@M
 \vfuzz\maxdimen
 \loopwhile{%
  \dimen@i=.5\dimen@i
  \outputdebug@sw{\saythe\dimen@\saythe\dimen@i\saythe\dimen@ii}{}%
  \setbox\z@\copy\@ne\setbox\tw@\vsplit\z@ to\dimen@
  \setbox\z@ \vbox{%
 \protected@write\@auxout{}{% 
 \string\@LN@col{\@ifnum{\pagegrid@cur=\@ne}{1}{2}}
      }%
   \unvcopy\z@
   \setbox\z@\vbox{\unvbox\z@ \setbox\z@\lastbox\aftergroup\vskip\aftergroup-\expandafter}\the\dp\z@\relax
  }%
  \setbox\tw@\vbox{%
   \unvcopy\tw@
   \setbox\z@\vbox{\unvbox\tw@\setbox\z@\lastbox\aftergroup\vskip\aftergroup-\expandafter}\the\dp\z@\relax
  }%
  \dimen@ii\ht\tw@\advance\dimen@ii-\ht\z@
  \@ifdim{\dimen@i>.5\p@}{%
   \advance\dimen@\@ifdim{\dimen@ii<\z@}{}{-}\dimen@i
   \true@sw
  }{%
   \@ifdim{\dimen@ii<\z@}{%
    \advance\dimen@\tw@\dimen@i
    \true@sw
   }{%
    \false@sw
   }%
  }%
 }%
 \outputdebug@sw{\saythe\dimen@\saythe\dimen@i\saythe\dimen@ii}{}%
\@ifdim{\ht\z@=\z@}{%
\@ifdim{\ht\tw@=\z@}{%
\true@sw
}{%
\false@sw
}%
}{%
\true@sw
}%
{%
}{%
\ltxgrid@info{Unsatifactorily balanced columns: giving up}%
\setbox\tw@\box#1%
\setbox\z@ \box#2%
}%
 \setbox\tw@\vbox{\unvbox\tw@\vskip\z@skip}%
 \setbox\z@ \vbox{\unvbox\z@ \vskip\z@skip}%
 \set@colroom
\dimen@\ht\z@\@ifdim{\dimen@<\ht\tw@}{\dimen@\ht\tw@}{}%
\@ifdim{\dimen@>\@colroom}{\dimen@\@colroom}{}%
 \outputdebug@sw{\saythe{\ht\z@}\saythe{\ht\tw@}\saythe\@colroom\saythe\dimen@}{}%
\setbox#1\vbox to\dimen@{\unvbox\tw@\unskip\raggedcolumn@skip}%
\setbox#2\vbox to\dimen@{\unvbox\z@ \unskip\raggedcolumn@skip}%
\outputdebug@sw{{\tracingall\scrollmode\showbox#1\showbox#2}}{}%
}%
\catcode`\@=12\relax

\usepackage{graphicx}
\usepackage{dcolumn}
\usepackage{array, longtable}
\usepackage{epsfig}
\usepackage{amsmath}
\usepackage{colordvi}
\usepackage{hhline}
\usepackage{color}
\newcommand{\BaBarYear}{22}
\newcommand{\BaBarNumber}{001}
\newcommand{\SLACPubNumber}{17694}

 \newcommand{\BaBarType}      {PUB}  % Journal publication
\input babarsym

\def\Ecm          {\ensuremath {E_{\rm c.m.}}\xspace}

\def\mgg  {\ensuremath {m(\gamma\gamma)}\xspace}
\def\KK {\ensuremath {K^+K^-}\xspace}
\def\Kpi {\ensuremath {K^{\pm}\pi^{\mp}}\xspace}

\long\def\inst#1{\par\nobreak\kern 4pt\nobreak
    {\it #1}\par\vskip 10pt plus 3pt minus 3pt}

\begin{document}

\begin{flushleft}
%\babar\ Analysis Document \# \BADNumber, Version 8 \\
%hep-ex/07xxxxx\\
\babar-\BaBarType-\BaBarYear/\BaBarNumber \\
SLAC-PUB-\SLACPubNumber \\
%To be Submitted to Physical Review D
\end{flushleft}

%\par\vskip 3cm

% Title of the paper
\title{\large \bf
\boldmath
Study of the reactions  $\epem\to K^+K^-\pi^0\pi^0\pi^0$,  $\epem\to
K^{0}_{S}K^{\pm}\pi^{\mp}\pi^0\pi^0$, and
$\epem\to K^{0}_{S}K^{\pm}\pi^{\mp}\pi^+\pi^-$  at
 center-of-mass energies from threshold to 4.5 GeV using initial-state
radiation
} % end title

\author{J.~P.~Lees}
\author{V.~Poireau}
\author{V.~Tisserand}
%\affiliation{Laboratoire d'Annecy-le-Vieux de Physique des Particules (LAPP), Universit\'e de Savoie, CNRS/IN2P3,  F-74941 Annecy-Le-Vieux, France}
\author{E.~Grauges}
%\affiliation{Universitat de Barcelona, Facultat de Fisica, Departament ECM, E-08028 Barcelona, Spain }
\author{A.~Palano}
%\affiliation{INFN Sezione di Bari, I-70126 Bari, Italy}
\author{G.~Eigen}
%\affiliation{University of Bergen, Institute of Physics, N-5007 Bergen, Norway }
\author{D.~N.~Brown}
\author{Yu.~G.~Kolomensky}
%\affiliation{Lawrence Berkeley National Laboratory and University of California, Berkeley, California 94720, USA }
\author{M.~Fritsch}
\author{H.~Koch}
% ...
%\affiliation{University of Wisconsin, Madison, Wisconsin 53706, USA }
\collaboration{The \babar\ Collaboration}
\noaffiliation

%\date{\today}

% Abstract
\begin{abstract}
  We study the processes
 $\epem\to K^+K^-\pi^0\pi^0\pi^0\gamma$,
 $K^{0}_{S}K^{\pm}\pi^{\mp}\pi^0\pi^0\gamma$, and 
$K^{0}_{S}K^{\pm}\pi^{\mp}\pi^+\pi^-\gamma$
in which an energetic
photon is radiated from the initial state. 
The data were collected with the \babar~ detector at the SLAC National
Accelerator Laboratory.
 About 1200, 2600, and 6000 events, respectively, are
selected from a data sample corresponding to an integrated
luminosity of 469\invfb.
The invariant mass of the hadronic final state defines the effective \epem
center-of-mass energy. The center-of-mass energies range from threshold to 4.5\gev.
From the mass spectra, 
the first ever measurements of
the $\epem\to K^+K^-\pi^0\pi^0\pi^0$,  $\epem\to
K^0_SK^{\pm}\pi^{\mp}\pi^0\pi^0$, and
$\epem\to K^0_SK^{\pm}\pi^{\mp}\pi^+\pi^-$
cross sections are  performed.
The contributions from the
intermediate  states that include $\eta$, $\phi$, $\rho$, $K^*(892)$, and
other resonances are presented. 
We observe the $J/\psi$ and $\psi(2S)$ in most of these final states and
measure the
corresponding branching fractions, many of them for the first time.
\end{abstract}

% PACS

\vfill
%\newpage
\maketitle

% reset footnote counter
\setcounter{footnote}{0}

%\begin{linenumbers}

% The body of the paper starts here
\section{Introduction}
\label{sec:Introduction}

Many precision Standard Model (SM) predictions require taking into account
the hadronic vacuum polarization (HVP) terms.
At a relatively large
momentum transfer, these terms are measured by studying the inclusive
hadronic production in  \epem annihilation and are relatively well calculated by pQCD.
However, in the energy region from the hadronic threshold to about 2 GeV,
the inclusive hadronic cross section cannot be measured or calculated 
reliably, and a sum of exclusive states must be used. It is particularly
important for the calculation of the muon anomalous
magnetic moment ($g_\mu-2$), which is most sensitive to the low-energy
region. Despite the large data set of \epem cross sections accumulated in the past years, and the
studies performed~\cite{dehz,theoryg2}, there is still  a discrepancy
between the SM calculation  and the experimental ($g_\mu-2$)
value. With the latest result of the ($g_\mu-2$)
experiment at Fermilab~\cite{fermilab}, this discrepancy increased to 4.2 sigma.

Electron-positron annihilation events with initial-state radiation
(ISR) can be used to study processes over a wide range of energies
below the nominal \epem center-of-mass (c.m.) energy (\Ecm),
as proposed in Ref.~\cite{baier}.
The possibility of exploiting ISR to make precise measurements of
low-energy cross sections at high-luminosity $\phi$ and $B$ factories
is discussed in Refs.~\cite{arbus, kuehn, ivanch},
and motivates the studies described in this paper.
In addition, studies of ISR events at $B$ factories 
are interesting in their own right, because they provide
information on resonance spectroscopy for masses up to the
charmonium region. 

Studies of hadron ($h$) production in the ISR process $\epem \to h\gamma$   have previously
been reported \cite{Druzhinin1,isr2pi,isr2k,isr2p,isr4pi,isr2k2pi,isr6pi,isr3pi,isr5pi,isr2pi2pi0,isr2pi3pi0,isrkkpi,isrkskl,isretapipi,isr4pi3pi0,isr2pi4pi0} by the \babar\ experiment at SLAC.
These studies consider up to seven hadrons with different combinations of
particles in the final state.
Nevertheless not all accessible states have yet been measured.
For  the $\epem\to K\overline K 3\pi$ process only the  
$\epem\to\KK\pipi\piz$
 reaction~\cite{isr5pi} has been studied, which is dominated by the $\phi(1020)\eta$ 
and $\omega\KK$ intermediate states.  Final states with neutral
kaons and/or combinations with two or three neutral pions have not been
measured. These cross sections could have a sizable 
 value below 2\gev; however  they have not yet been
included in the HVP calculation ~\cite{dehz}.  A direct
measurement of these channels can improve the reliability of the HVP calculation.
It is also important to extract the contribution of the intermediate
resonances, because  the 
total cross section calculation depends on
their decay rate to the measured final states.
 
 This paper reports on the \babar\ data analyses of the
$ K^+K^-\pi^0\pi^0\pi^0$,  $\KS\Kpi\pi^0\pi^0$, and
$\KS\Kpi\pi^+\pi^-$
 final states produced in $\epem$ collisions in
 conjunction with an energetic photon, assumed to result from ISR.
While the \babar\ data cover effective c.m.\@ energies up to
10.58\gev, 
this analysis is restricted to 
energies below 4.5\gev to minimize the backgrounds from $\Upsilon(4S)$
decays. 
We extract the 
contributions of  intermediate states, including the  $\phi(1020)$,
$\omega(782)$, $\rho(770)$, $\eta$, and $K^*(892)^{\pm,0}$ resonances, and  present the 
 corresponding cross sections.
Signals for the  $J/\psi$ and $\psi(2S)$ states are observed in most of the studied
intermediate states,
and the corresponding  branching fractions are measured.

\section{\boldmath The \babar\ detector and data set}
\label{sec:babar}

The data used in this analysis were collected with the \babar\ detector at
the \pep2\ asymmetric-energy \epem\ storage ring. 
The total integrated luminosity used is 468.6\invfb~\cite{lumi}, 
which includes data collected at the $\Upsilon(4S)$
resonance (424.7\invfb) and at a c.m.\ energy 40\mev below this
resonance (43.9\invfb). 

The \babar\ detector is described in detail elsewhere~\cite{babar}. 
Charged particles are reconstructed using the \babar\ tracking system,
which is comprised of the silicon vertex tracker (SVT) and the drift
chamber (DCH), both located 
inside a 1.5 T solenoid.
Separation of pions and kaons is accomplished by means of the detector of
internally reflected Cherenkov light (DIRC) and energy-loss measurements in
the SVT and DCH. 
Photons and \KL mesons are detected in the electromagnetic calorimeter
(EMC).
Muon identification is provided by the instrumented flux return (IFR).

The ISR events with detection of the ISR photon in the EMC are characterized by good
reconstruction efficiency and by well understood kinematics,
demonstrated in the above references. 
The \babar~ detector performance
         (tracking, particle identification (PID), \piz, \KS, and
 \KL reconstruction) is well suited to the study of ISR processes.

To evaluate the detector acceptance and efficiency, 
we have developed a special package of Monte Carlo (MC) simulation programs for
radiative processes based on 
the approach of K\"uhn  and Czy\.z~\cite{kuehn2}.  
Multiple collinear soft-photon emission from the initial \epem state 
is implemented with the structure function technique~\cite{kuraev,strfun}, 
while additional photon radiation from final-state particles is
simulated using the PHOTOS package~\cite{PHOTOS}.  
The precision of the radiative simulation is such that it contributes less than 1\% to
the uncertainty in the measured hadronic cross sections.

We simulate $\epem\to K^+K^-\ppz\piz\gamma$ events assuming production
through the $\phi(1020)\eta$ intermediate channel,
with decay of  the $\phi$ to charged kaons and
decay of the $\eta$ to all  its measured decay modes~\cite{PDG}.
For the $\epem\to \KS\Kpi\ppz\gamma$ and $\epem\to \KS\Kpi\pipi\gamma$ reactions we use a phase space model for
the hadronic states. As was shown in our previous studies,  events
with  hard ISR photon detection are characterized by a weak model
dependence, which does not exceed 5\%,
in the efficiency calculation.

A sample of about 300\,000 simulated events is generated  for 
each reaction  and is
processed through the detector response simulation, based on the GEANT4 package~\cite{GEANT4}. These
events are reconstructed using
the same software chain as the data. 
Most of the experimental events contain additional soft photons due to machine background
or interactions in the detector material. Variations in the detector
and background conditions are included in the simulation.

For the purpose of background estimation,  large samples of events from the
main relevant ISR processes ($5\pi\gamma$, $\omega\ppz\gamma$, 
$K^+K^-\ppz\gamma$, $\KS\Kpi\piz\gamma$) are simulated.  
To evaluate the background from the relevant
 non-ISR processes, namely $\epem\to\qqbar$ $(q=u, d, s)$ and
 $\epem\to\tau^+\tau^-$, 
 simulated samples with integrated luminosities similar
that of the data are generated 
using the \textsc{jetset}~\cite{jetset}
and \textsc{koralb}~\cite{koralb} programs,
respectively.
The cross sections for the above processes
are  known with an accuracy slightly better than
10\%, which is sufficient for the present purpose.

\begin{figure*}[tbh]
\begin{center}
\includegraphics[width=0.34\linewidth]{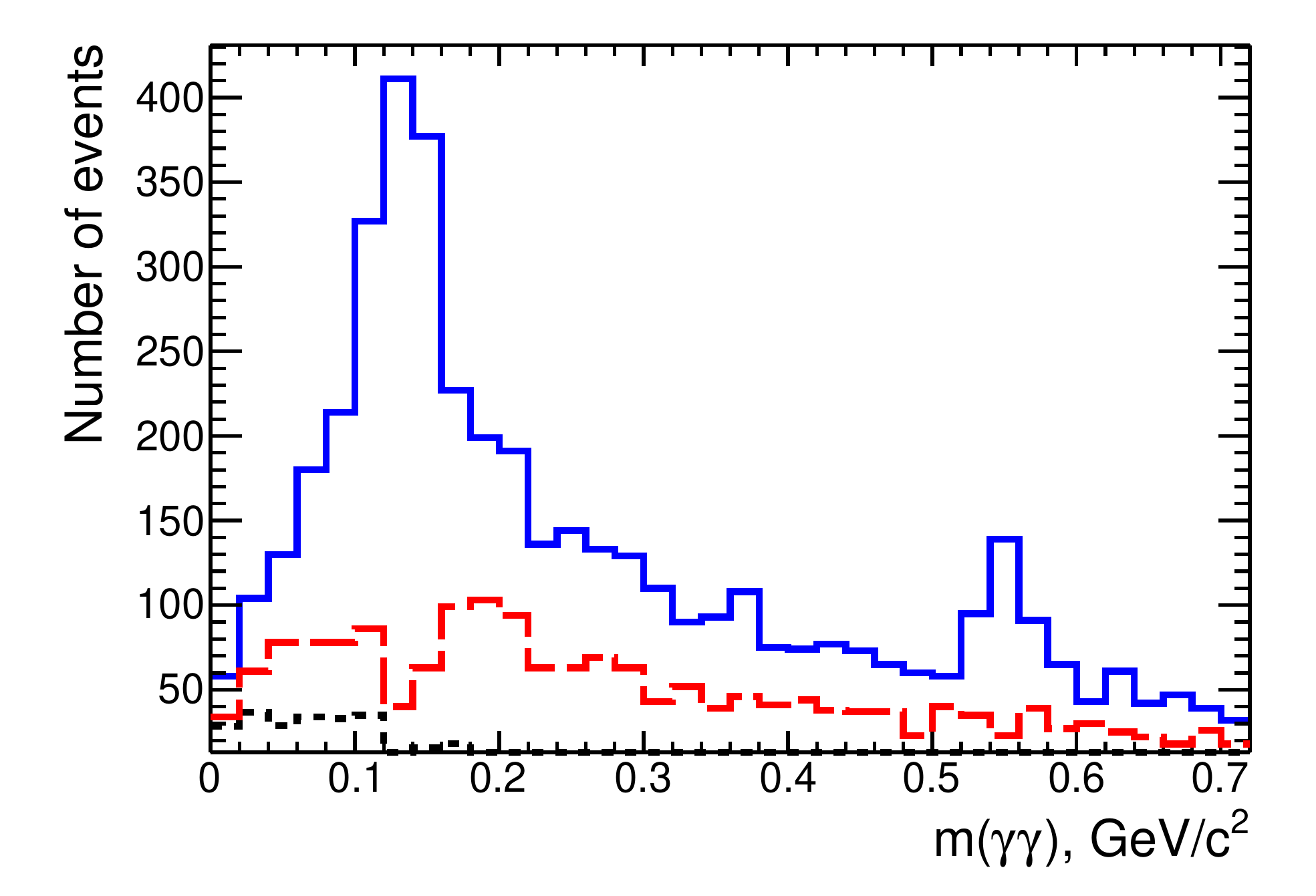}
\put(-50,90){\makebox(0,0)[lb]{\bf(a)}}
\includegraphics[width=0.34\linewidth]{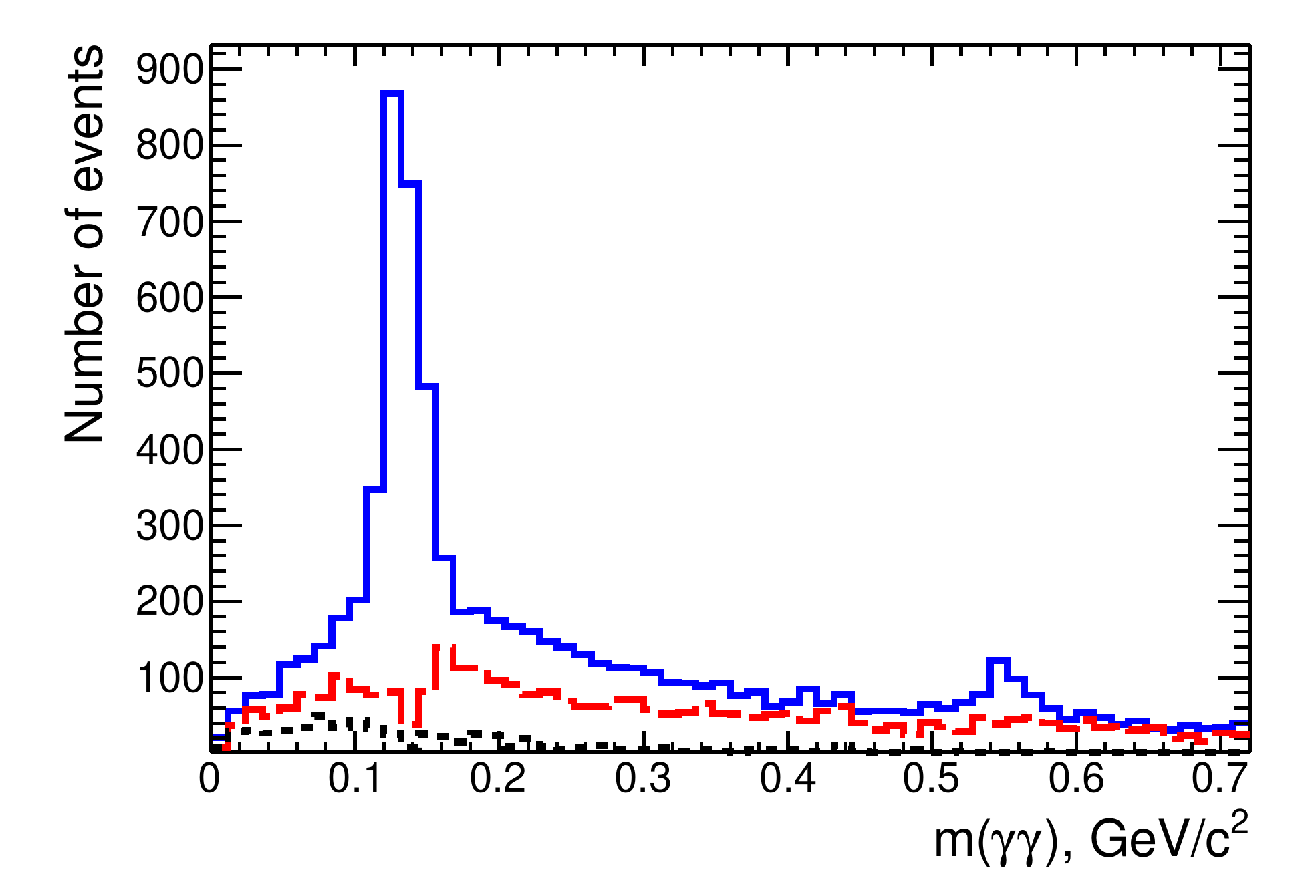}
\put(-50,90){\makebox(0,0)[lb]{\bf(b)}}
\includegraphics[width=0.34\linewidth]{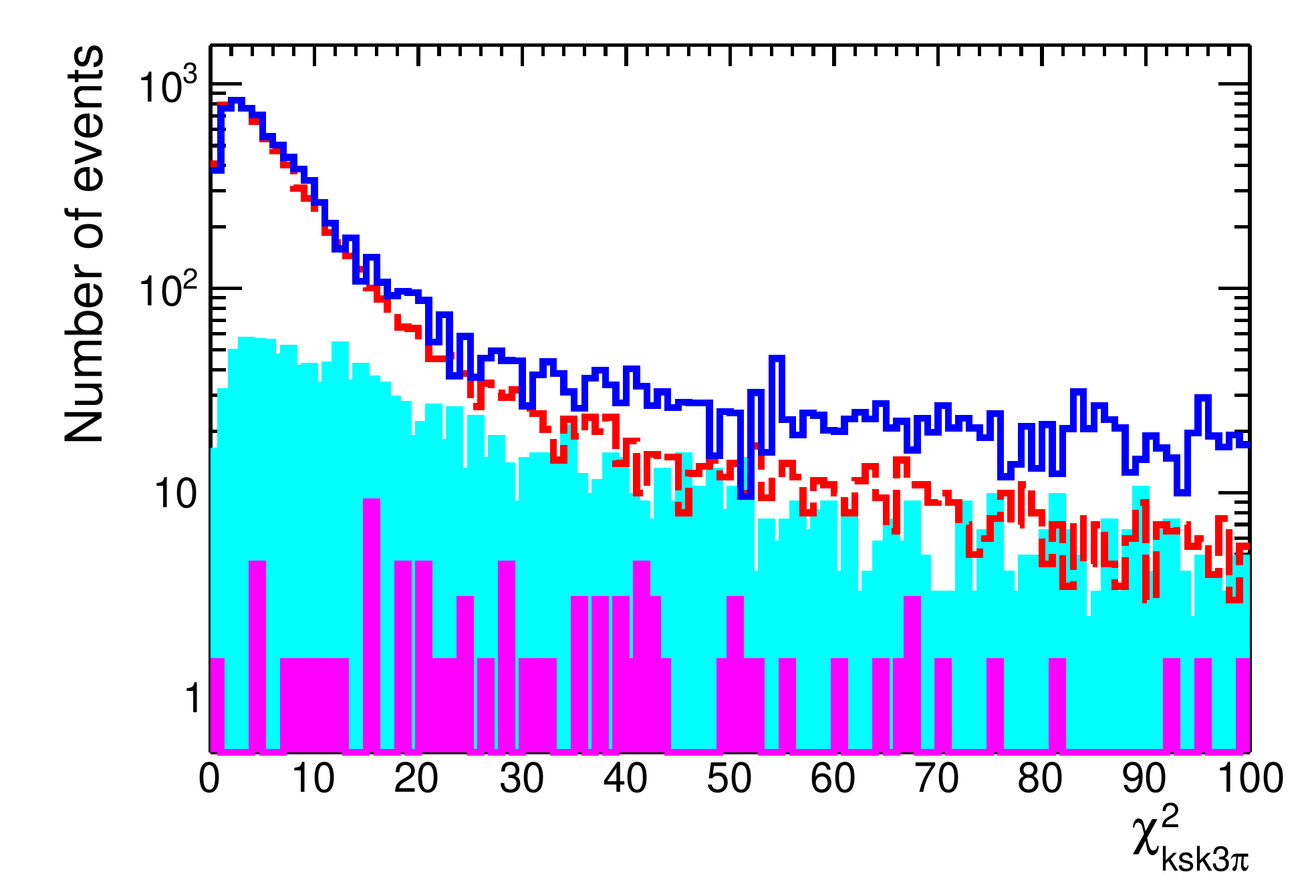}
\put(-50,90){\makebox(0,0)[lb]{\bf(c)}}
\vspace{-0.5cm}
\caption{
 The $m(\gamma\gamma)$ invariant mass for the third photon pair (a)
  in the $\epem\to\KK \ppz\gamma\gamma\gamma_{ISR}$ hypothesis
  and (b) in the $\epem\to\KS\Kpi\piz\gamma\gamma\gamma_{ISR}$ hypothesis
  for the  $\chi^2$ signal regions.
  The dashed histograms are for events from the $\chi^2$ control  regions. The
  dotted histograms are for the remaining  $\epem\to\KK
  \ppz\gamma_{ISR}$  or  $\epem\to\KS\Kpi\piz\gamma_{ISR}$
  backgrounds from simulation;
(c) The $\chi^2_{\KS K3\pi}$ distribution in the $\epem\to\KS\Kpi\pipi\gamma_{ISR}$ hypothesis  for data (solid histogram) and
for the MC simulation (dashed). The shaded histogram is for the $uds$
backgrounds and single bins are for the remaining contribution from the
$\epem\to\KS\Kpi\gamma_{ISR}$ process.
}
\label{mgg_all}
\end{center}
\end{figure*}

\section{\boldmath Event Selection and Kinematic Fit}
\label{sec:Fits}

A relatively clean sample of  ISR-related events is selected by 
requiring that there be charged tracks reconstructed in the DCH,
SVT, or both, and some number of  photons (sometimes up to 20), with an energy above
0.02\gev in the EMC.
We  assume the photon with the highest energy to be the
 ISR photon, and we require its c.m. energy to be larger
 than 3\gev.

The event selections and procedures are  based on the methods
described in our previous analyses for the
$\epem\to\pipi\ppz\piz$~\cite{isr2pi3pi0} and $\epem\to
\KS\KS\pipi$~\cite{isrkskl} channels. 

We require either exactly two or exactly four tracks in an event
with zero total charge that extrapolate to
within 0.25 cm of the beam axis and 3.0 cm 
of the nominal collision point along that axis.
If there are two such tracks
we require either that both be identified as kaons or, if
only one is identified as a kaon, that a $\KS$ candidate
be present. We detect \KS using  $\KS\to\pipi$ decays with pions not
from the collision region,  and require
the decay point  to be within 0.2 to 40 cm from the collision point.
If there are  four  tracks from the collision region we require one of them to
be identified as a kaon and require the presence of a $\KS$ candidate.
We also allow the presence of one extra track to capture the
relatively small fraction of signal events that contain a
background track.
The tracks that satisfy the extrapolation criteria to the collision
region are fit to a vertex, which is used as the point of origin
in the calculation of the photon direction.

We subject each candidate event to a set of constrained 
kinematic fits and use the fit results,
along with charged-particle identification,
to select the final states of interest and evaluate backgrounds
from other processes.
 The kinematic fits make use of the
four-momenta and covariance matrices of the initial $e^+$, $e^-$, and the
set of selected tracks, \KS candidates, and photons.
The fitted three-momenta of each track, \KS, and photon are then used 
in further calculations.

Excluding the photon with the highest c.m.\  energy,
which is assumed to arise from ISR, we consider all independent sets of
six (four) other photons, and   combine them into three (two) pairs.
For each set of six (four) photons, we test all
possible independent combinations of three (two) photon pairs.
 For the next stage we select those combinations in which the di-photon mass
 of at least two (one) pairs lies within $\pm$35\mevcc ($\pm 3\sigma$ of
 the resolution) of the \piz mass, $m_{\piz}$~\cite{PDG}.
 
The selected combinations are subjected to a fit
 in which the di-photon masses of the two (one) pairs with
$|m(\gamma\gamma)-m_{\pi^0}|<35\mevcc$
 are constrained to $m_{\piz}$.
For  the signal hypothesis,
$\epem\to K^+K^- \ppz\gamma\gamma\gamma_{ISR}$, with the constraints
due to four-momentum conservation, there are thus six
constraints (6C) in the fit. For the $\epem\to\KS\Kpi\piz\gamma\gamma\gamma_{ISR}$ hypothesis there are  five
constraints (5C) in the fit. For the
$\epem\to\KS\Kpi\pipi\gamma_{ISR}$ hypothesis we use
the 4C fit with only four-momentum constraints.
The photons in the remaining (``third'' or ``second'') pair are treated as being independent.
If all three (two) photon pairs in the combination
 satisfy $|m(\gamma\gamma)-m_{\pi^0}|<35\mevcc$,
 we rotate the combinations,
  allowing each of the  di-photon pairs in turn
 to be the third (second) pair, i.e., the pair without the
 $m_{\piz}$ constraint.
The combination with the smallest \chisq is retained, along with the obtained
$\chisq_{2K2\piz\gamma\gamma}$ ($\chisq_{6C}$),
$\chisq_{\KS K\pi\piz\gamma\gamma}$ ($\chisq_{5C}$), and
$\chisq_{\KS K3\pi}$ ($\chisq_{4C}$)  values and the fitted three-momenta of each
particle and photon.

Each retained event is also subjected to a 6C (5C) fit under the
$\epem\to\KK \ppz\gamma_{ISR}$ ($\epem\to\KS\Kpi\piz\gamma_{ISR}$) background hypothesis, and the smallest
 values of $\chisq_{2K2\piz}$ and $\chisq_{\KS K\pi\piz}$  from all photon combinations are retained.  
These processes have a comparable
cross section to the  signal processes and can
contribute to the background when two or more background photons are present.

\section{Additional selection criteria}
\label{selections}

The results of the kinematic fits      
are used to perform the final selection of  signal events. 
We require the tracks to lie within the fiducial
region of the DCH (0.45-2.40 radians) and to be
inconsistent with being a  muon.
The photon candidates are required to lie within the
fiducial region of the EMC (0.35--2.40 radians) and to have
an energy larger than 0.035 GeV.
A requirement that there be no charged
tracks within 1 radian of the ISR photon reduces the $\tau^+\tau^-$ background
to a negligible level. 
A requirement that any extra photons in an event each
have an energy below 0.7 GeV slightly reduces the multi-photon
background. 
We use the  \chisq values for the signal selection and
for the background evaluation.

We require $\chisq_{2K2\piz\gamma\gamma}< 65$ to select the signal for
the $\KK\ppz\piz$ events, and use events in the control region,  $65 < \chisq_{2K2\piz\gamma\gamma}< 130$, for the
background estimate.
We apply a $\chi_{2K2\piz}^2 >30$ condition if these events also
satisfy  the $\KK \ppz$ background hypothesis. 
This requirement  reduces the contamination due to    
 $\KK \ppz$ events from ~30\% to about 1\%--2\% while reducing
 the signal efficiency by only 5\%.
 Figure~\ref{mgg_all} (a) shows the invariant mass $\mgg$ of the third
photon pair for the $\epem\to\KK\ppz\gamma\gamma\gamma_{ISR}$ hypothesis
for the signal  and control  regions of $\chisq_{2K2\piz\gamma\gamma}$. Clear $\piz$
and $\eta$ peaks are visible as well as a relatively smooth
background, exceeding the  level of events from the \chisq control
region but with a similar shape. Because of the constraint to the best 
photon pairs, the third photon pair is
sometimes formed from photon candidates that are less
well measured and have a dip in the distribution, explained in Ref.~\cite{isr2pi3pi0}.

Figure~\ref{mgg_all} (b)
shows the $\mgg$ distribution  after the $\chisq_{\KS K\pi\piz\gamma\gamma}< 70$ 
requirement has been applied in the $\epem\to\KS
\Kpi\piz\gamma\gamma\gamma_{ISR}$ hypothesis (solid histogram).
The dashed histogram is for the events in the $70 <
\chisq_{\KS K\pi\piz\gamma\gamma}< 140$  control region, while the dotted
histogram is for a remaining $\KS K^{\pm}\pi^{\mp}\piz$ background
estimated from the simulation.

\begin{figure*}[tbh]
\begin{center}
\includegraphics[width=0.34\linewidth]{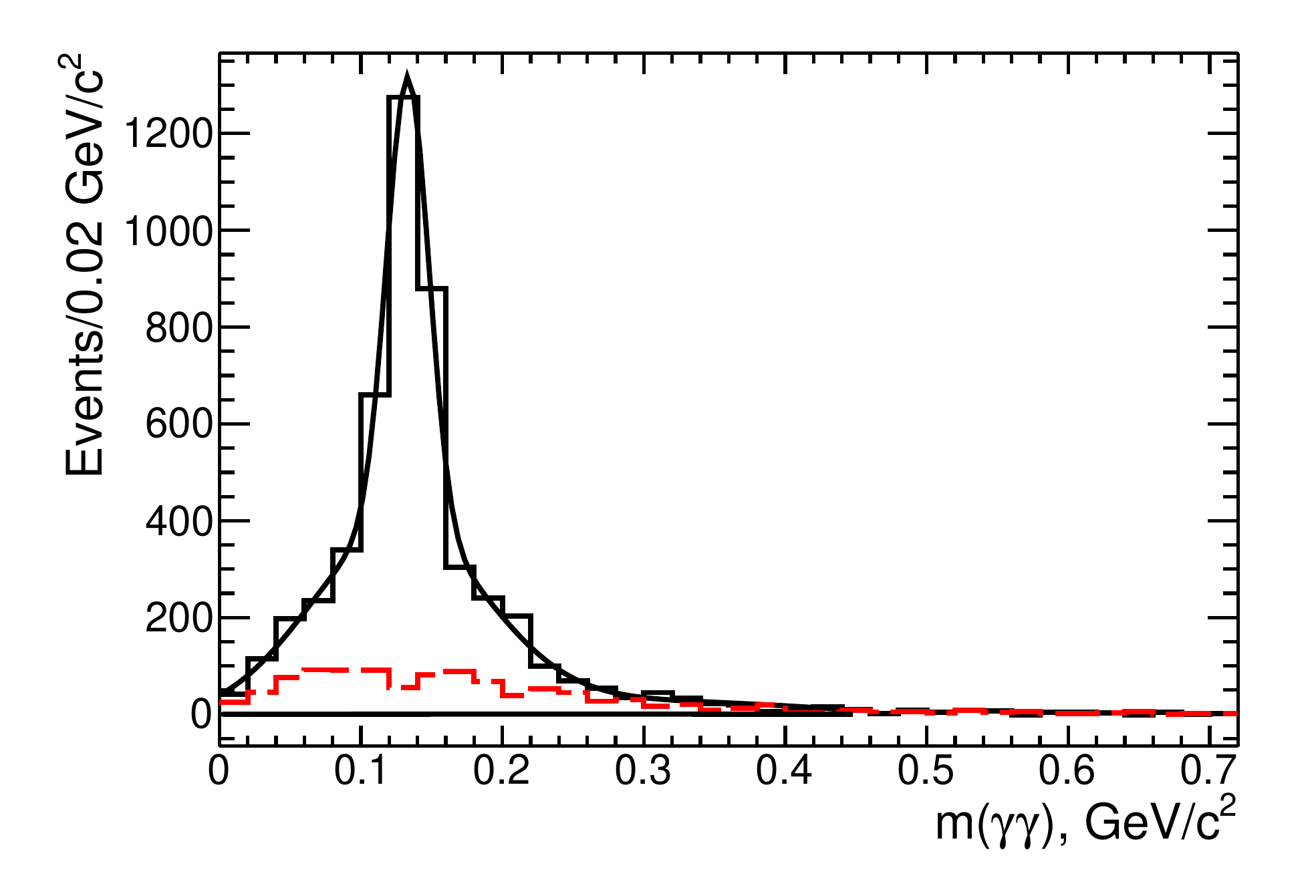}
\put(-50,90){\makebox(0,0)[lb]{\bf(a)}}
\includegraphics[width=0.34\linewidth]{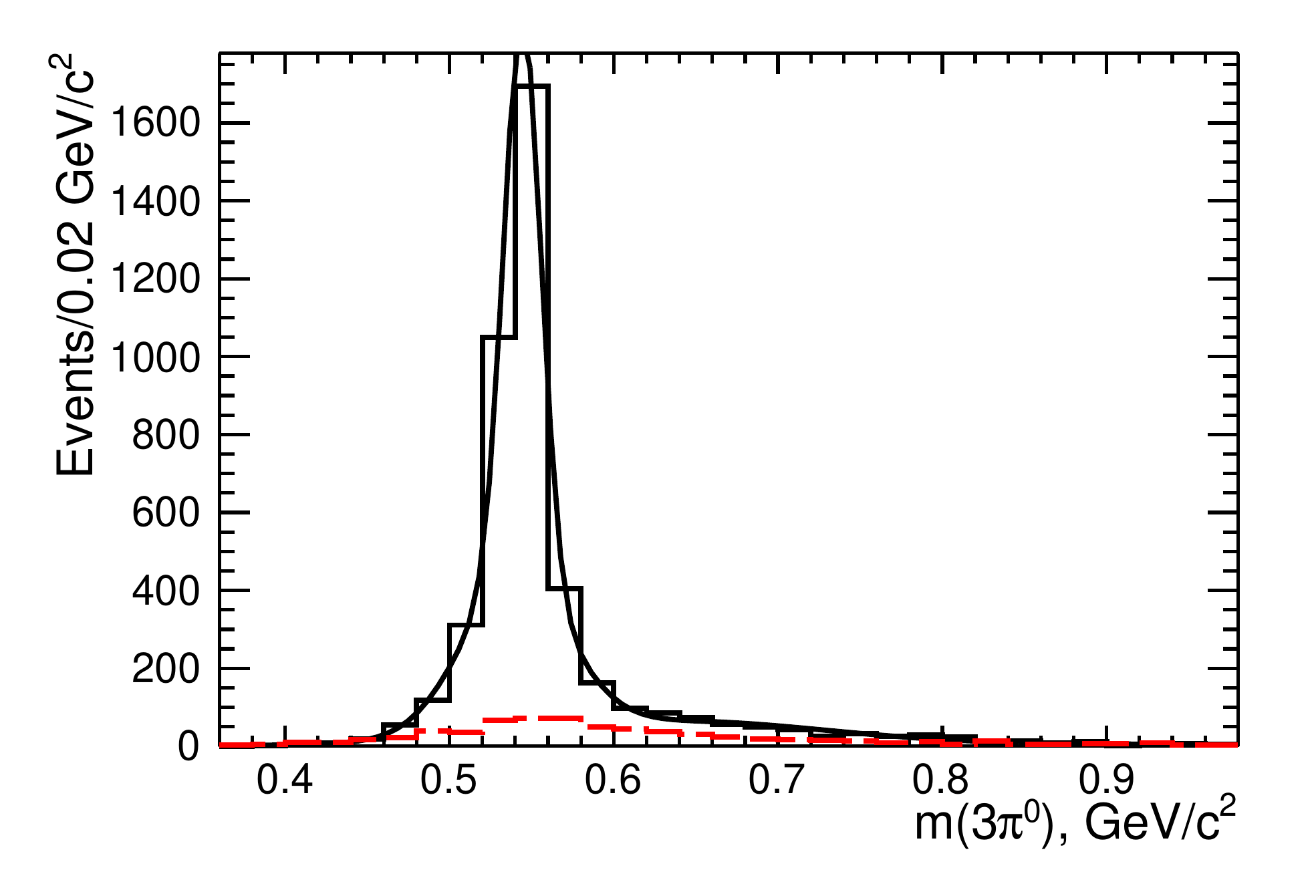}
\put(-50,90){\makebox(0,0)[lb]{\bf(b)}}
\includegraphics[width=0.34\linewidth]{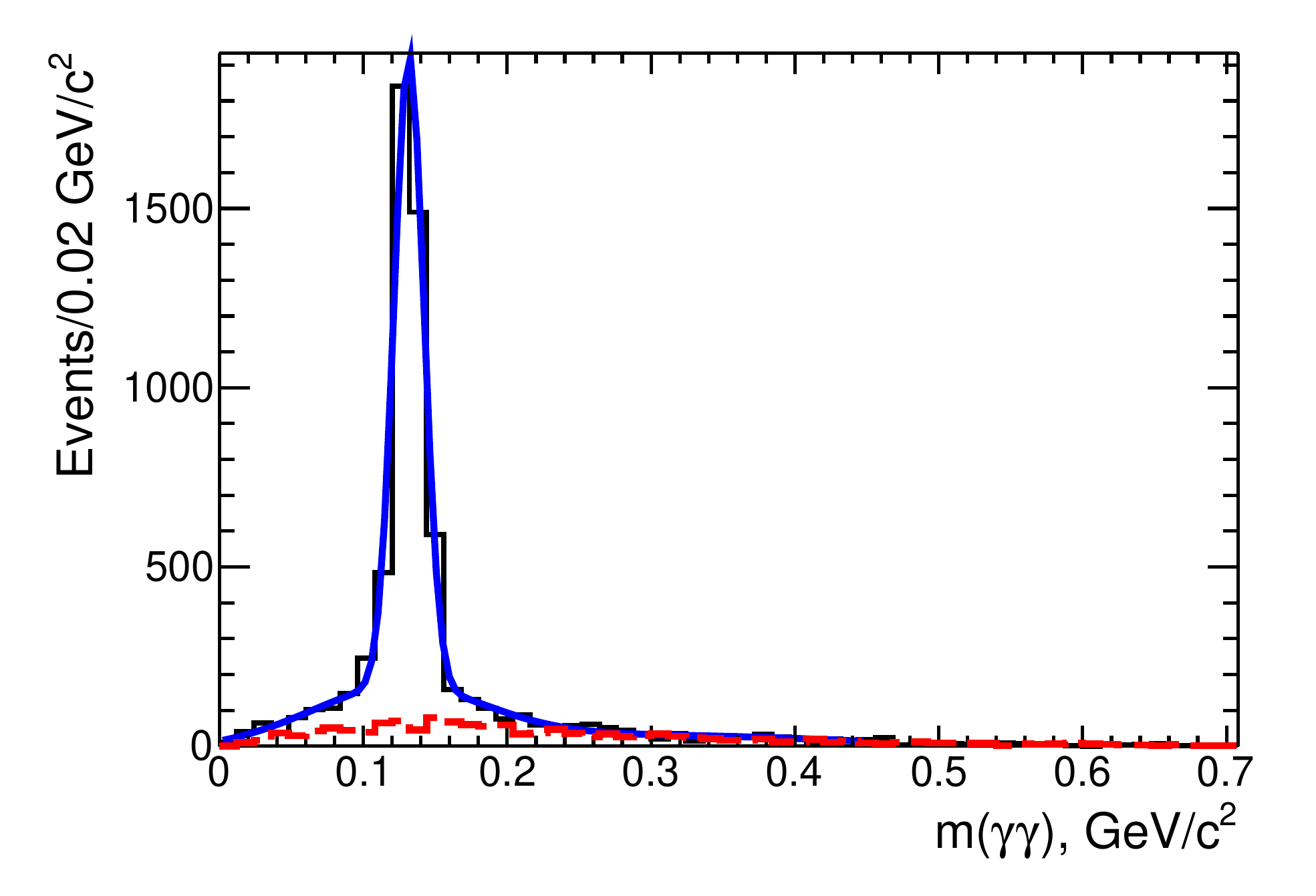}
\put(-50,90){\makebox(0,0)[lb]{\bf(c)}}
\vspace{-0.5cm}
\caption{
(a) The third photon pair \mgg invariant mass for the MC simulated
$\phi\eta$ events after applied selections and background subtraction (solid histogram).  The dashed histogram shows the
contribution of the subtracted events from the  \chisq control
region. The curve is for the three-Gaussian fit
to the $\piz$ signal;
(b) The background-subtracted $\ppz\piz$ invariant mass distribution (solid histogram) for
the MC simulated
$\phi\eta$ final state, and the events from the \chisq control region
 (dashed histogram).
The solid curve is fit to  the $\eta$ signal;
(c) The background-subtracted second photon pair invariant mass (solid histogram) for the $\KS\Kpi \ppz$
final state with the three-Gaussian fit function. The dashed histogram
shows the contribution from the \chisq control region.
}
\label{mgg_mc}
\end{center}
\end{figure*}
\begin{figure*}[tbh]
  \begin{center}
    \vspace{-0.5cm}
\includegraphics[width=0.34\linewidth]{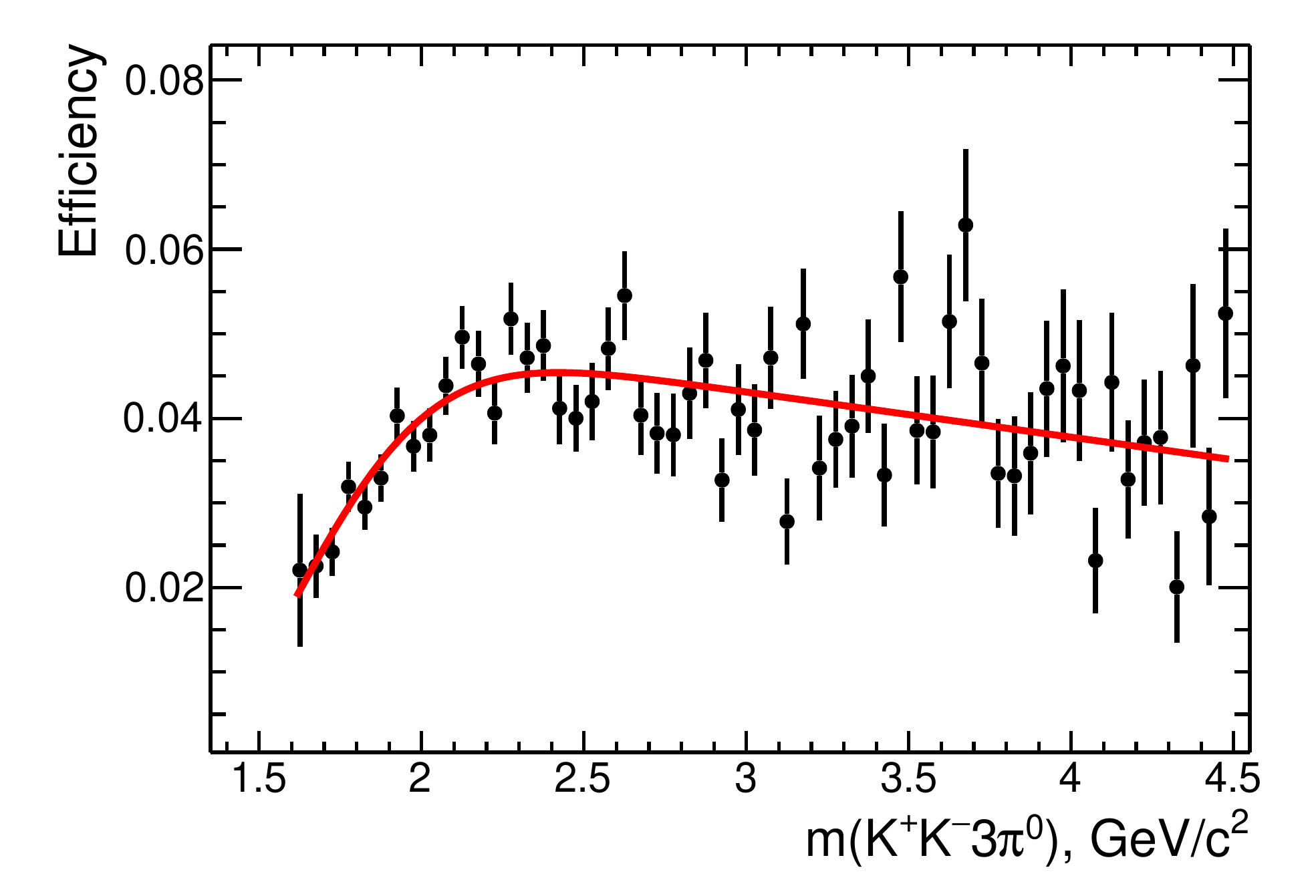}
\put(-135,90){\makebox(0,0)[lb]{\bf(a)}}
\includegraphics[width=0.34\linewidth]{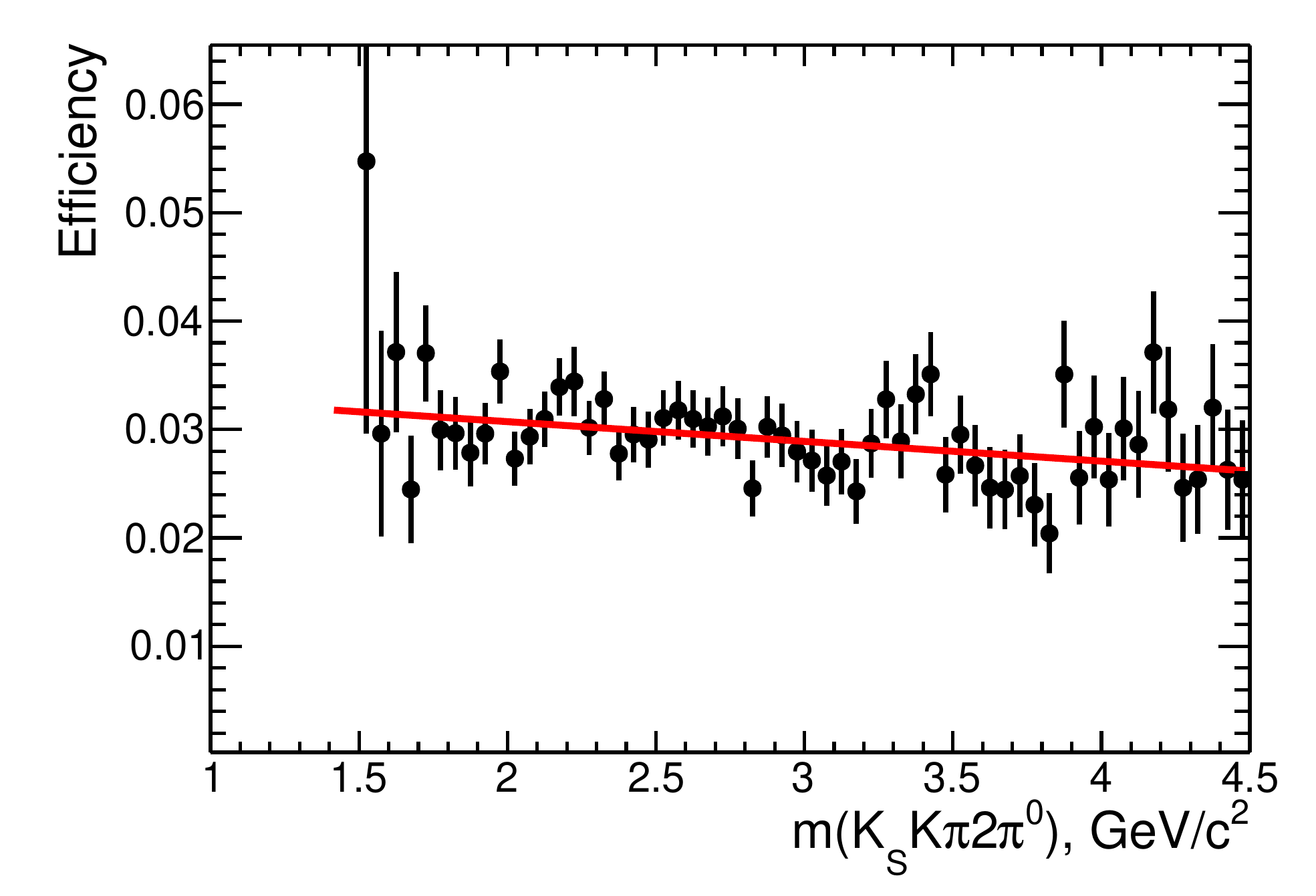}
\put(-50,90){\makebox(0,0)[lb]{\bf(b)}}
\includegraphics[width=0.34\linewidth]{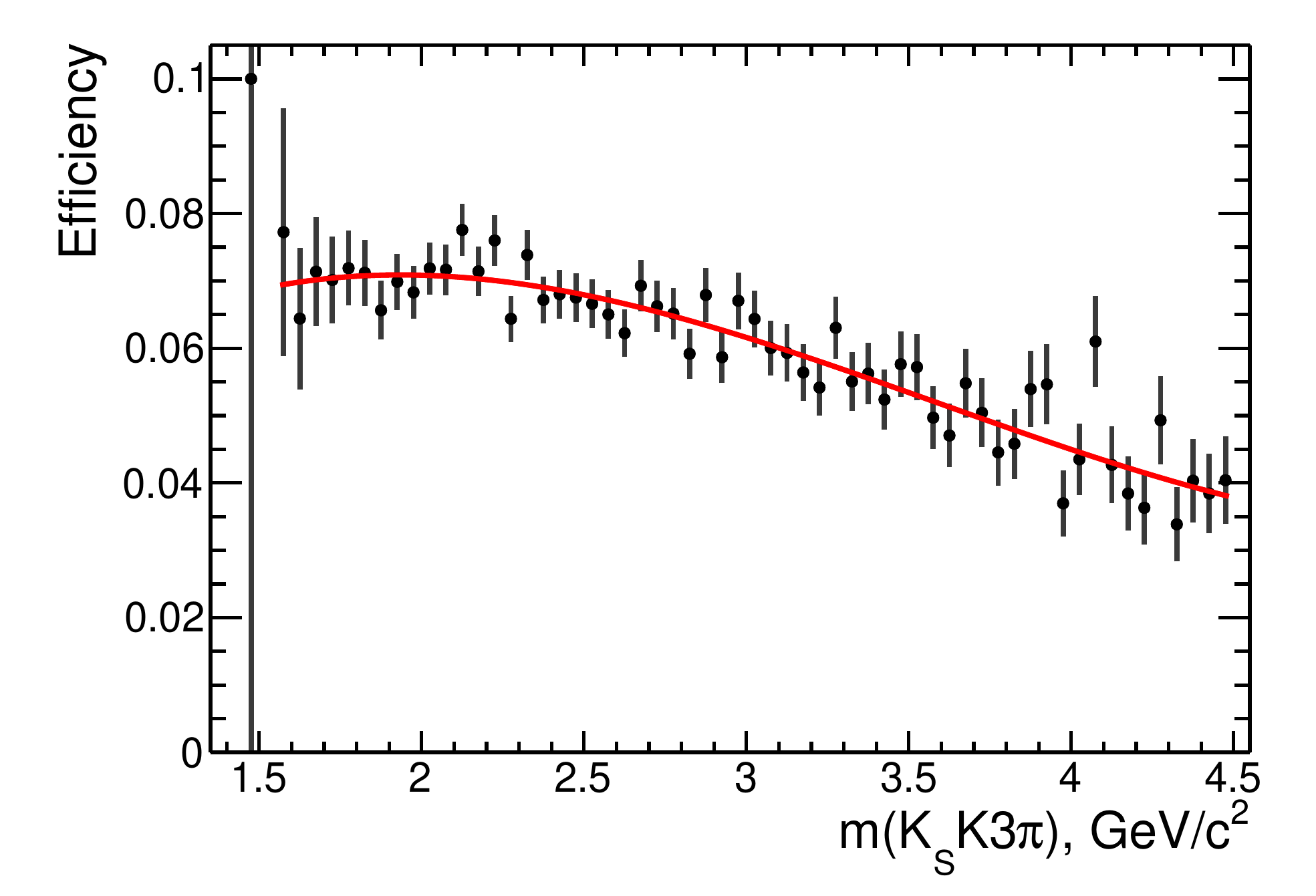}
\put(-50,90){\makebox(0,0)[lb]{\bf(c)}}
\vspace{-0.5cm}
\caption{ The hadronic invariant-mass-dependent reconstruction efficiency for (a)
  the  $\epem\to\KK\ppz\piz\gamma_{ISR}$ events, (b) the $\epem\to\KS\Kpi \ppz\gamma_{ISR}$
  events, and (c) for the $\epem\to\KS\Kpi\pipi\gamma_{ISR}$ events.
The curves show the fit results, which are used in the cross
section calculation.
}
\label{mc_eff}
\end{center}
\end{figure*}

Our strategy to extract the signals for the $\epem\to\KK\ppz\piz$
and $\KS\Kpi\ppz$ processes 
is to perform a fit to the $\piz$  yields
in intervals of 0.05\gevcc in the distributions of
the  invariant masses $m(\KK2\piz\gamma\gamma)$ and $m(K_S
\Kpi\piz\gamma\gamma)$. The procedure is described in detail in Ref.~\cite{isr2pi3pi0}.

For the $\epem\to\KS\Kpi\pipi\gamma_{ISR}$ process we use the
procedure described in Ref.~\cite{isrkskl}, based on the \chisq
distribution study. The signal events are
selected by the requirement $\chisq_{\KS K3\pi}<40$ while the events in the
control region, $40<\chisq_{\KS K3\pi}<80$,
are used for the background evaluation.
Figure~\ref{mgg_all}(c) shows the $\chisq_{\KS K3\pi}$  
distribution for data (solid histogram) in comparison with the simulation
(dashed), normalized to the first five bins where the contribution from
the background is small. The non-ISR background is
shown by the shaded histogram (see next section), while the remaning
$\KS\Kpi$ events are at a negligible level (individual (pink)
bins). 

\section{\boldmath Detection efficiency}\label{sec:efficiency}
\subsection{Number of signal events in simulation}

The selection procedure applied to the data is also applied to the
MC-simulated events.  Figure~\ref{mgg_mc}
shows the two-photon and three-pion invariant mass distributions, which are  used to extract the number of signal events.

The $\piz$ signal for the simulation is 
not Gaussian because of the photon pair selection, which was described  in the previous section.
It also includes a combinatoric
background arising from the combination of background photons, included in the
simulation, with the photons from the signal reactions. This
combinatoric background can be subtracted using events from the \chisq
control region, shown by the dashed histogram in Fig.~\ref{mgg_mc}(a).
The solid histogram in Fig.~\ref{mgg_mc}(a) corresponds to the
two-photon mass distribution obtained from the \chisq signal region
for MC simulation of the $\phi\eta$ final state after the combinatoric background
subtraction. The  background  is subtracted assuming a scale
factor, which is varied to estimate the uncertainty in its
contribution.
The signal yield is then extracted by fitting the \piz\ peak of this
distribution with a sum of three Gaussian functions for the signal plus a
second-order polynomial function to account for a residual combinatoric background.
If a scale factor 1.5  is used, the background level becomes
negligible, and we can determine and fix all parameters for the signal function.
If then  we change the scale factor to 1.0 or to 0.0 in the fit, the fitted
signal yield does not change by more than 3\%.
The result, for a scale factor of 1.0, is shown by the smooth solid
curve in Fig.~\ref{mgg_mc}(a).
%The fitted signal yields $2639\pm66$ events.
We apply a similar fitting procedure in every
0.05\gevcc interval of the $m(\KK 3\piz)$ invariant mass
distribution.

As a cross check, for
the $\phi\eta$ events, we determine the number of
events by fitting the
$\eta$ signal from  $\eta\to\ppz\piz$ decay: the simulated
distribution is shown in Fig.~\ref{mgg_mc}(b) after combinatoric
background subtraction with a scale factor 1.0. 
 The fit functions are the sum of three
 Gaussian functions and a polynomial for the combinatoric background.
 No difference in the number of events is observed.
%This fit yields $2569\pm79$ events in total.
The $\KK \ppz\piz$
mass distribution is also obtained 
in every 0.05\gevcc interval.

The same approach is used for the $\KS\Kpi\ppz$ final
state. Figure~\ref{mgg_mc}(c) shows the second photon pair invariant
mass distribution after combinatorial background subtraction with a
scale factor of 1.0. The dashed
histogram shows the level of this background. The
curve shows the fit function used to determine the number of events.

For the $\KS K\pi\pipi$ final state we use simulated events in the
\chisq signal region for the efficiency evaluation. There is no
combinatorial background for this final state.

\begin{figure*}[tbh]
\begin{center}
\includegraphics[width=0.34\linewidth]{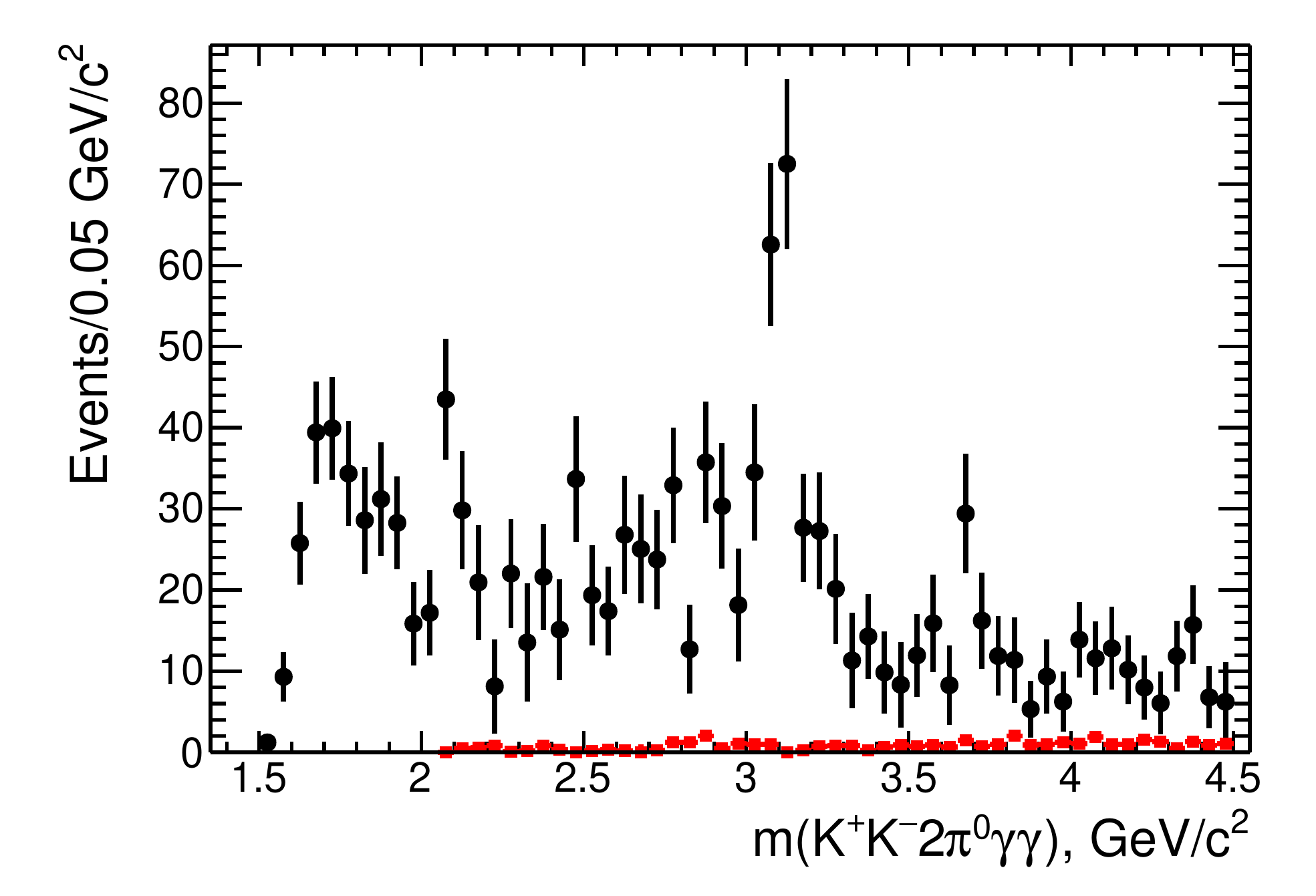}
\put(-50,90){\makebox(0,0)[lb]{\bf(a)}}
\includegraphics[width=0.34\linewidth]{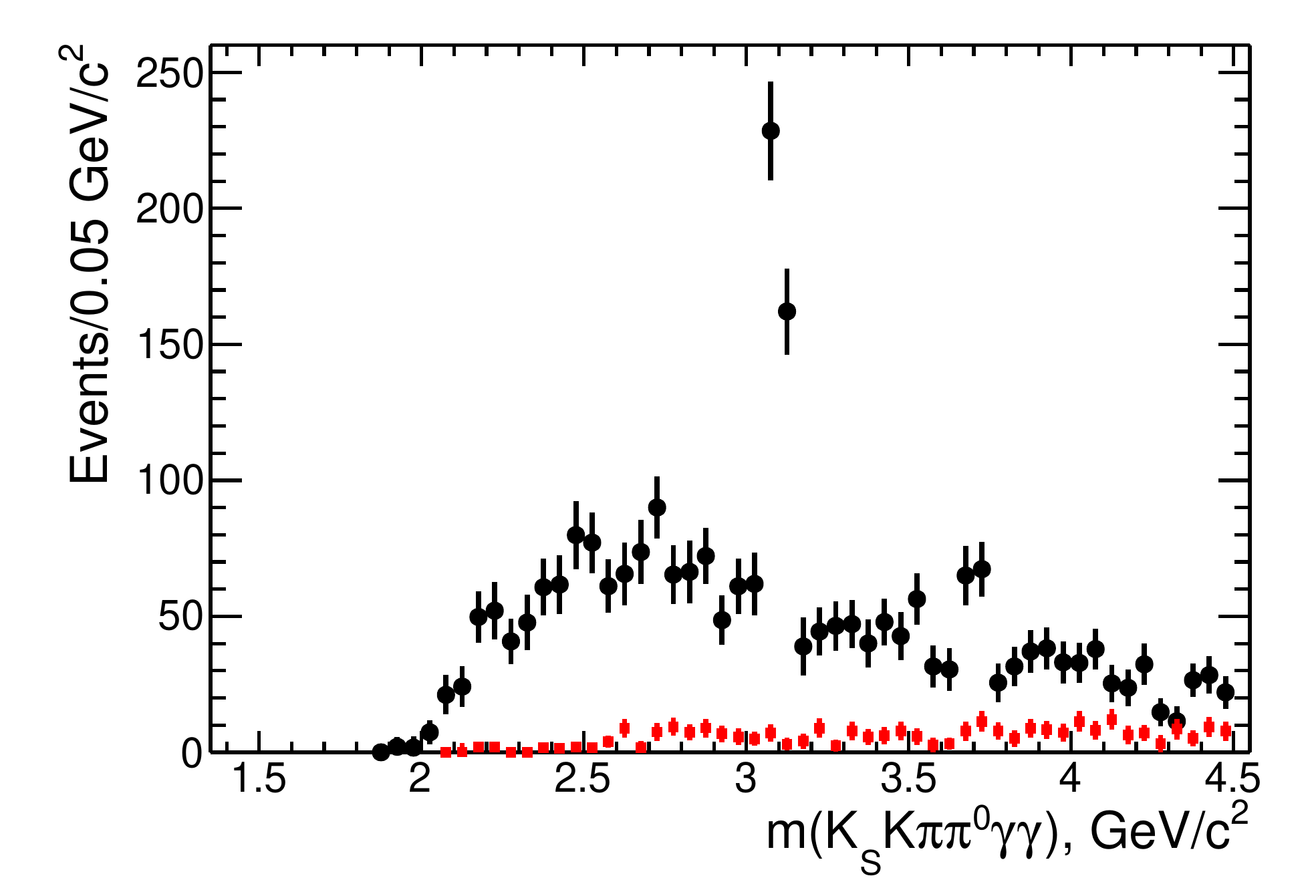}
\put(-50,90){\makebox(0,0)[lb]{\bf(b)}}
\includegraphics[width=0.34\linewidth]{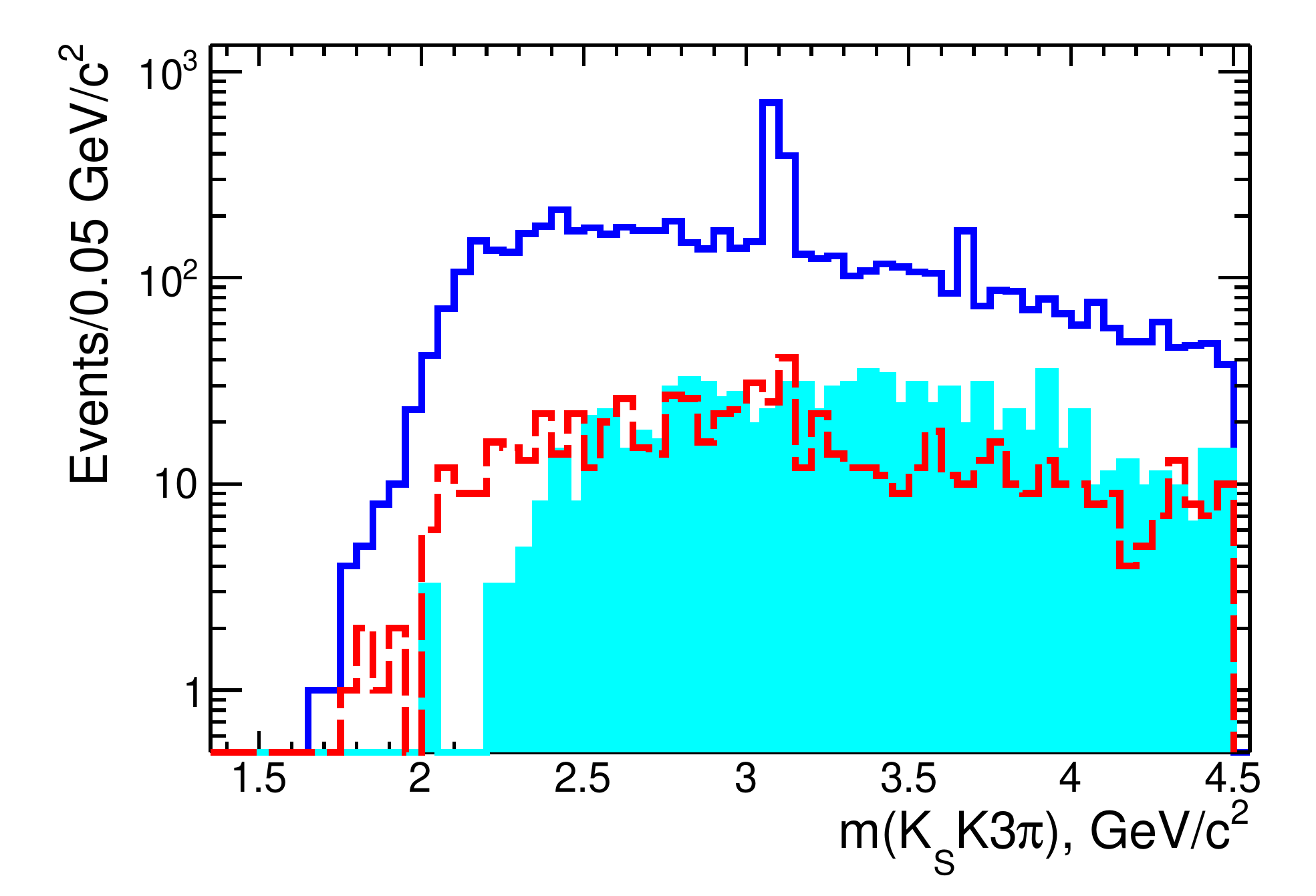}
\put(-50,90){\makebox(0,0)[lb]{\bf(c)}}
\vspace{-0.5cm}
\caption{ The number of events determined from the $\piz$ fit vs the hadronic invariant mass
 for (a)  the  $\epem\to\KK\ppz\piz$  and (b)  $\epem\to\KS K\pi \ppz$
  reactions. The contributions from  $uds$ events are shown by
 (red) squares.
 (c) The number of $\epem\to\KS K\pi\pipi$ events in the \chisq
 signal (solid histogram) and control (dashed) regions. The shaded
 histogram shows the contribution from the $uds$ events.
}
\label{nevents_data}
\end{center}
\end{figure*}

\subsection{Efficiency evaluation}
The mass-dependent  detection
efficiency is obtained by dividing the number of fitted MC
events in each 0.05\gevcc mass interval by the number generated in          
the same interval.
We determine that the total efficiency does not change by more than 
5\% because of variations of the functions used to extract the number
of events or the use of different background subtraction procedures. This
value is taken as an estimate of the systematic uncertainty in the efficiency
associated with the simulation model used and with the fit procedure.
We obtain the  efficiency in each 0.05\gevcc mass interval for
the $\KK \ppz\piz$ final state and
fit the result with a third-order polynomial function, shown in
Fig.~\ref{mc_eff}(a). Although the signal simulation accounts for all
$\eta$ decay modes, the efficiency calculation
considers only the $\eta\to\ppz\piz$ decay mode.
From Fig.~\ref{mc_eff}(b) it is seen that the reconstruction
efficiency for the $\KS K\pi \ppz$ final state is about 3\%, roughly independent of mass.
The result of the linear fit is used for the cross section
calculation. Figure~\ref{mc_eff}(c) shows the detection efficiency for
the $\KS K\pi\pipi$ final state with the fit function used for the
cross section calculation. 

This efficiency estimate takes into account the geometrical acceptance of the detector 
for the final-state photons and the charged pions and kaons, the inefficiency of 
the  detector subsystems, and  the event loss due to additional
soft-photon  emission from the initial and final states.
Corrections to the efficiency that account for data-MC differences are discussed below.

\begin{figure*}[tbh]
\begin{center}
\includegraphics[width=0.34\linewidth]{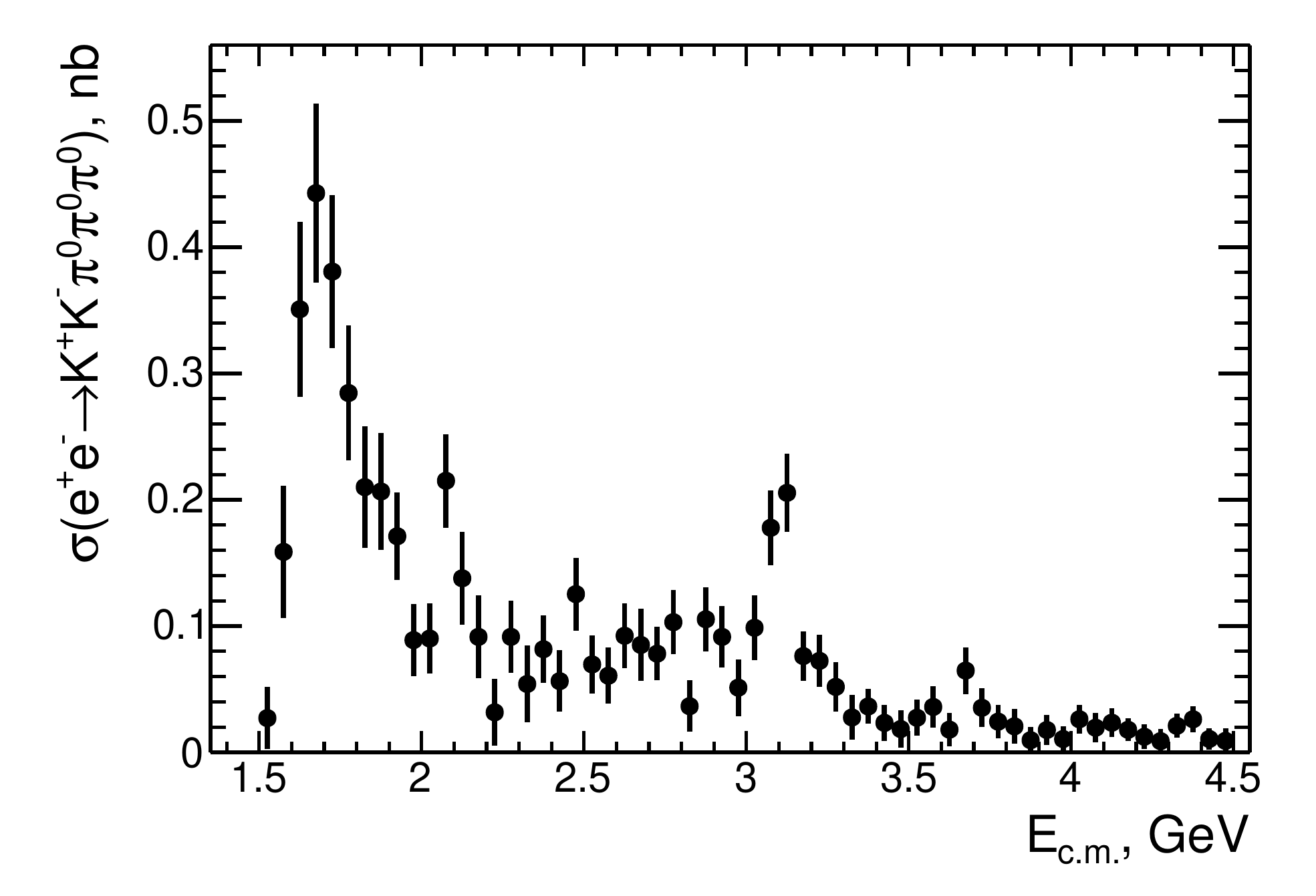}
\put(-50,90){\makebox(0,0)[lb]{\bf(a)}}
\includegraphics[width=0.34\linewidth]{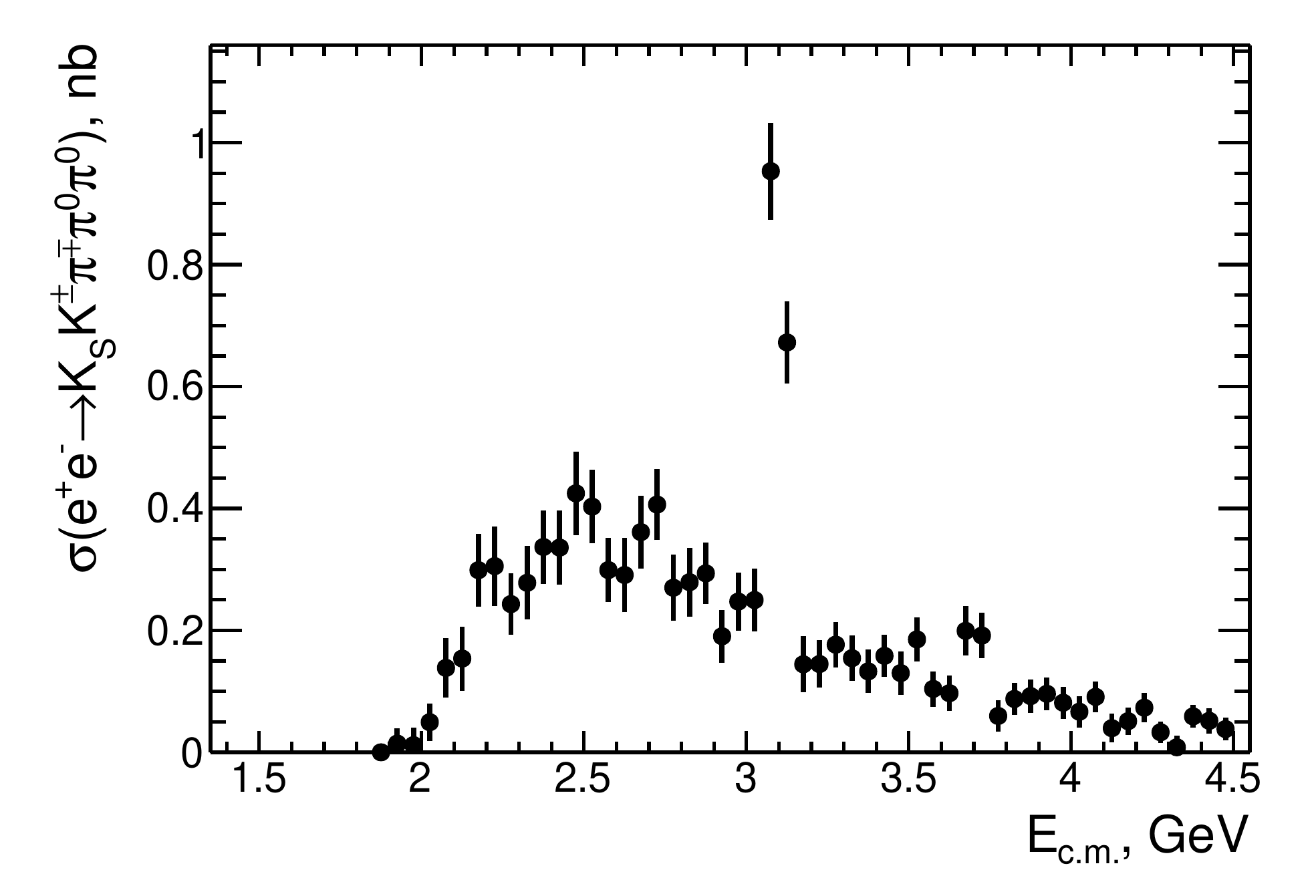}
\put(-50,90){\makebox(0,0)[lb]{\bf(b)}}
\includegraphics[width=0.34\linewidth]{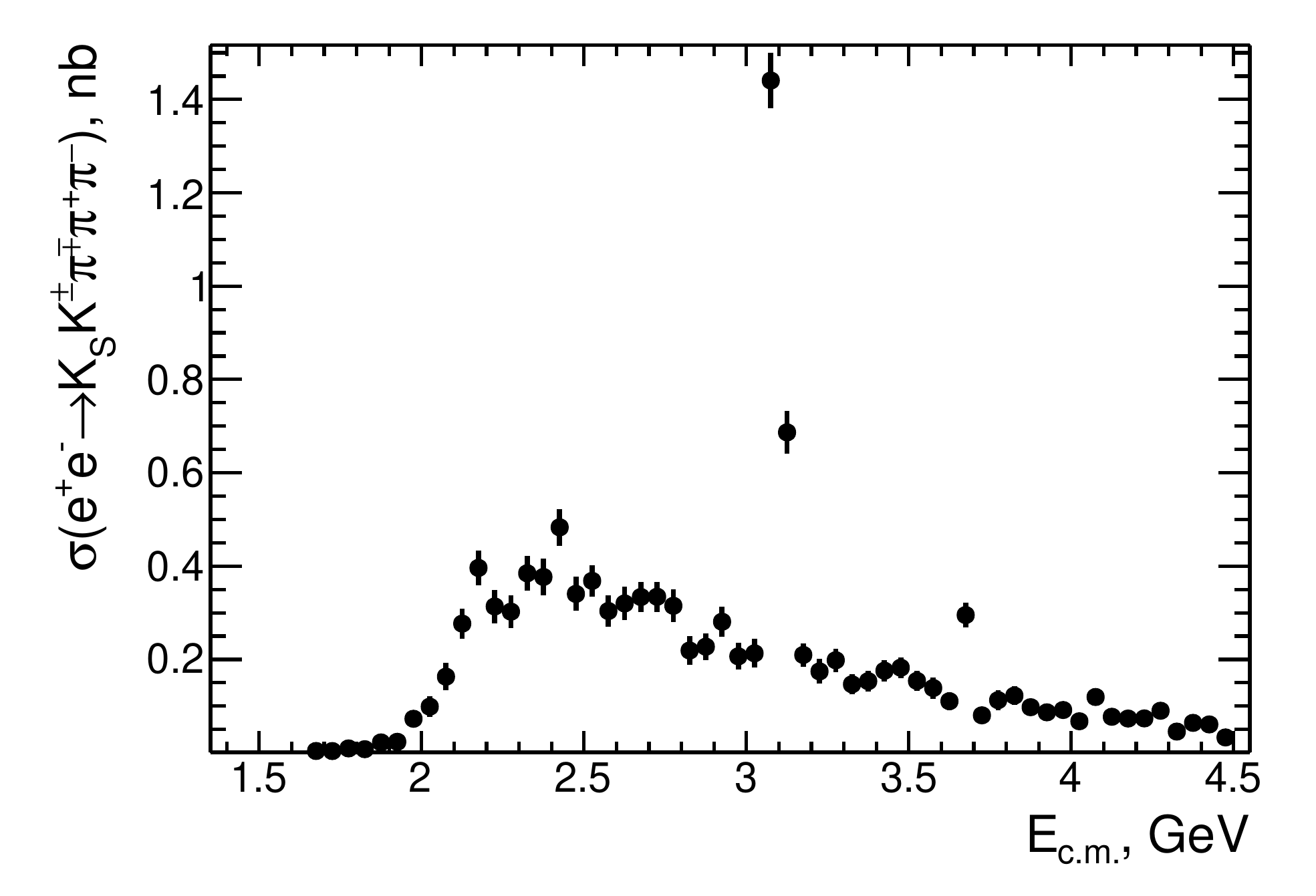}
\put(-50,90){\makebox(0,0)[lb]{\bf(c)}}
\vspace{-0.5cm}
\caption{
The  measured (a) $\epem\to\KK\ppz\piz$, (b) $\epem\to\KS
K^{\pm}\pi^{\mp}\ppz$, and (c) $\epem\to\KS K^{\pm}\pi^{\mp}\pipi$  cross sections.
The uncertainties are statistical only.
}
\label{xs_bab}
\end{center}
\end{figure*} 

\section{Cross section calculations}
\label{xs_all}
\subsection{\boldmath Number of signal events}\label{sec:signal}

The solid histograms in Fig.~\ref{mgg_all}  show the $\mgg$
invariant mass distributions for two photons for data in the \chisq signal
region for the ISR processes $\epem\to\KK\ppz\gamma\gamma$
(a) and $\epem\to\KS\Kpi\piz\gamma\gamma$ (b), while the dashed
histograms show the distribution of data from the  
\chisq control region. The dotted
histograms are the estimated contribution from the
remaining background from other ISR-related processes using simulation. 
No evidence for a peaking background is seen  in either
of the two background distributions. The background includes not only
the combinatorial part as modelled in MC simulation, but also a general background
from  $B$ hadron decays and other processes at the nominal c.m. energy.
We subtract the background evaluated using the \chisq control region with the  scale factor 1.0,
and fit the data  with a combination of three Gaussian
signal functions and a
background function, taken to be a third-order polynomial.
All parameters of the Gaussians are fixed to values taken from
simulation fits except the number of events.
The fit is performed in the $\mgg$ mass range from 0.0 to 0.45\gevcc.
In total $1230\pm168$ and $2658\pm65$  events are obtained for the
$\KK\ppz\piz$ and $\KS\Kpi\ppz$ channels, respectively.
Note that these numbers include a relatively small
peaking background component, due to $\qqbar$ events,
which is discussed in Sect.~\ref{sec:udsbkg}.
The same fit is applied to the corresponding \mgg
distribution in each 0.05\gevcc interval in the
$\KK \ppz\piz\gamma\gamma$ and $\KS \Kpi\piz\gamma\gamma$  invariant mass.
The resulting numbers of  $\KK \ppz\piz$ and $\KS \Kpi\ppz$ event
candidates, including the peaking $\qqbar$ background,
are shown as a function of hadronic mass by the
points in Fig.~\ref{nevents_data}(a,b). We vary the  fitting procedure by
releasing the resolution and the position of the main Gaussian function or
varying the scale factor.  
A variation of about 7\% in the number of events is taken as the estimate of
the systematic uncertainty.

For the $\KS\Kpi\pipi$ final state we obtain 6582 event candidates from
the signal and 737 events from the control region in the \chisq
distribution of Fig.~\ref{mgg_all}(c), shown by solid and dashed
histograms in Fig.~\ref{nevents_data}(c).

\subsection{Peaking background}\label{sec:udsbkg}
The major background producing signal-like events
following the application of the
selection criteria of Sect.~\ref{selections} is
from non-ISR \qqbar events, the most important
channels being $\epem\to\KK\ppz\ppz$, $\epem\to\KS\Kpi\piz\ppz$, and $\epem\to\KS\Kpi\pipi\piz$ 
in which one of the
neutral pions decays asymmetrically, 
yielding a high energy photon
that mimics an ISR photon. 
We apply all our selection criteria and fit procedures to
the non-ISR light quark $\qqbar$ ($uds$) simulation.
Indeed we observe a $\piz$ peak in the \mgg invariant mass distributions
for the $\KK \ppz\piz\gamma\gamma$ and $\KS\Kpi\piz\gamma\gamma$
candidate events. Also we have 459 $uds$ events in the \chisq signal region for
the $\KS\Kpi\pipi$ final state, shown by the shaded histogram in Fig.~\ref{nevents_data}(c).

To normalize the $uds$ simulation, we form the di-photon invariant
mass distribution of the ISR candidate with each of the other
photons in the event. 
  A $\piz$ peak is observed, with approximately the same
number of events in data and simulation, leading
to a normalization factor of $1.0\pm0.1$.
The resulting $uds$ background is shown
in Fig.~\ref{nevents_data}(a,b) by squares: the $uds$ background is negligible
below 2\gevcc and increases slightly with energy from 2 to 4.5\gevcc. We subtract this
background for the cross section calculation.

\begin{figure*}[bh]
\begin{center}
\includegraphics[width=0.34\linewidth]{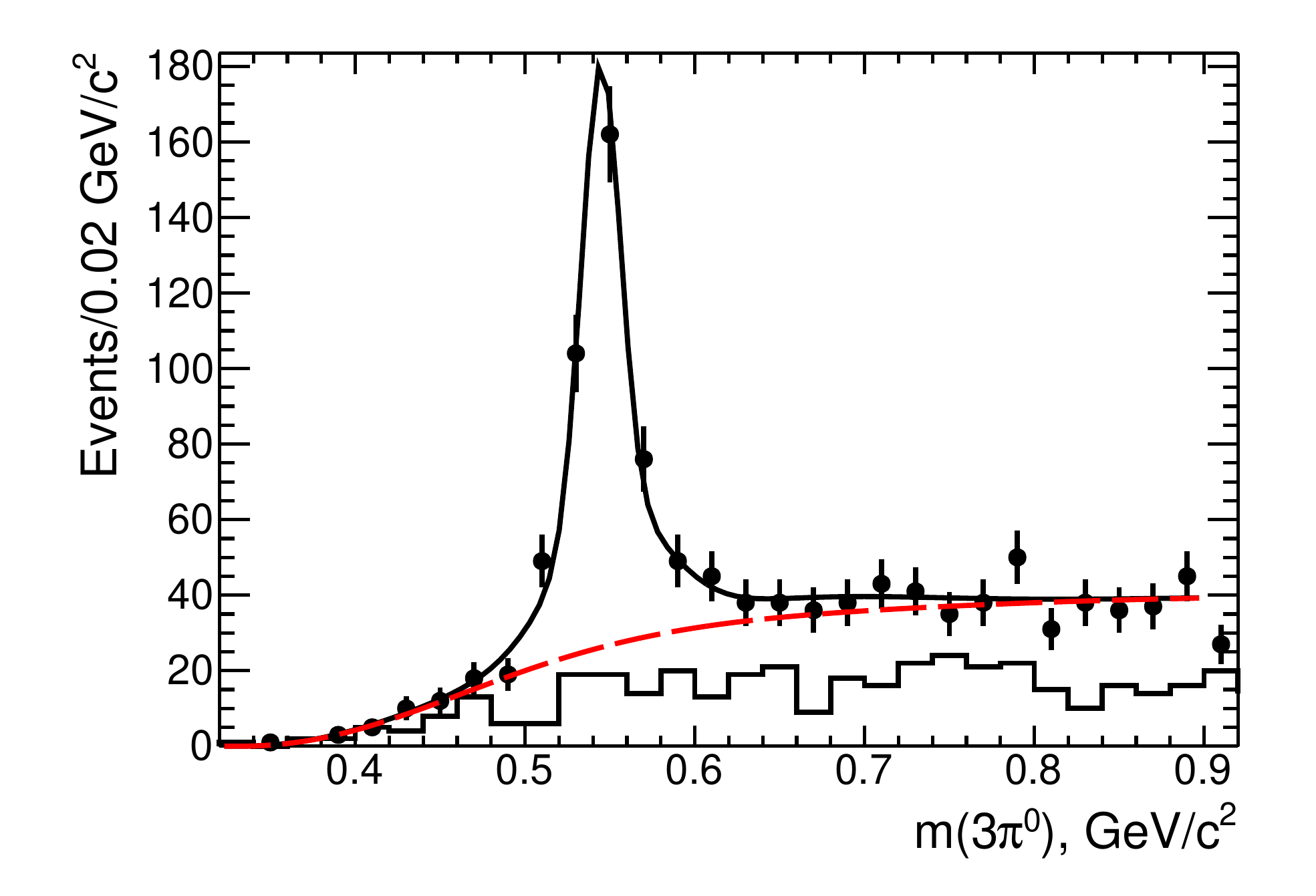}
\put(-50,90){\makebox(0,0)[lb]{\bf(a)}}
\includegraphics[width=0.34\linewidth]{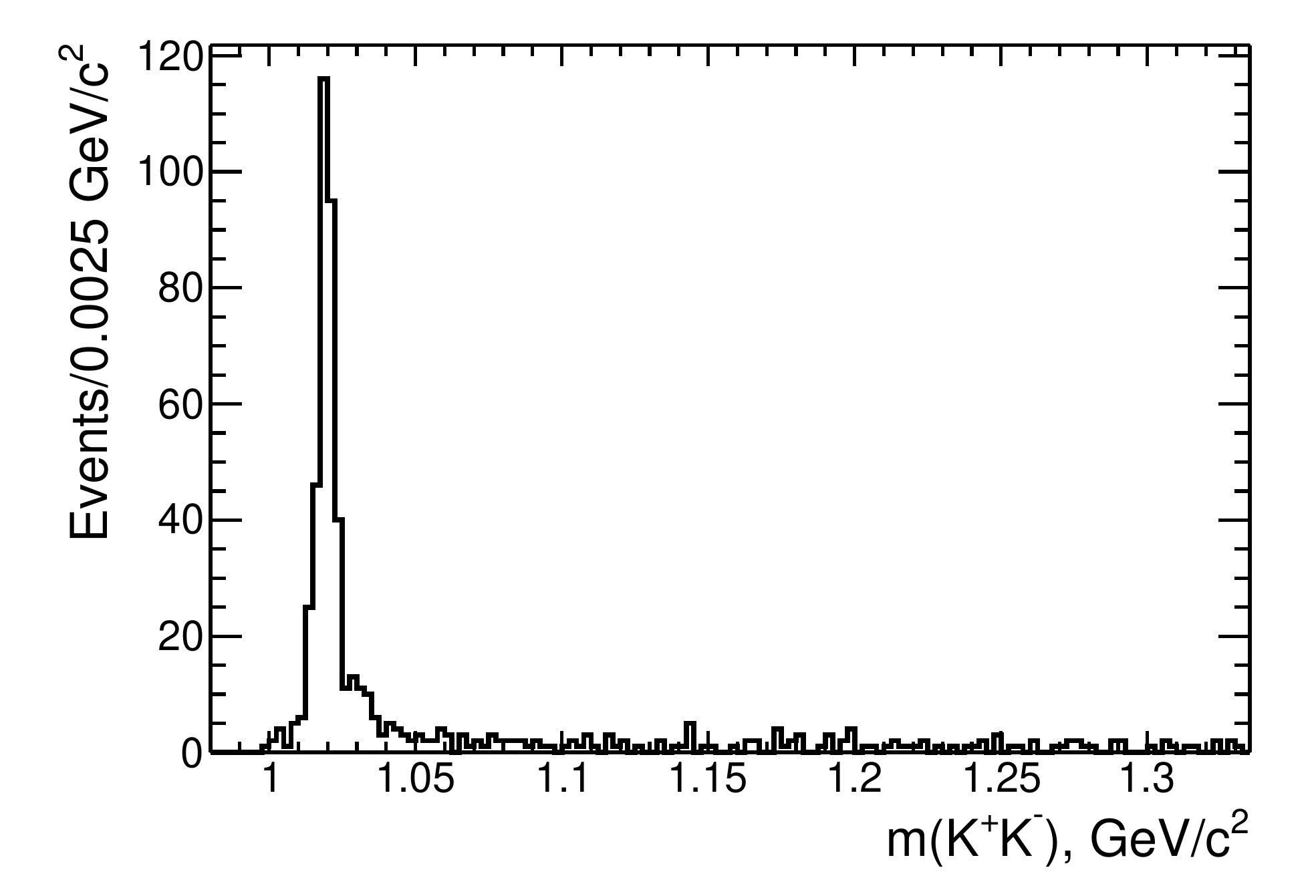}
\put(-50,90){\makebox(0,0)[lb]{\bf(b)}}
\includegraphics[width=0.34\linewidth]{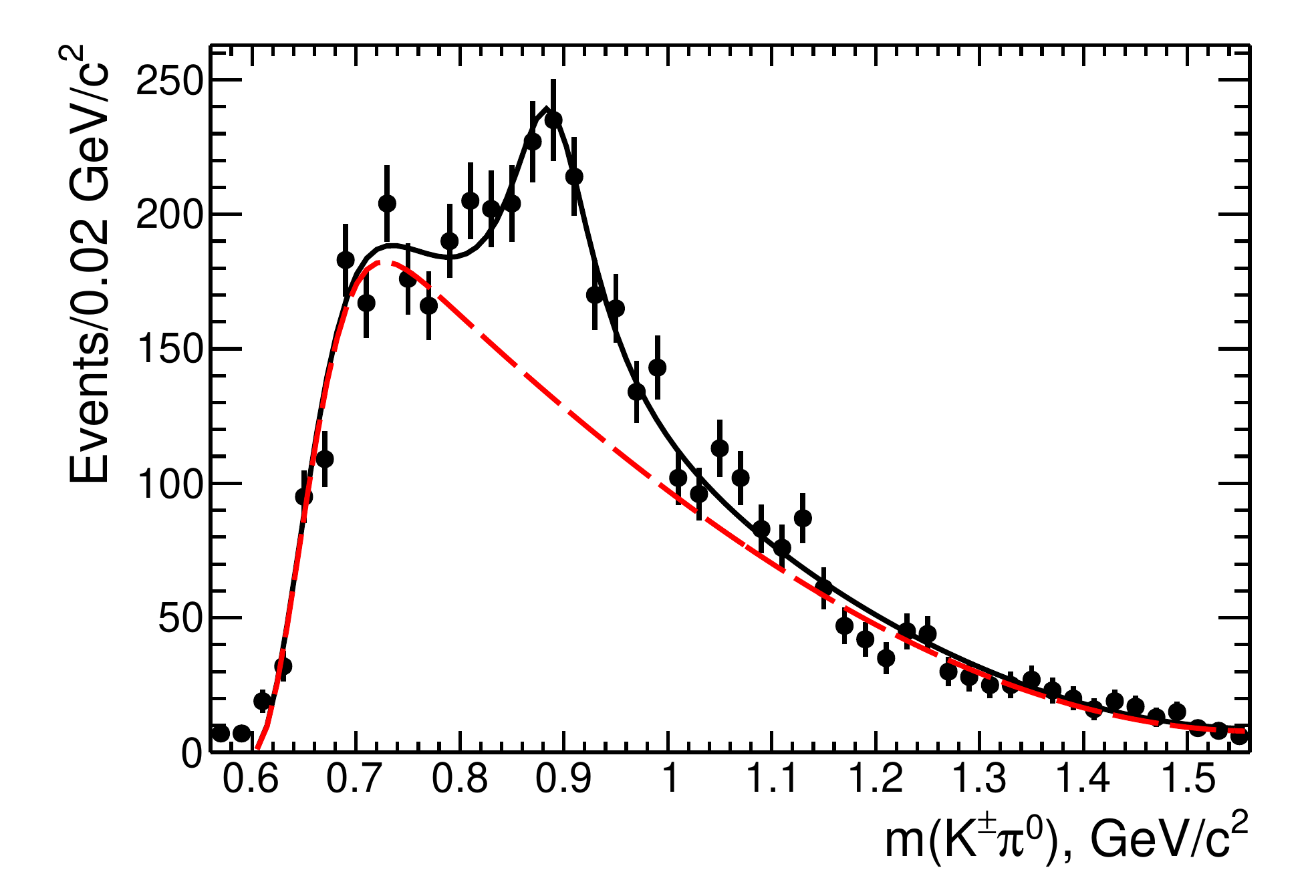}
\put(-50,90){\makebox(0,0)[lb]{\bf(c)}}
\vspace{-0.5cm}
\caption{For the $\epem\to\KK\ppz\piz$ reaction: (a) The $m(3\piz)$ invariant mass distribution for the
  \chisq signal (dots) and control (histogram) regions. The curve
  shows a  fit to the $\eta\to \ppz\piz$ signal. The dashed line shows the combinatorial background.
(b) The $m(\KK)$ invariant mass distribution for events with
$m(3\piz)<0.7$ from (a).
 (c) The $m(K^{\pm}\piz)$ invariant mass distribution (six
 entries/event). The curve shows a fit to the $K^{*}(892)$ signal.
}
\label{2k3pi0_inter}
\end{center}
%\end{figure*}
%

%
%\begin{figure*}[tbh]
  \begin{center}
    \vspace{-0.5cm}
\includegraphics[width=0.34\linewidth]{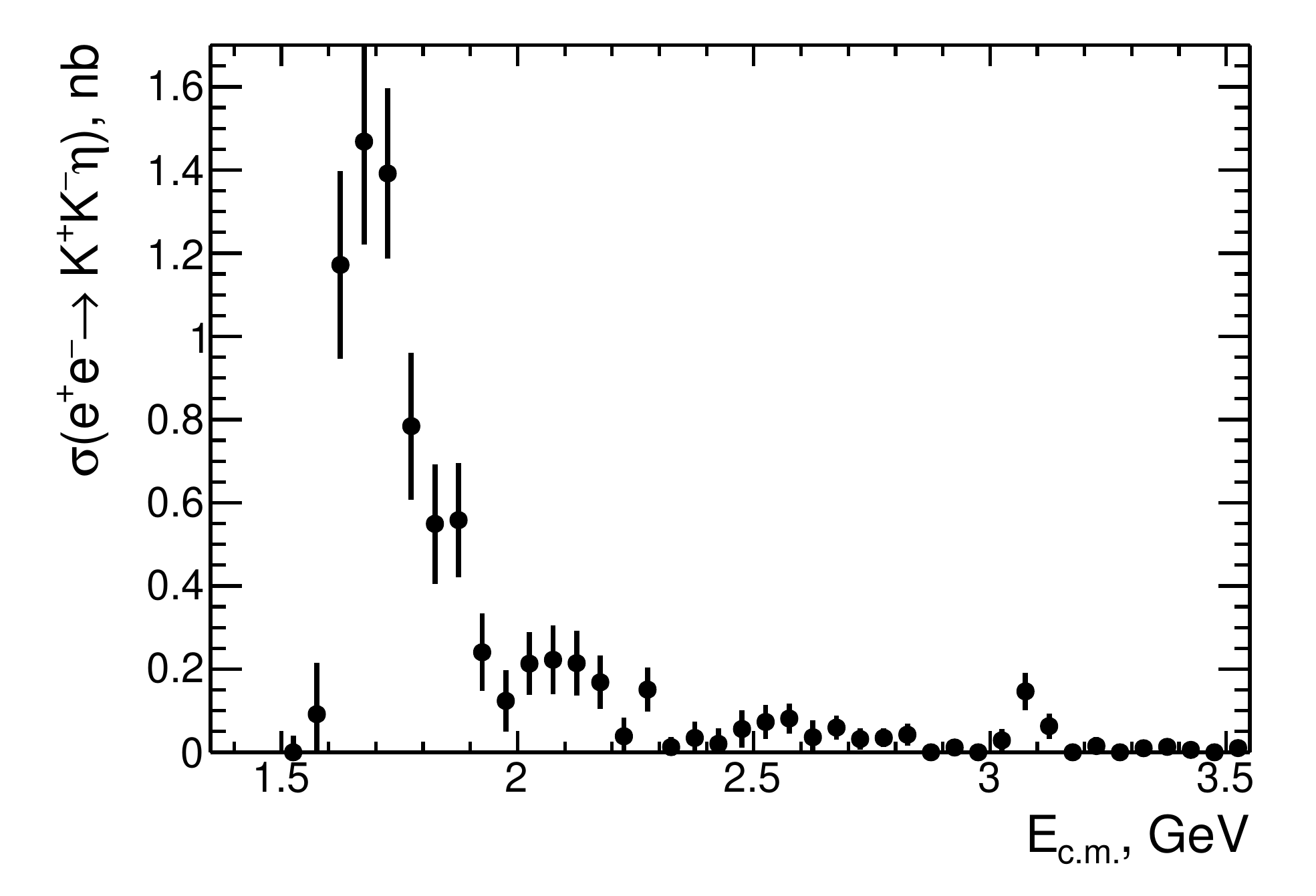}
\put(-50,90){\makebox(0,0)[lb]{\bf(a)}}
\includegraphics[width=0.34\linewidth]{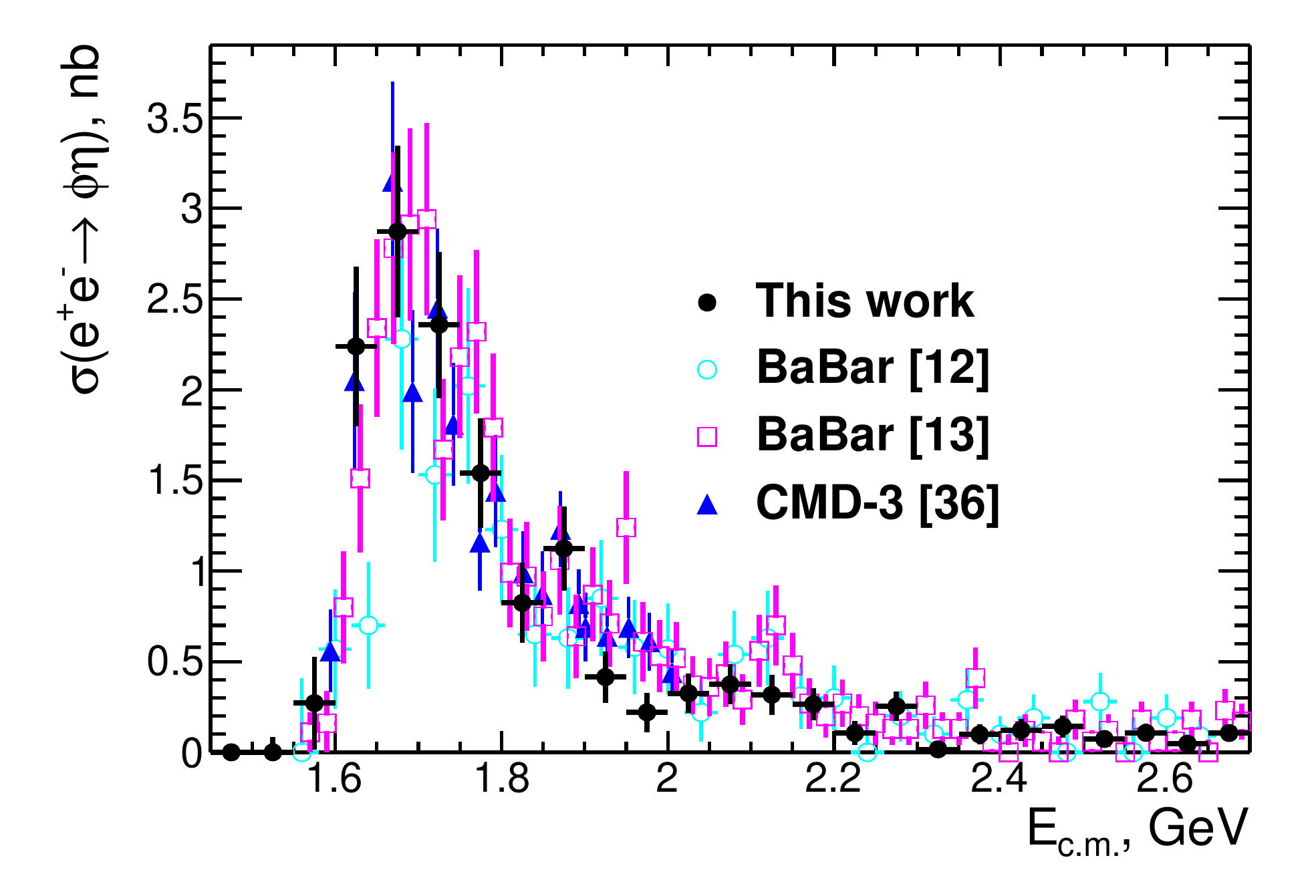}
\put(-50,90){\makebox(0,0)[lb]{\bf(b)}}
\includegraphics[width=0.34\linewidth]{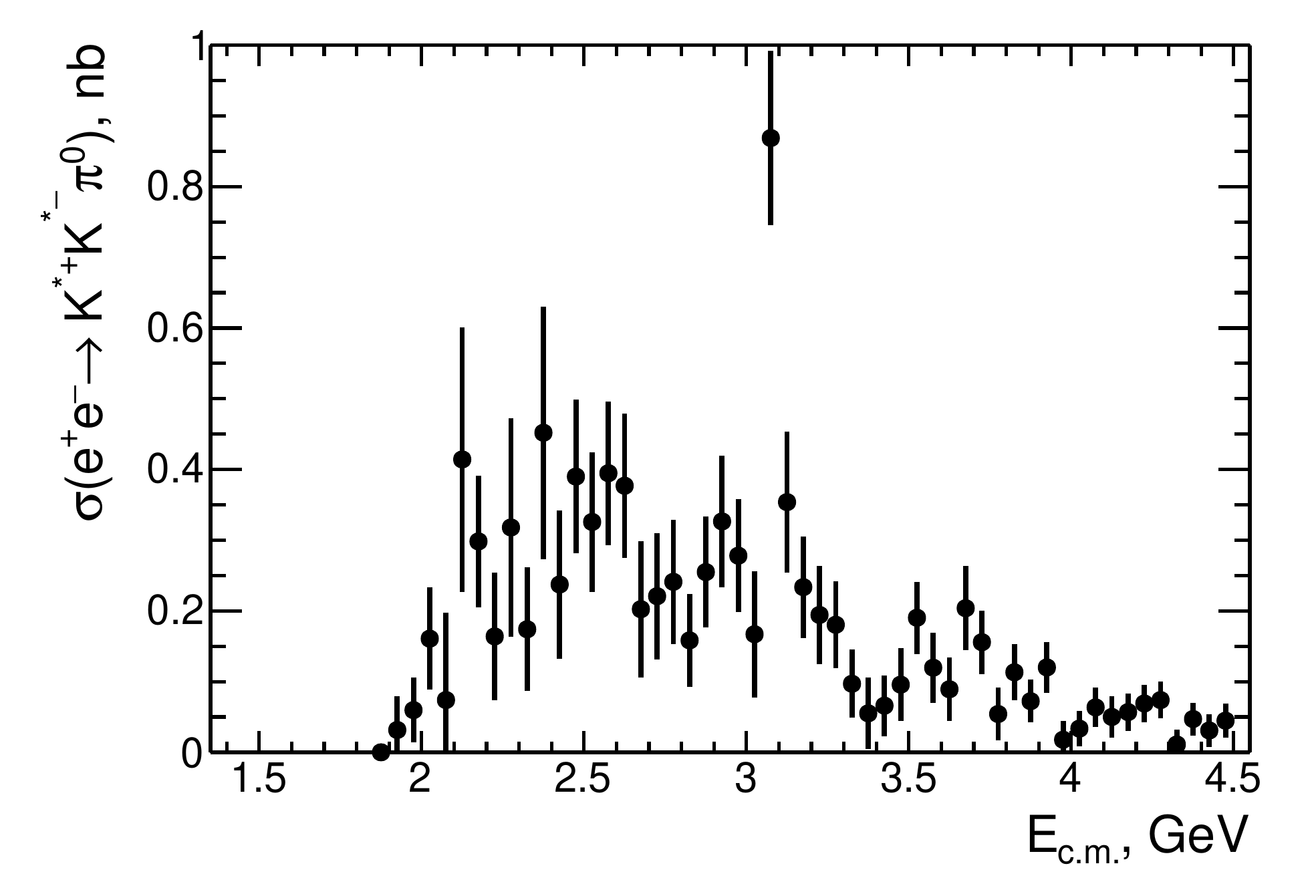}
\put(-50,90){\makebox(0,0)[lb]{\bf(c)}}
\vspace{-0.5cm}
\caption{
The  measured (a) $\epem\to\KK\eta$, (b) $\epem\to\phi\eta$, and (c)
$\epem\to K^{*+}K^{*-}\piz$  cross sections.
The uncertainties are statistical only.
}
\label{xs_KK3pi0}
\end{center}
\end{figure*} 

\subsection{\boldmath Cross section evaluation}
\label{xs_babar}

The $\epem\to$~hadrons Born cross section is  determined from
\begin{equation}
  \sigma(\rm had)(\Ecm)
  = \frac{dN_{\rm had}(\Ecm)}
         {d{\cal L}(E_{\rm c.m.})\epsilon_{\rm had}^{\rm corr}
          \epsilon_{\rm had}^{\rm MC}(E_{\rm c.m.})(1+\delta_{\rm R})}\ ,
\label{xseq}
\end{equation}
where $\Ecm$ is the invariant mass of
the hadronic system, $dN_{\rm had}$ is the background-subtracted number of selected
signal events in the interval $dE_{\rm c.m.}$,  and
$\epsilon_{\rm had}^{\rm MC}(E_{\rm c.m.})$ is the corresponding detection
efficiency from simulation. The factor $\epsilon_{\rm had}^{\rm corr}$ 
accounts for the difference between data and
simulation in the tracking
(1.0\%$\pm$1.0\%/per track)~\cite{isr4pi} and $\piz$
(3.0\%$\pm$1.0\% per pion)~\cite{isr2pi2pi0} reconstruction efficiencies.  
The ISR differential luminosity, $d{\cal L}$, is calculated using the 
total integrated \babar~ luminosity of 469 fb$^{-1}$~\cite{isr3pi}.
The initial- and final-state soft-photon emission is accounted for
by the radiative correction factor $(1+\delta_{\rm R})$,
which lies within 1\% of unity for our selection criteria.
The cross section results contain the effect of
 vacuum polarization because this effect is not accounted for in
the luminosity calculation.

Our results for the $\epem\to\KK\ppz\piz$ cross section
are shown in Fig.~\ref{xs_bab}(a).  The cross section exhibits a
structure around 1.7\gev with  a peak value of about 0.4~nb, 
followed by a monotonic decrease toward higher energies, perturbed
by the $J/\psi$ and $\psi(2S)$ signals. 
Because we present our data in bins of width 0.05\gevcc, compatible   
with the experimental resolution, we do not apply an unfolding procedure to the data.

Figure~\ref{xs_bab} also shows the $\epem\to\KS
K^{\pm}\pi^{\mp}\ppz$ (b) and $\epem\to\KS K^{\pm}\pi^{\mp}\pipi$ (c)
cross sections. Some possible structures are seen above  2\gev and
more prominent signals from $J/\psi$ and $\psi(2S)$ are observed.
Numerical values for the cross sections are presented in
Tables~\ref{2k3pi0_tab},\ref{kskpi2pi0_tab}, and~\ref{ksk3pi_tab}.
The $J/\psi$ region is discussed later.

Our results represent the first measurements of these cross sections.
\subsection{\boldmath Summary of  systematic studies}
\label{sec:Systematics}
The systematic 
uncertainties, presented in the previous sections, are summarized in
Table~\ref{error_tab},  along 
with the corrections that are applied to the measurements.
\begin{table}[tbh]
\caption{
Summary of  correction factors and  systematic uncertainties in the $\epem\to
\KK \ppz\piz (\KS\Kpi\ppz, \KS\Kpi\pipi)$  cross section measurements. The total uncertainly is computed assuming no correlations.
}
\label{error_tab}
%\begin{center}
%%\begin{ruledtabular}
\begin{tabular}{l c c} 
\hline
Source & Correction & Uncertainty\\
%\hline
\hline
Luminosity  &  --  &  $1\%$ \\
%\hline
  MC-data difference in:\\
  ISR photon efficiency & +1.5\%  & $1\%$\\
  Track losses, PID & $+2(3,5)\%$ & $2(3,3)\%$ \\
$\pi^0$ losses & $+9(6,0)\%$ & $4(2,0)\%$ \\
\chisq cut uncertainty & -- & $3\%$\\
%\hline
Fit and background subtraction & -- &  $7(7,0)\%$ \\
%\hline 
%\hline
Radiative corrections accuracy & -- & $1\%$ \\
%\hline
Efficiency from MC \\(model-fit-dependent) & -- & $5\%$  \\
%\hline
\hline
Total    &  $+12.5(10.5,6.5)\%$   & $10(10,8)\%$  \\
%\hline
\end{tabular}
%%\end{ruledtabular}
%\end{center}
\end{table}

The three corrections applied to the cross sections sum
up to 12.5\%, 10.5\%, and 6.5\% for the $\epem\to
\KK\ppz\piz$, $\epem\to\KS K^{\pm}\pi^{\mp}\ppz$, and $\epem\to\KS K^{\pm}\pi^{\mp}\pipi$ cross
sections, respectively, with the corresponding
systematic uncertainties estimated as
 10\%, 10\%, and 8\%. 
The largest systematic uncertainty arises from the fitting
and background subtraction procedures of the $\piz$ signal.
This is estimated by varying the background levels and the parameters
of the functions used.

\section{\boldmath Intermediate structures in the $\KK\ppz\piz$ final
  state}
\label{inter_2k3pi0}
As we assumed from the beginning, the $\epem\to\KK\ppz\piz$ reaction
has a significant contribution from the $\phi(1020)\eta$ intermediate
state. Indeed, Fig.~\ref{2k3pi0_inter}(a)
exhibits a clear $\eta$ meson peak in the three-pion
invariant mass $m(3\piz)$. The histogram shows a
background contribution from the \chisq control region.
The fit, with a
two-Gaussian function for the signal and a polynomial function for the
background, yields
$353\pm28$ events for the $\KK\eta$ intermediate state. The cross
section for the $\epem\to\KK\eta$ reaction is shown in
Fig.~\ref{xs_KK3pi0}(a) and listed in Table~\ref{2k3pi0_2keta_tab},
accounting for the $\eta\to \ppz\piz$ branching ratio.
If we restrict the three pion mass by the
requirement $m(3\piz)<0.7$\gevcc, the $m(\KK)$ invariant mass
exhibits a $\phi(1020)$ resonance, shown in Fig.~\ref{2k3pi0_inter}(b). With the  $m(\KK)<1.05$\gevcc
selection we determine the cross section for the $\epem\to\phi\eta$
process, shown as solid dots in Fig.~\ref{xs_KK3pi0}(b), in comparison with
other measurements by ~\babar~\cite{isr5pi} (open squares), ~\babar~\cite{isrkkpi}  (open circles),
and CMD-3~\cite{cmd3phieta} (triangles). The decay rates $\phi\to\KK$
and $\eta\to \ppz\piz$ are taken into account. The result is listed in Table~\ref{2k3pi0_phieta_tab}.
\begin{figure*}[tbh]
\begin{center}
\includegraphics[width=0.34\linewidth]{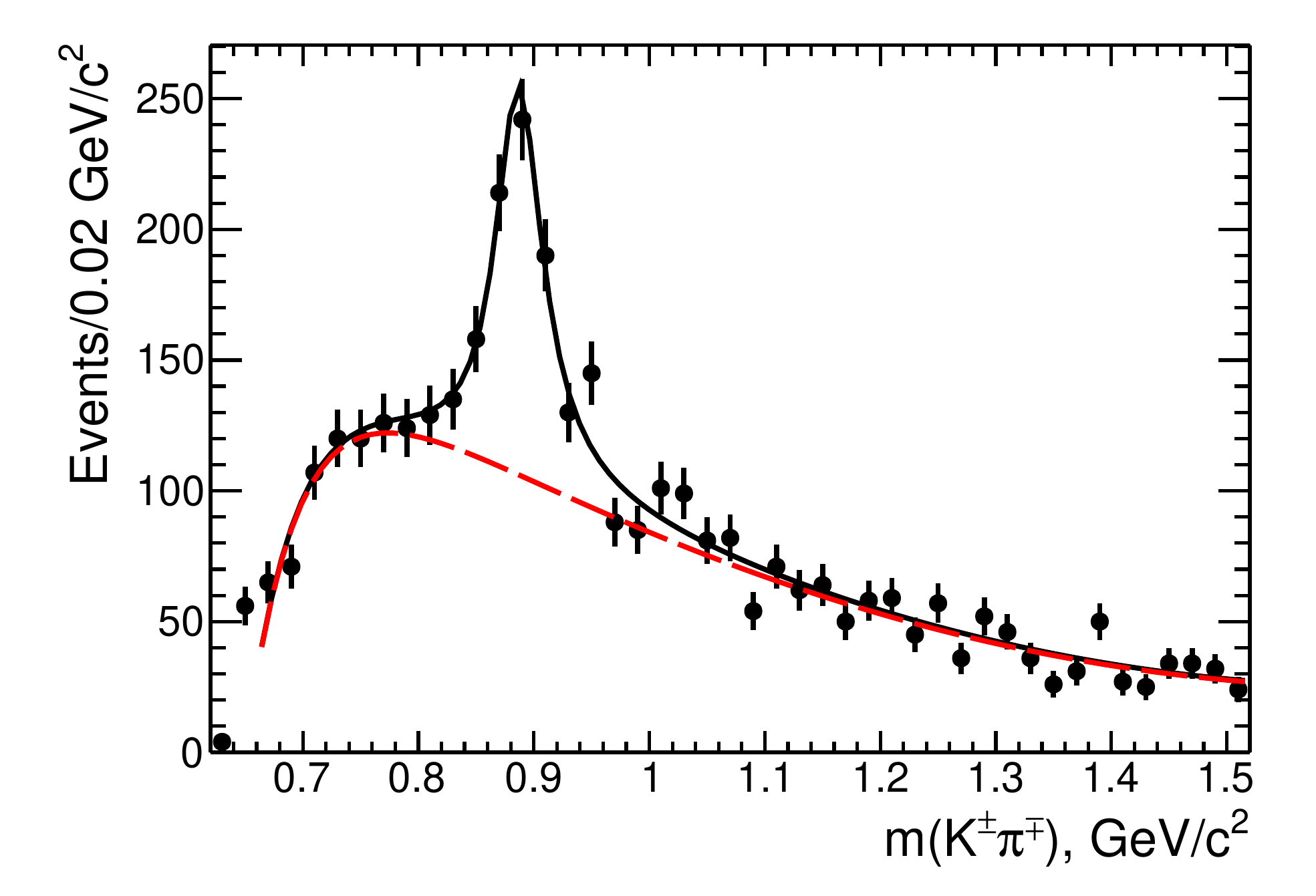}
\put(-50,90){\makebox(0,0)[lb]{\bf(a)}}
\includegraphics[width=0.34\linewidth]{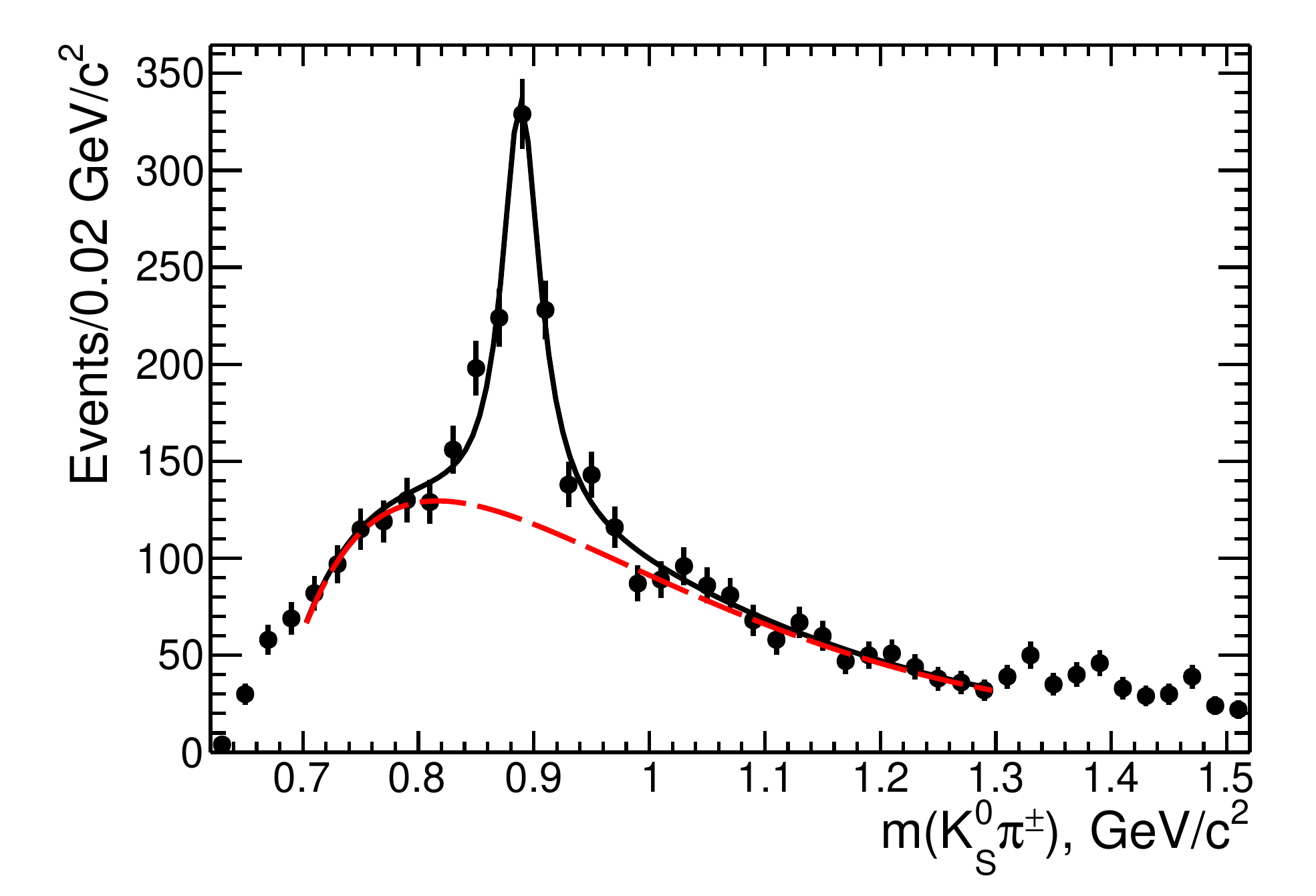}
\put(-50,90){\makebox(0,0)[lb]{\bf(b)}}
\includegraphics[width=0.34\linewidth]{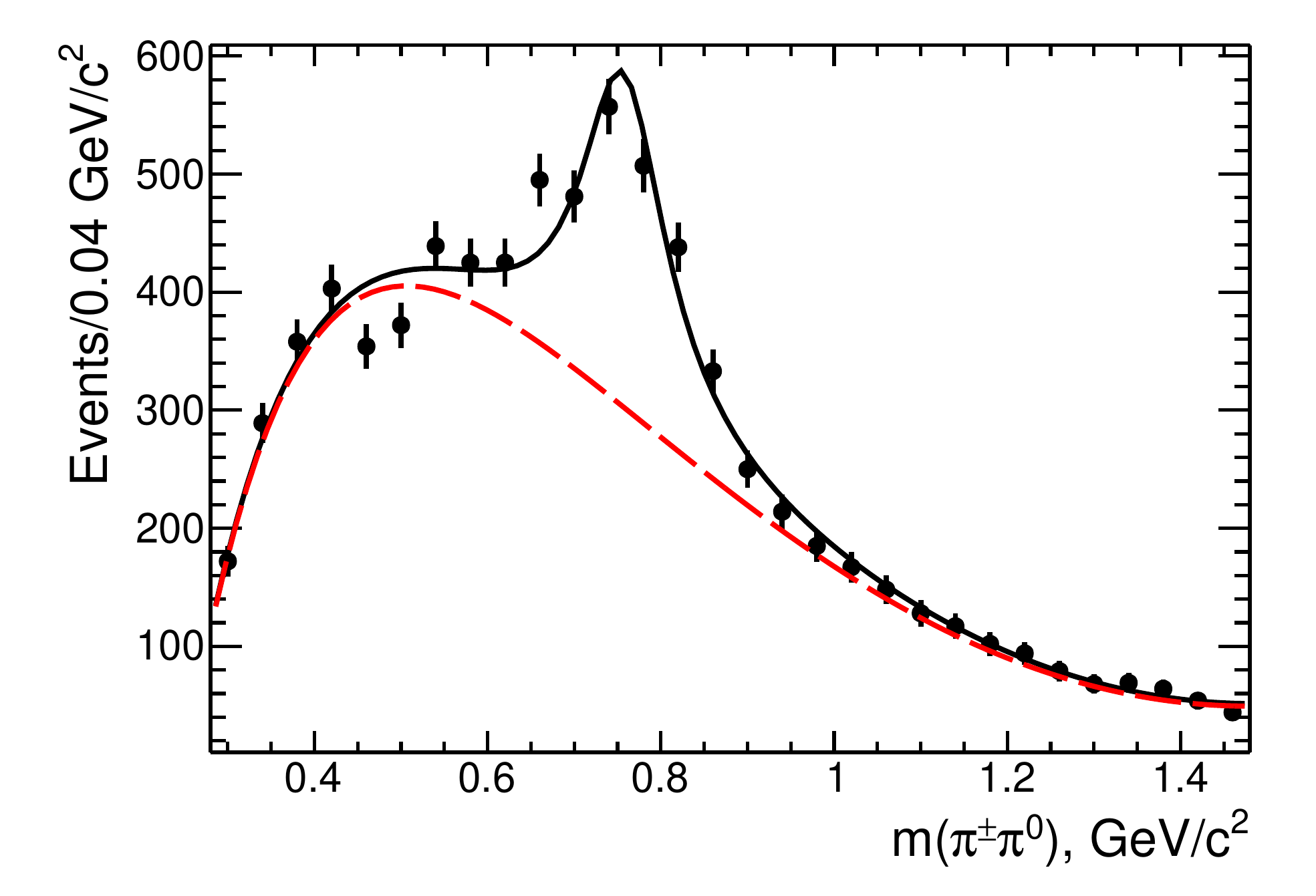}
\put(-50,90){\makebox(0,0)[lb]{\bf(c)}}
\vspace{-0.5cm}
\caption{ The (a) $m(K^{\pm}\pi^{\mp})$, (b)  $m(\KS\pi^{\pm})$, and (c)
  $m(\pi^{\pm}\piz)$ invariant mass distributions for the 
 $\KS K^{\pm}\pi^{\mp}\ppz$  events.
The curves show the fit to the $K^{*}(892)$ and $\rho(770)$ signals
with the combinatoric background contribution shown by the dashed curves.
}
\label{kskpi2pi0_inter}
\end{center}
\end{figure*}
\begin{figure*}[tbh]
\begin{center}
\includegraphics[width=0.34\linewidth]{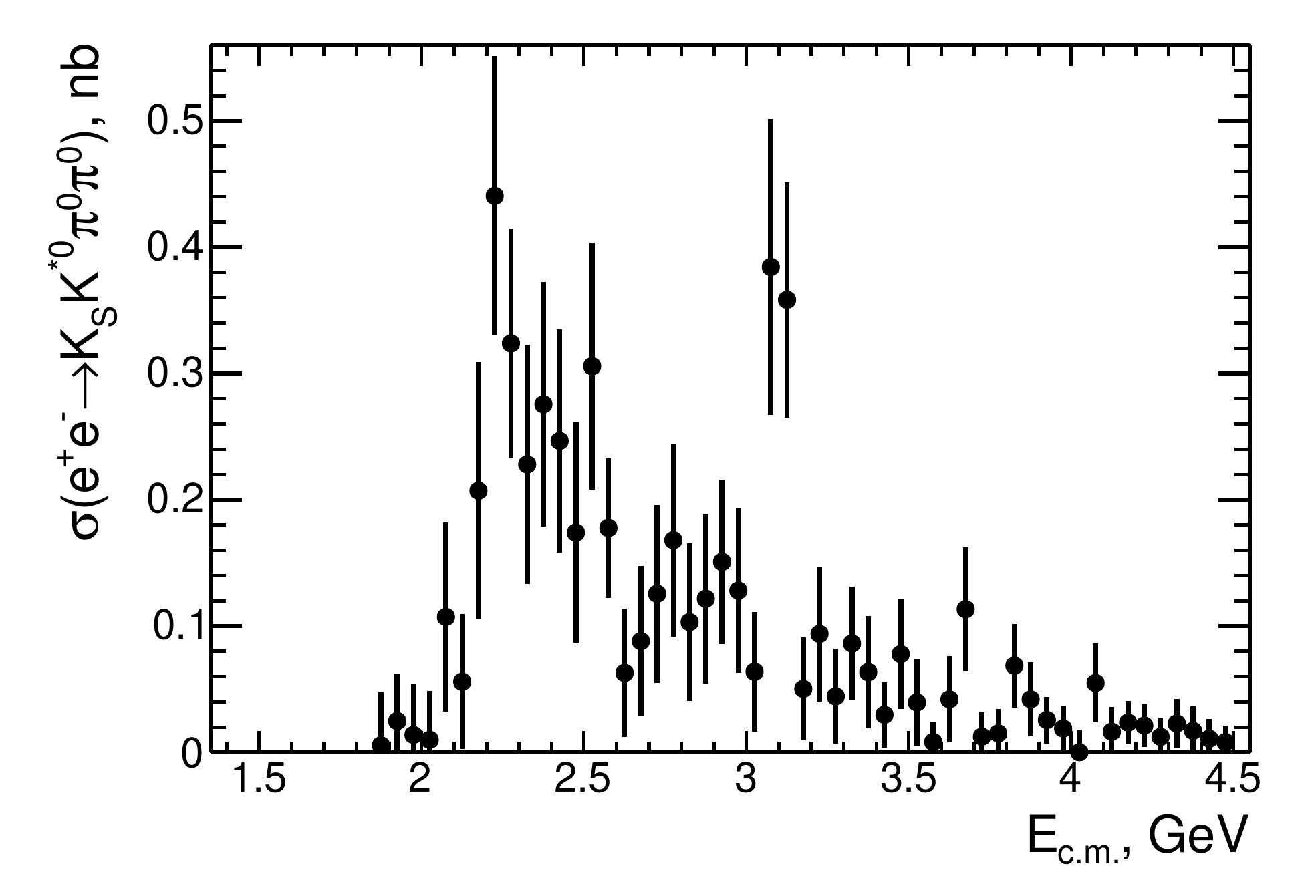}
\put(-50,90){\makebox(0,0)[lb]{\bf(a)}}
\includegraphics[width=0.34\linewidth]{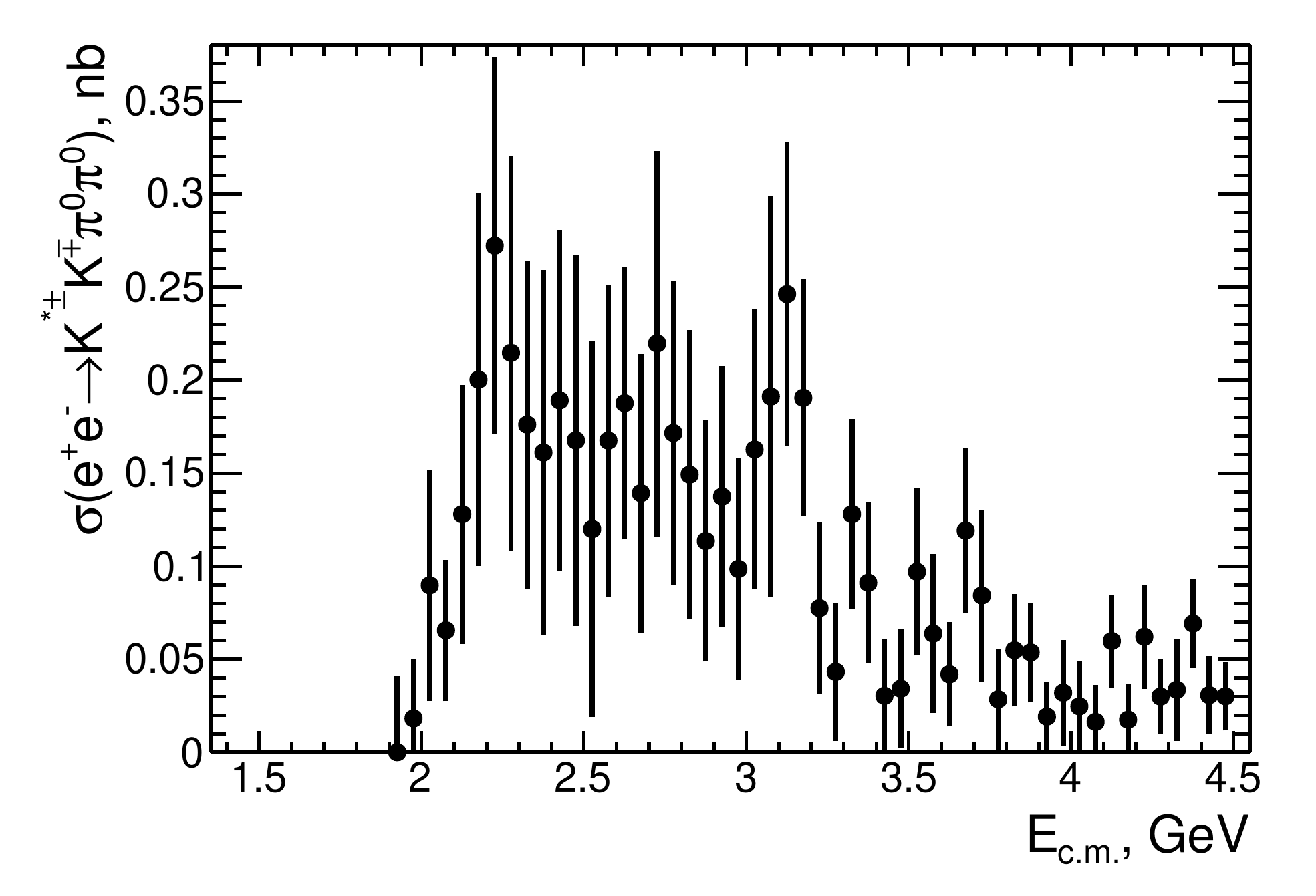}
\put(-50,90){\makebox(0,0)[lb]{\bf(b)}}
\includegraphics[width=0.34\linewidth]{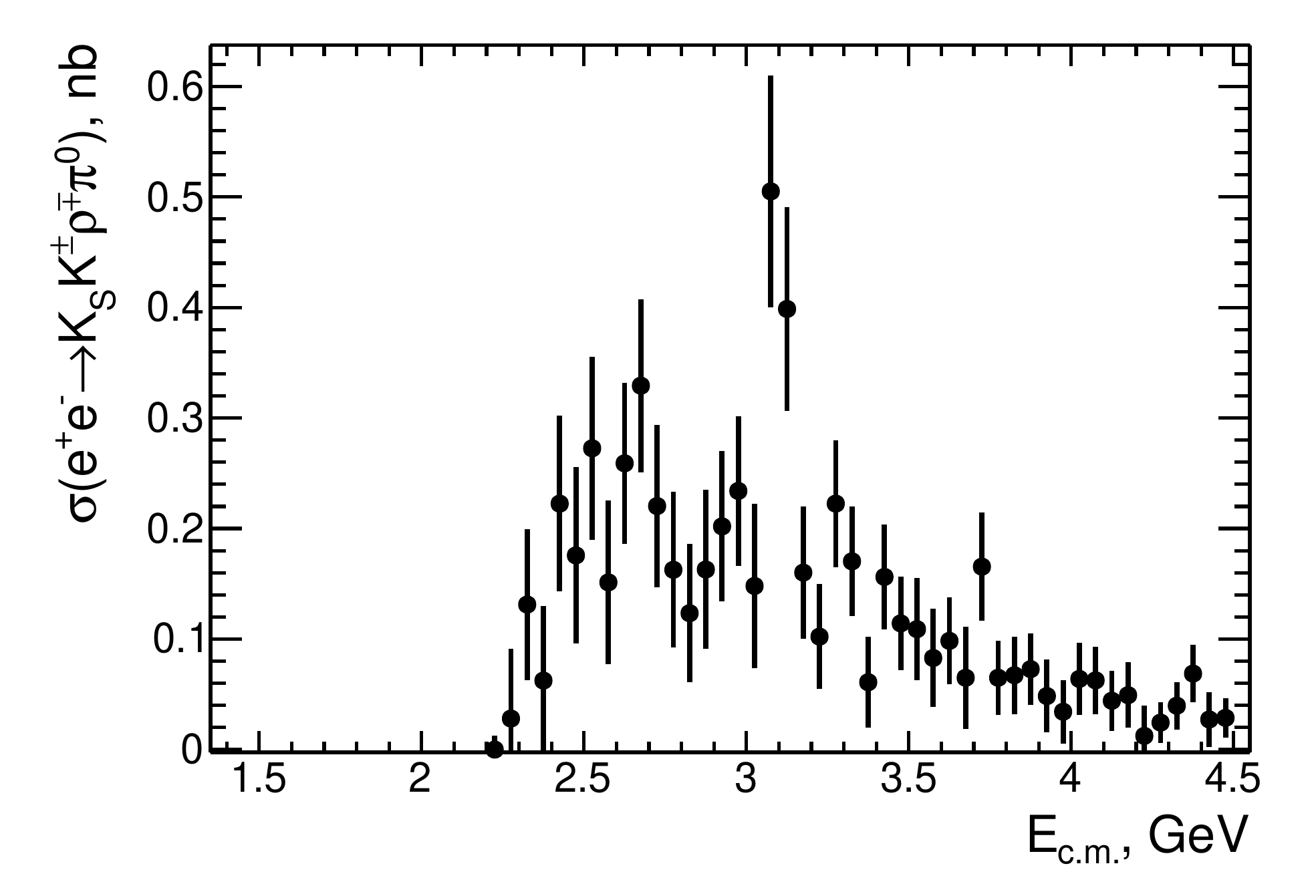}
\put(-50,90){\makebox(0,0)[lb]{\bf(c)}}
\vspace{-0.5cm}
\caption{
The  measured (a) $\epem\to\KS K^{*0}\ppz$, (b) $\epem\to K^{*\pm}K^{\mp}\ppz$, and
(c) $\epem\to \KS K^{\pm}\rho^{\mp}\piz$  cross sections.
The uncertainties are statistical only.
}
\label{xs_KsKpi2pi0}
\end{center}
\end{figure*} 
\begin{figure*}[tbh]
\begin{center}
\includegraphics[width=0.34\linewidth]{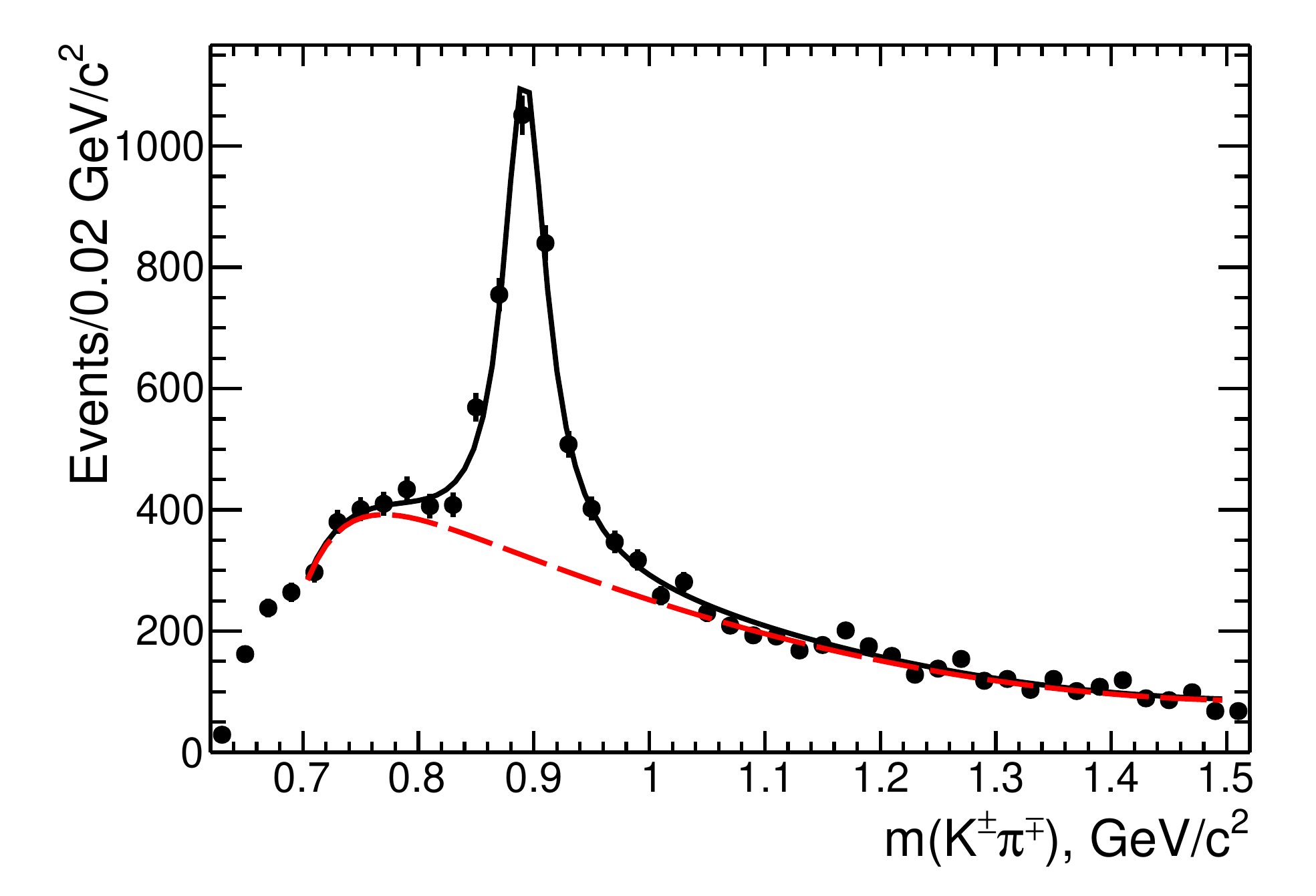}
\put(-50,90){\makebox(0,0)[lb]{\bf(a)}}
\includegraphics[width=0.34\linewidth]{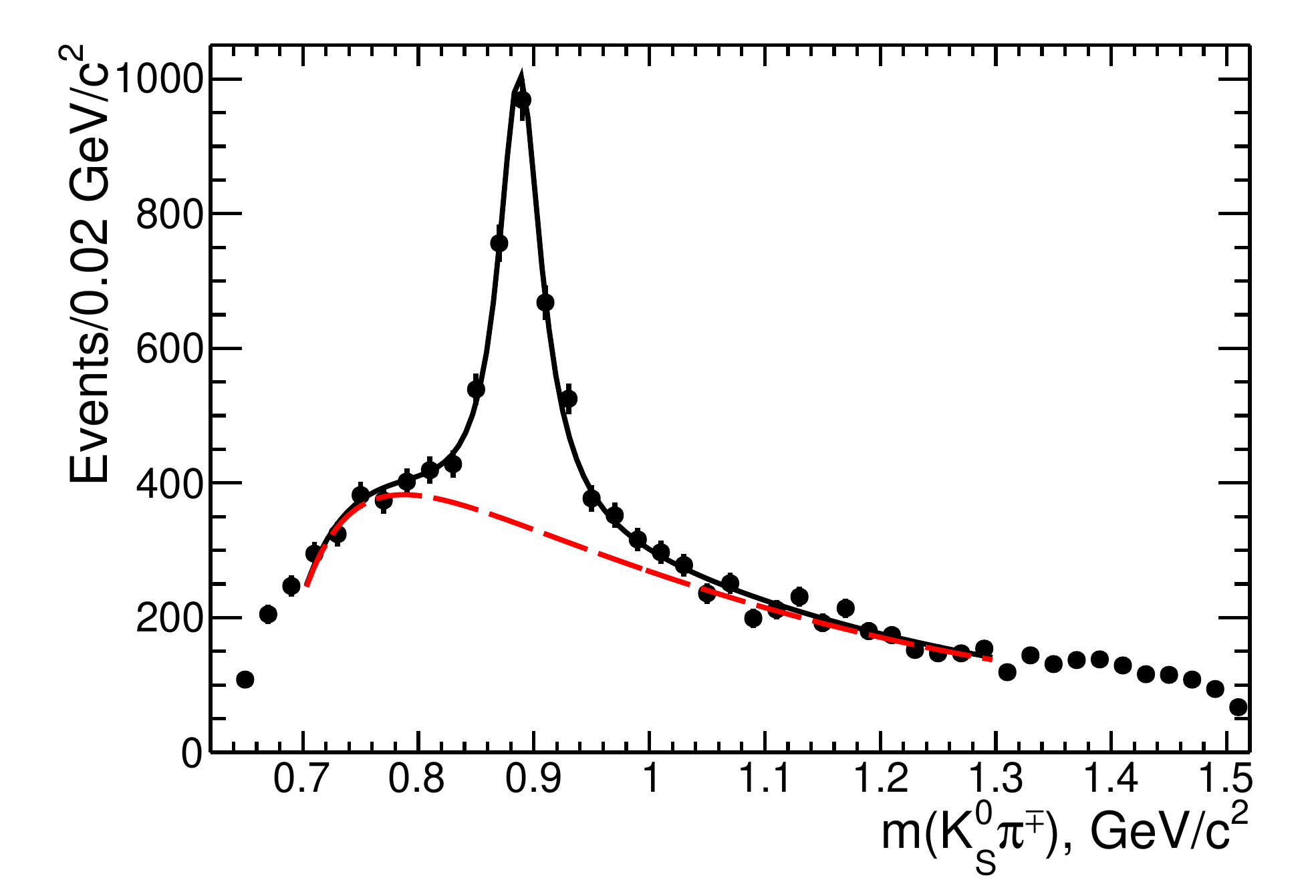}
\put(-50,90){\makebox(0,0)[lb]{\bf(b)}}
\includegraphics[width=0.34\linewidth]{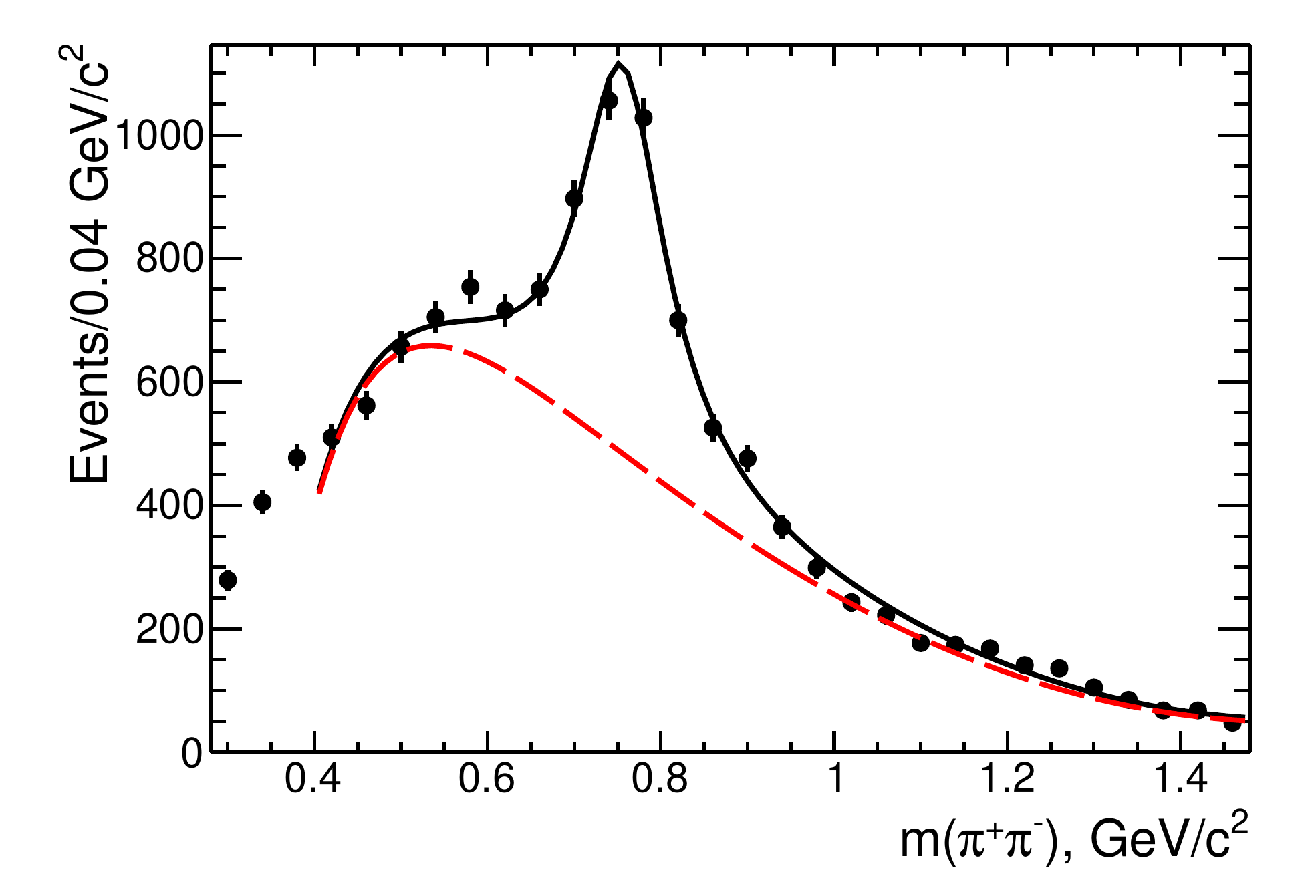}
\put(-50,90){\makebox(0,0)[lb]{\bf(c)}}
\vspace{-0.5cm}
\caption{ The (a) $m(K^{\pm}\pi^{\mp})$, (b)  $m(\KS\pi^{\pm})$, and (c)
  $m(\pi^{\pm}\piz)$ invariant mass distributions for the 
 $\KS K^{\pm}\pi^{\mp}\pipi$  events.
The curves show fits to (a,b) the  $K^{*}(892)$ signal   and to (c) the
$\rho(770)$  signal. The dashed curves indicate the combinatoric background.
}
\label{ksk3pi_inter}
\end{center}
\end{figure*}
\begin{figure*}[tbh]
\begin{center}
\includegraphics[width=0.34\linewidth]{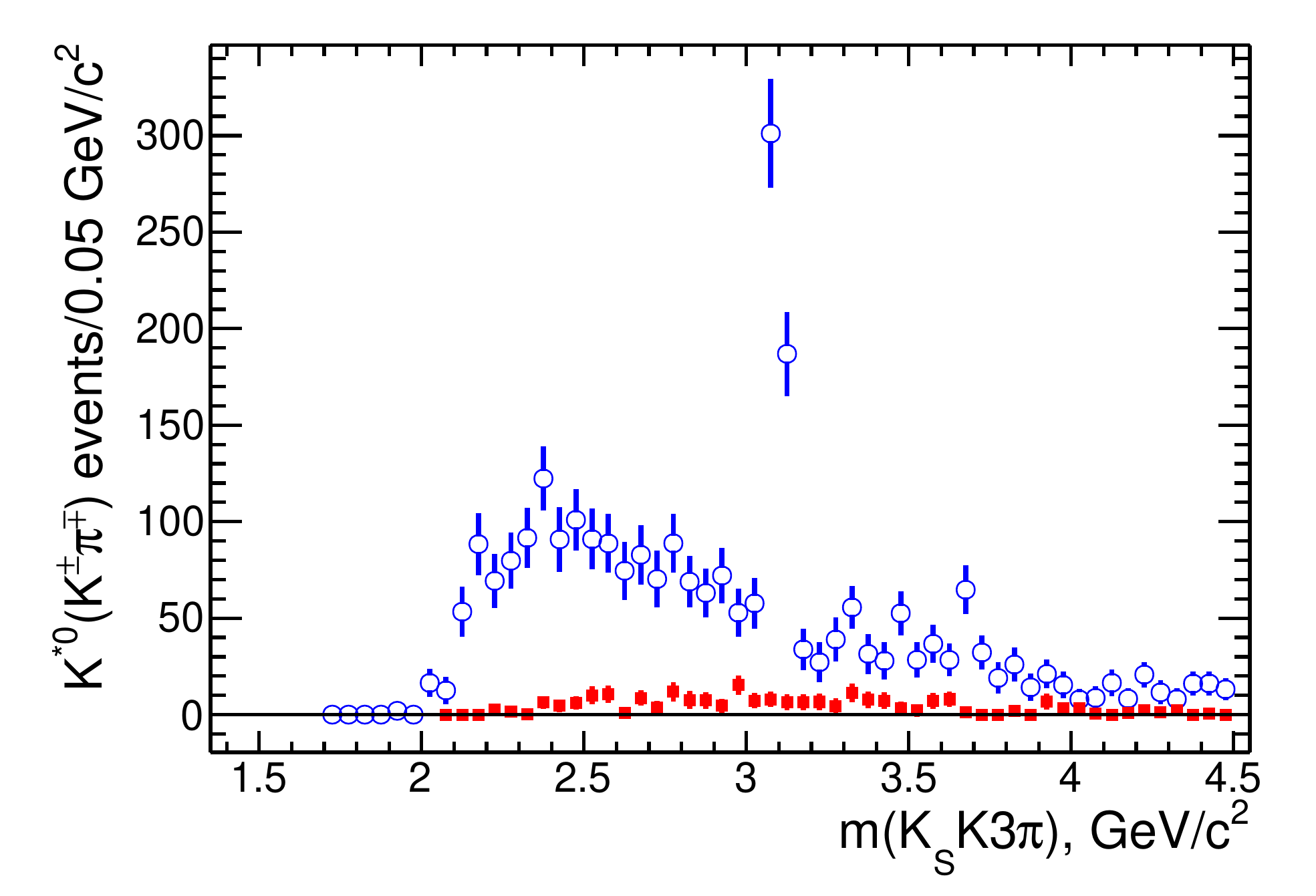}
\put(-50,90){\makebox(0,0)[lb]{\bf(a)}}
\includegraphics[width=0.34\linewidth]{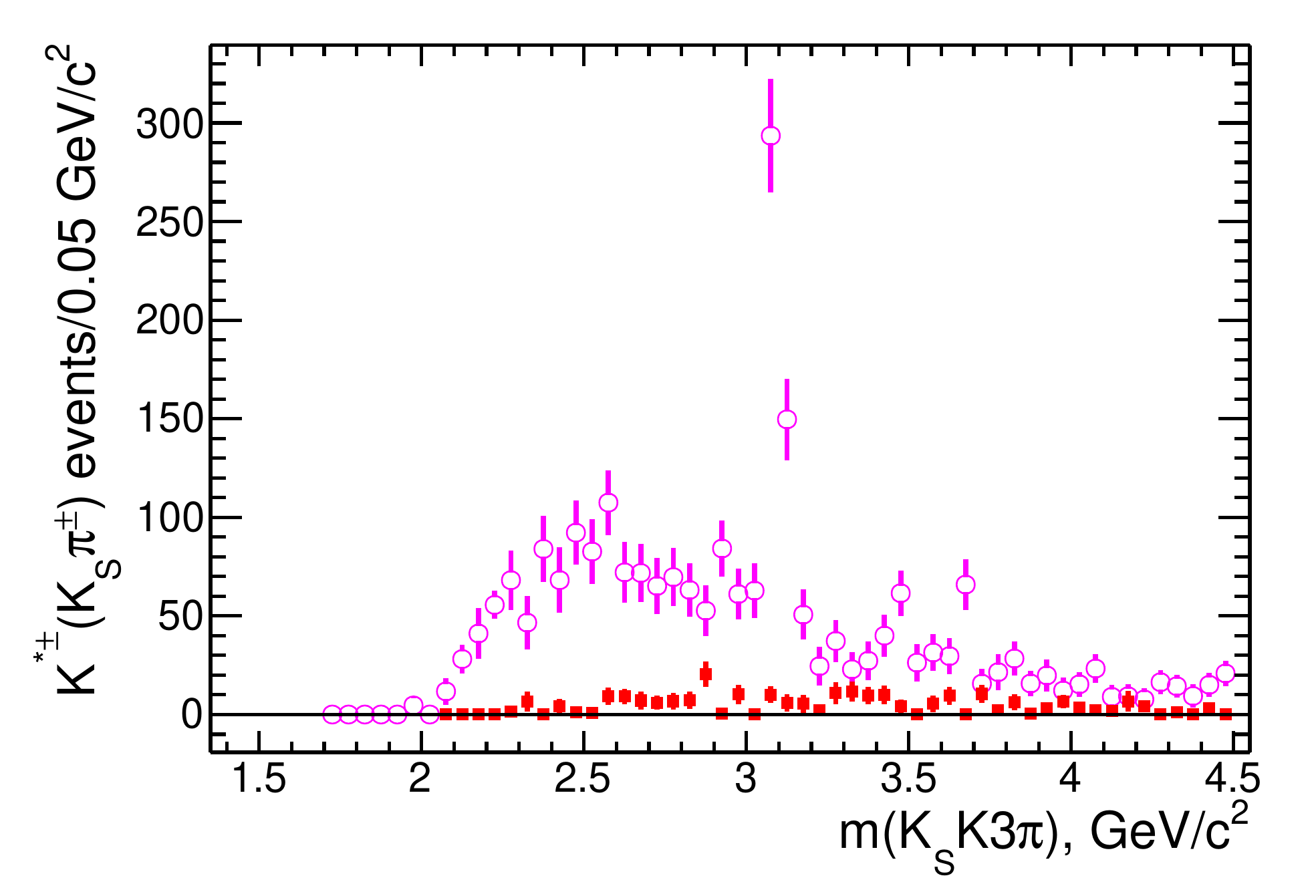}
\put(-50,90){\makebox(0,0)[lb]{\bf(b)}}
\includegraphics[width=0.34\linewidth]{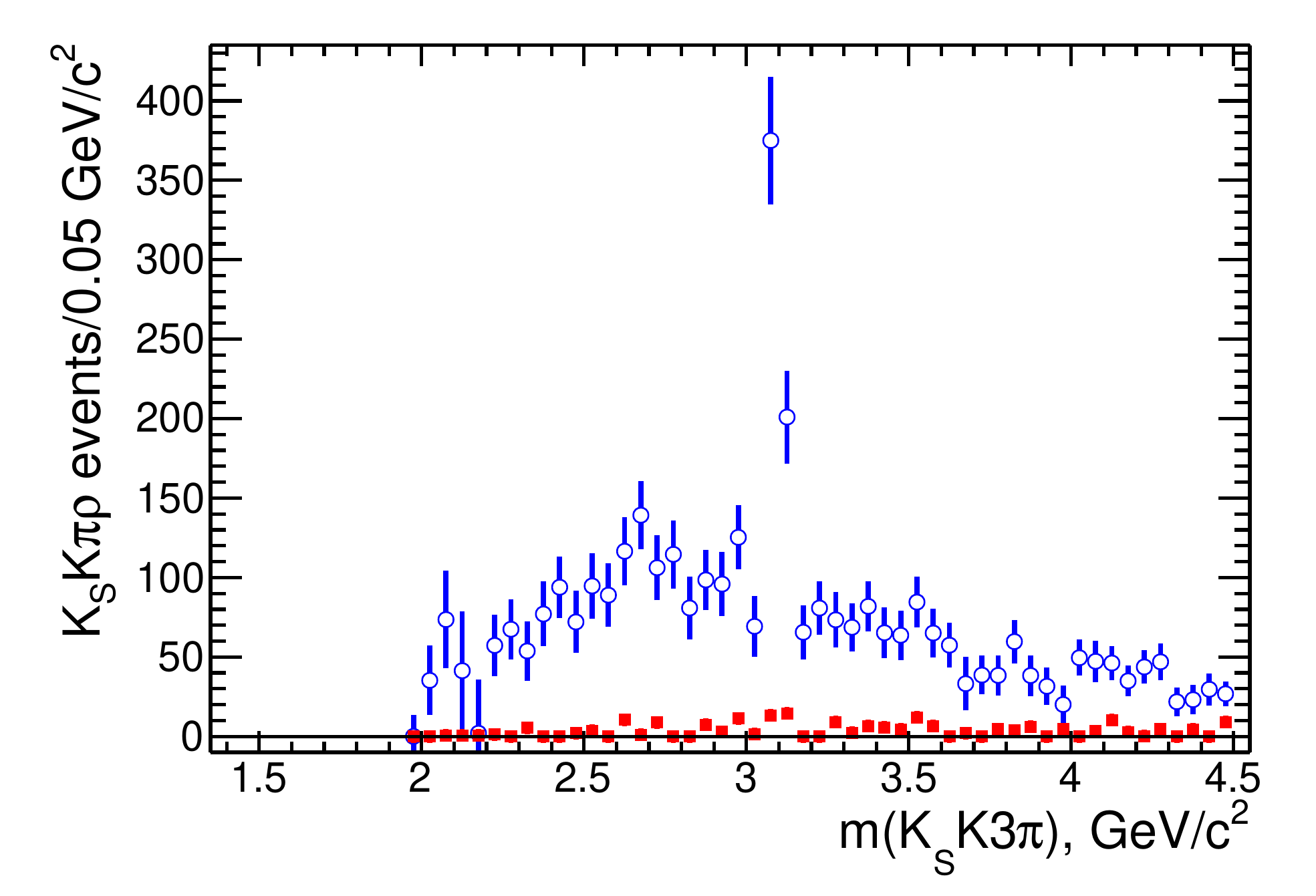}
\put(-50,90){\makebox(0,0)[lb]{\bf(c)}}
\vspace{-0.5cm}
\caption{
The observed mass dependence  for (a) the  $\epem\to\KS K^{*0}\pipi$, (b) $\epem\to K^{*\pm}K^{\mp}\pipi$, and
(c) $\epem\to \KS K^{\pm}\rho^{0}\pi^{\mp}$  reactions.  The contribution
from the $uds$ reaction is shown by squares.
}
\label{xs_KsK3pi}
\end{center}
\end{figure*}

Figure~\ref{2k3pi0_inter}(c) shows (dots) the $m(K^{\pm}\piz)$ invariant mass
(six entries/event).
This distribution exhibits a clear signal from $K^*(892)^{\pm}$. We fit
this distribution with a Breit-Wigner (BW) function and combinatorial
background, yielding $1506\pm84$ signal events. 
 We conclude that the $K^{*\pm}$ signal arises from
 $\epem\to K^{*+}K^{*-}\piz$ production and calculate the corresponding cross section,
which is shown in Fig.~\ref{xs_KK3pi0}(c) and tabulated in
Table~\ref{2k3pi0_kstar_tab}.
The cross section accounts for the 50\% branching ratio
of $K^{*\pm}\to K^{\pm}\piz$. The final state with only one
$K^*$  does not exceed 10\%, which is within the statistical uncertainty.

\section{\boldmath Intermediate structures in the $\KS\Kpi\ppz$ final
  state}
\label{inter_kskpi2pi0}
Figures~\ref{kskpi2pi0_inter}(a,b) show the $m(K^{\pm}\pi^{\mp})$ and
$m(\KS\pi^{\pm})$ invariant mass distributions for the  $\KS K^{\pm}\pi^{\mp}\ppz$ final
  state. Clear signals from $K^*(892)^0$ and $K^*(892)^{\pm}$ are
  seen. A fit based on a BW and a combinatorial background function
yields $593\pm 53$ and $674\pm55$ events, respectively. We perform
  similar fits for every 0.05\gevcc interval in the hadronic invariant
  mass and calculate the corresponding cross sections, shown in
  Figs.~\ref{xs_KsKpi2pi0}(a,b) and listed in
  Tables~\ref{kskpi2pi0_kstarn_tab} and~\ref{kskpi2pi0_kstarc_tab}. They
  are very similar in shape and 
  values. We have extracted the correlated production yields of the $K^*$ mesons, with
  $115\pm45$ events for the $\epem\to K^{*0}\overline K^{*0}\piz$ and
  $339\pm45$ events for the $\epem\to K^{*+}K^{*-}\piz$ reactions. The
  corresponding cross section for  the latter is in agreement with that in Fig.~\ref{xs_KK3pi0}(c). 

  Figure~\ref{kskpi2pi0_inter}(c) shows the $m(\pi^{\pm}\piz)$
  invariant mass distribution.
A clear peak from the $\rho(770)$  is visible.
A fit based on a BW and a combinatorial background function yields 
$1535\pm84$  $\epem\to\KS  K^{\pm}\rho^{\mp}\piz$ events. 
  The corresponding cross section is shown in
  Fig.~\ref{xs_KsKpi2pi0}(c) and listed in Table~\ref{kskpi2pi0_rho_tab}. A correlated production study yields
$194\pm62$ and $170\pm59$ events for the $\epem\to \KS
K^{*\pm}\rho^{\mp}$ and $\epem\to K^{*0} K^{\pm}\rho^{\mp}$ reactions,
respectively. The number of events is too low to present the cross sections for
these reactions.

\section{\boldmath Intermediate structures in the $\KS K^{\pm}\pi^{\mp}\pipi$ final
  state}
\label{inter_ksk3pi}
\subsection{States with $K^{*0}$, $K^{*\pm}$ , or  $\rho^0(770)$.}
The $\epem\to\KS K^{\pm}\pi^{\mp}\pipi$ reaction is dominated by
$K^{*}(892)$ in the intermediate
states. Figures~\ref{ksk3pi_inter}(a,b) show the $m(K^{\pm}\pi^{\mp})$
and $m(\KS\pi^{\pm})$ invariant mass with fit functions yielding $2587\pm86$
of $K^{*}(892)^{0}$ and $2407\pm85$ of $K^{*}(892)^{\pm}$ events.  The $m(\pipi)$ invariant
mass, shown in Fig.~\ref{ksk3pi_inter}(c), exhibits a large fraction of
events with $\rho(770)$ in the intermediate state, with a fit yielding  $3583\pm140$ such
events. The sum of the three yields exceeds the total number of the $\KS
K^{\pm}\pi^{\mp}\pipi$ events, indicating a correlated production of
above resonances.  Because of the many possible correlations we do not
extract the cross sections for the intermediate
states. Figure~\ref{xs_KsK3pi} shows the mass dependence for the
number of the $\KS K^{\pm}\pi^{\mp}\pipi$ events that include (a)
$K^{*0}$, (b) $K^{*\pm}$, or (c)  $\rho^0(770)$. All distributions
demonstrate relatively large signals from the $J/\psi$ and $\psi(2S)$ resonances.
The $uds$ non-ISR background events are shown by (red) squares. The
mass-dependent behavior for the $K^*(892)$ production is very close
to that for the total $\epem\to\KS K^{\pm}\pi^{\mp}\pipi$ cross
section in Fig.~\ref{xs_bab}(c). By studying the correlations we estimate
that about 40\% of events,
$1024\pm77$,  correspond to  the $\epem\to K^{*0} K^{*\pm}\pi^{\mp}$
reaction. The mass-dependent number of $\rho^0$ events exhibits some
structure, as discussed below. We estimate that $165\pm110$ of $\rho^0$
 events are correlated with the $K^{*0}$ production and $402\pm116$ events
are correlated with the $K^{*\pm}$ production.

\begin{figure}[tbh]
\begin{center}
\includegraphics[width=0.49\linewidth]{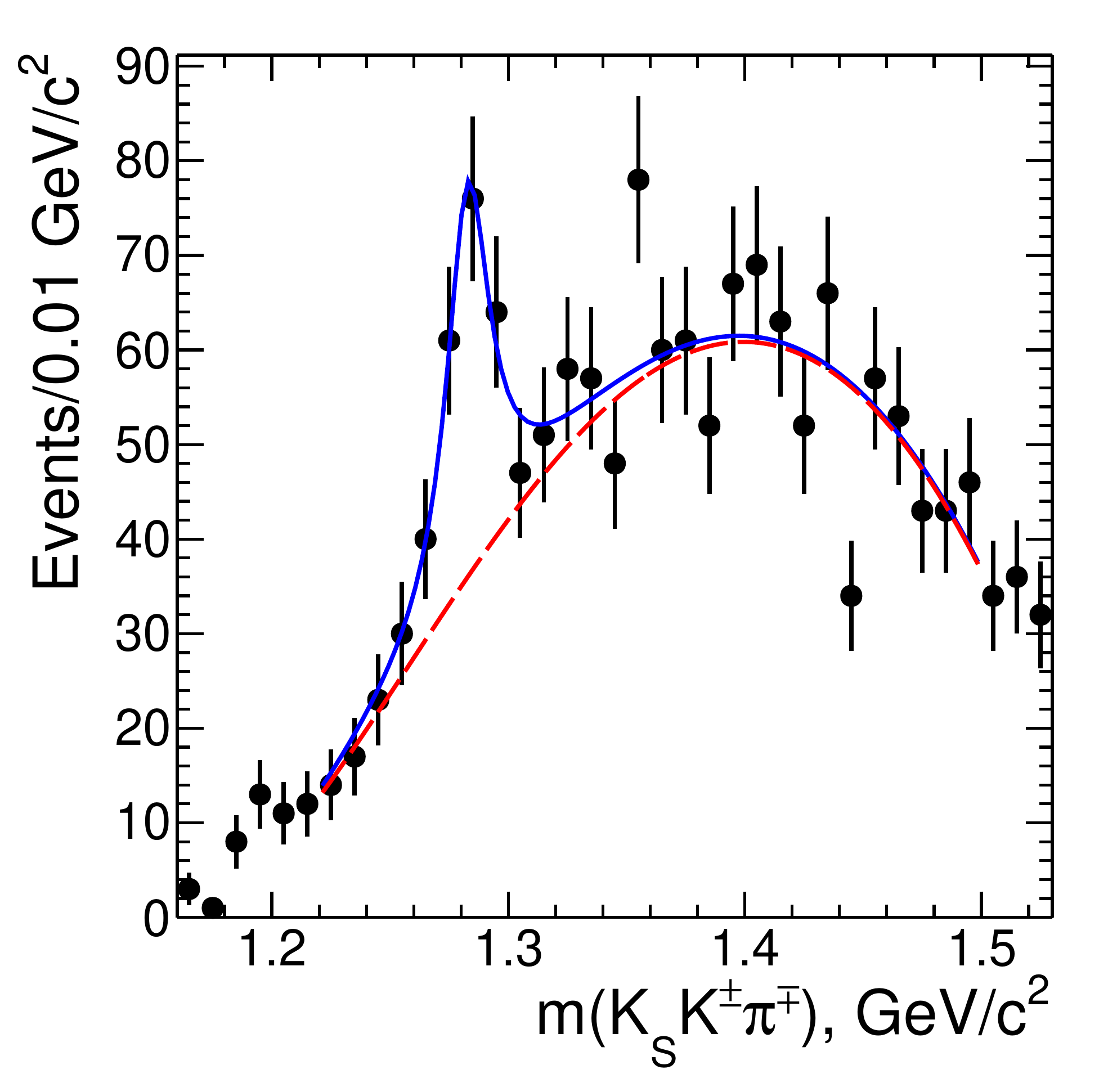}
\put(-25,90){\makebox(0,0)[lb]{\bf(a)}}
\includegraphics[width=0.5\linewidth]{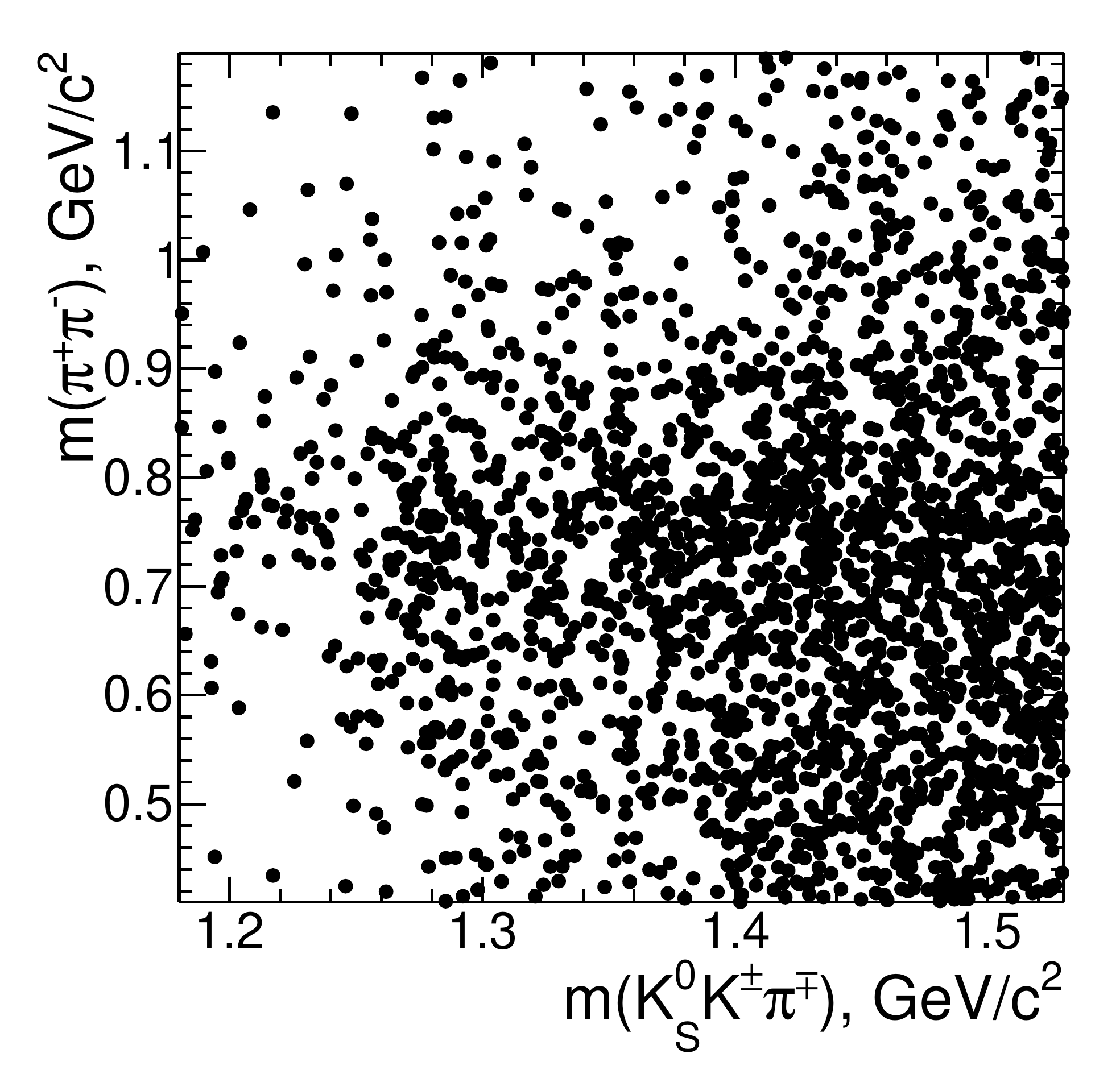}
\put(-95,90){\makebox(0,0)[lb]{\bf(b)}}
\vspace{-0.5cm}
\caption{
(a) The $m(\KS K^{\pm}\pi^{\mp})$ invariant mass distribution.  The solid
curve is the fit to the $f_1(1285)$ signal with the combinatorial background,
shown by the dashed curve. (b) The scatter plot for the $m(\pipi)$ vs $m(\KS
K^{\pm}\pi^{\mp})$  invariant mass. 
}
\label{mkskpi}
\end{center}
\end{figure}

\begin{figure}[tbh]
\begin{center}
\includegraphics[width=0.9\linewidth]{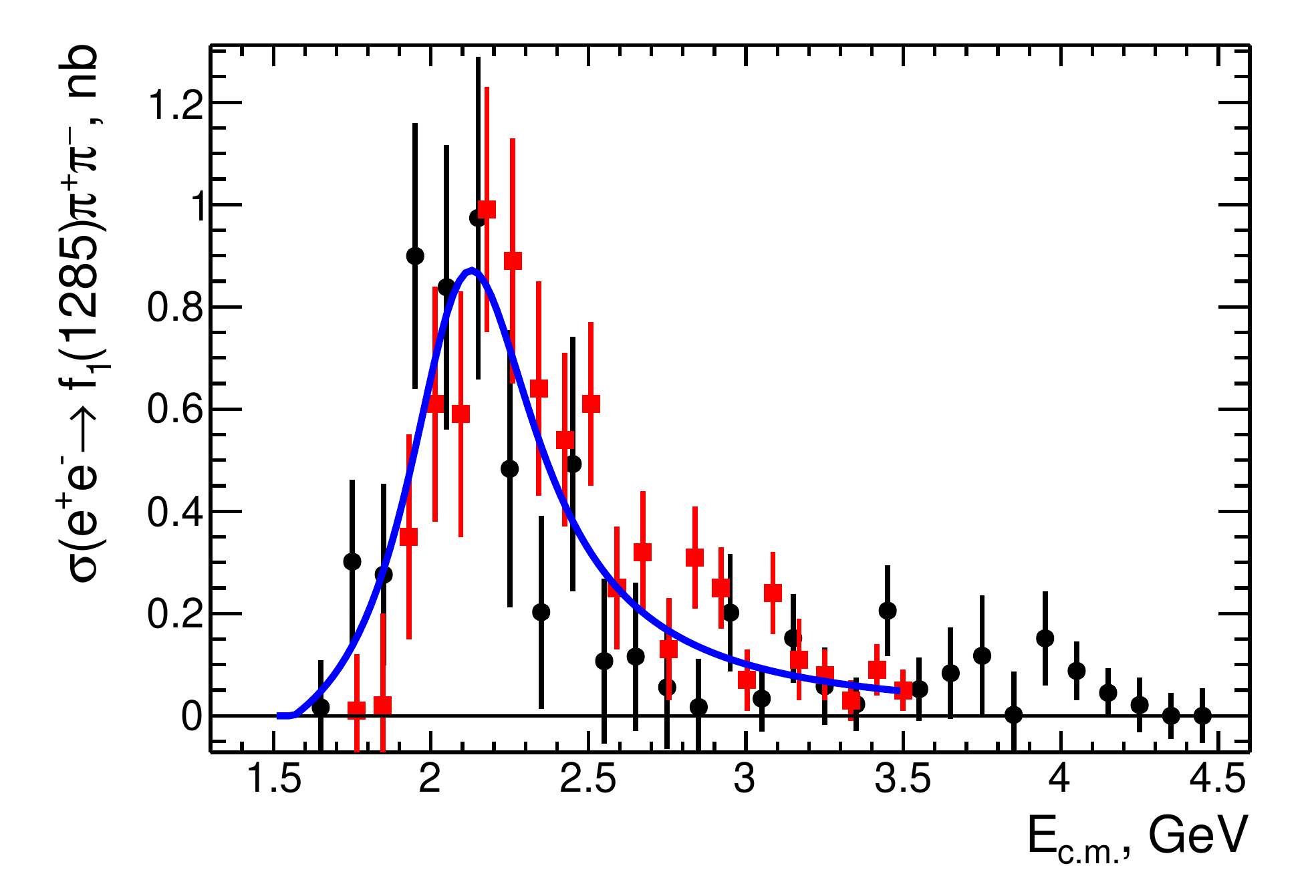}
\vspace{-0.5cm}
\caption{
 The measured $\epem\to f_1(1285)\pipi$ cross section from the present
 analysis (dots) in comparison with previous measurement (squares)~\cite{isr5pi}.
 The solid curve is fit explained in the text.
}
\label{xs_f1pipi}
\end{center}
\end{figure}

 \subsection{The $\epem\to f_1(1285)\rho$ reaction}

 Figure~\ref{mkskpi}(a) shows the $m(\KS K^{\pm}\pi^{\mp})$ invariant
 mass from the $\epem\to\KS K^{\pm}\pi^{\mp}\pipi$ reaction with two
 entries per event. We fit
 the observed peak with a BW function and the combinatorial
 background with a third-order polynomial,
 and obtain $m = 1.283\pm0.002$\gevcc for the mass and $\Gamma =
 0.022\pm0.007$\gev for the width of the resonance. This is interpreted
 as $f_1(1285)$ production in the $\epem\to f_1(1285)\rho$
 reaction. The presence of $\rho^0(770)$ is seen from the scatter
 plot of $m(\pipi)$ vs $m(\KS K^{\pm}\pi^{\mp})$ shown in Fig.~\ref{mkskpi}(b).
 The $\epem\to f_1(1285)\pipi$ cross section was measured
 for the first time by \babar~\cite{isr5pi} where $f_1(1285)$ was
 observed in the $f_1(1285)\to\eta\pipi$ decay. We extract the number of the
 $f_1(1285)$ events in 0.1\gevcc bins of the $\KS K\pi\pipi$ invariant
 mass and calculate the energy-dependent cross section for the $\epem\to
 f_1(1285)\pipi$ reaction shown as dots in Fig.~\ref{xs_f1pipi} 
and listed in Table~\ref{ksk3pi_f1_tab}. The number of events are
 corrected by a factor of three for the missing kaonic channels
 and for the branching fraction of $f_1(1285)\to K\overline K\pi$, 0.09, taken from
 Ref.~\cite{PDG}. Using our results and data and  the BW function suggested in
 Ref.~\cite{isr5pi} we perform a combined fit, and obtain the following
 parameters for the resonance:\\
 ~~$~~~\sigma_0 = 0.85 \pm 0.12 ~{\rm nb}$, \\
 ~~$~~~m = 2.09 \pm 0.03 \gevcc$,\\
 ~~$~~~\Gamma = 0.50 \pm 0.06\gev$,\\
consistent with that in Ref~\cite{isr5pi} with better statistical
accuracy. This structure is included in the PDG~\cite{PDG} as the $\rho(2150)$ resonance.

\begin{figure}[tbh]
\begin{center}
\includegraphics[width=0.50\linewidth]{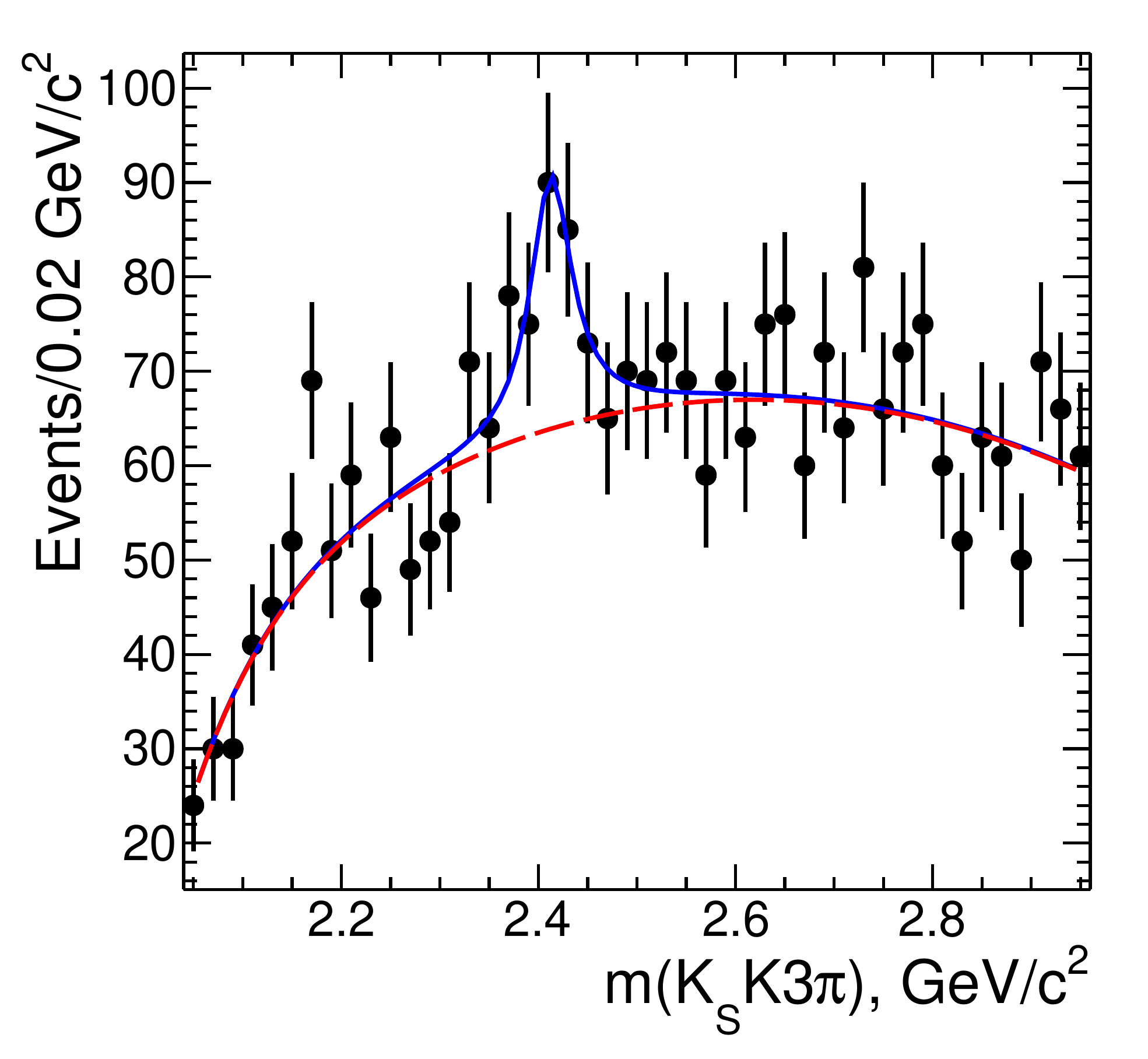}
\put(-25,90){\makebox(0,0)[lb]{\bf(a)}}
\includegraphics[width=0.47\linewidth]{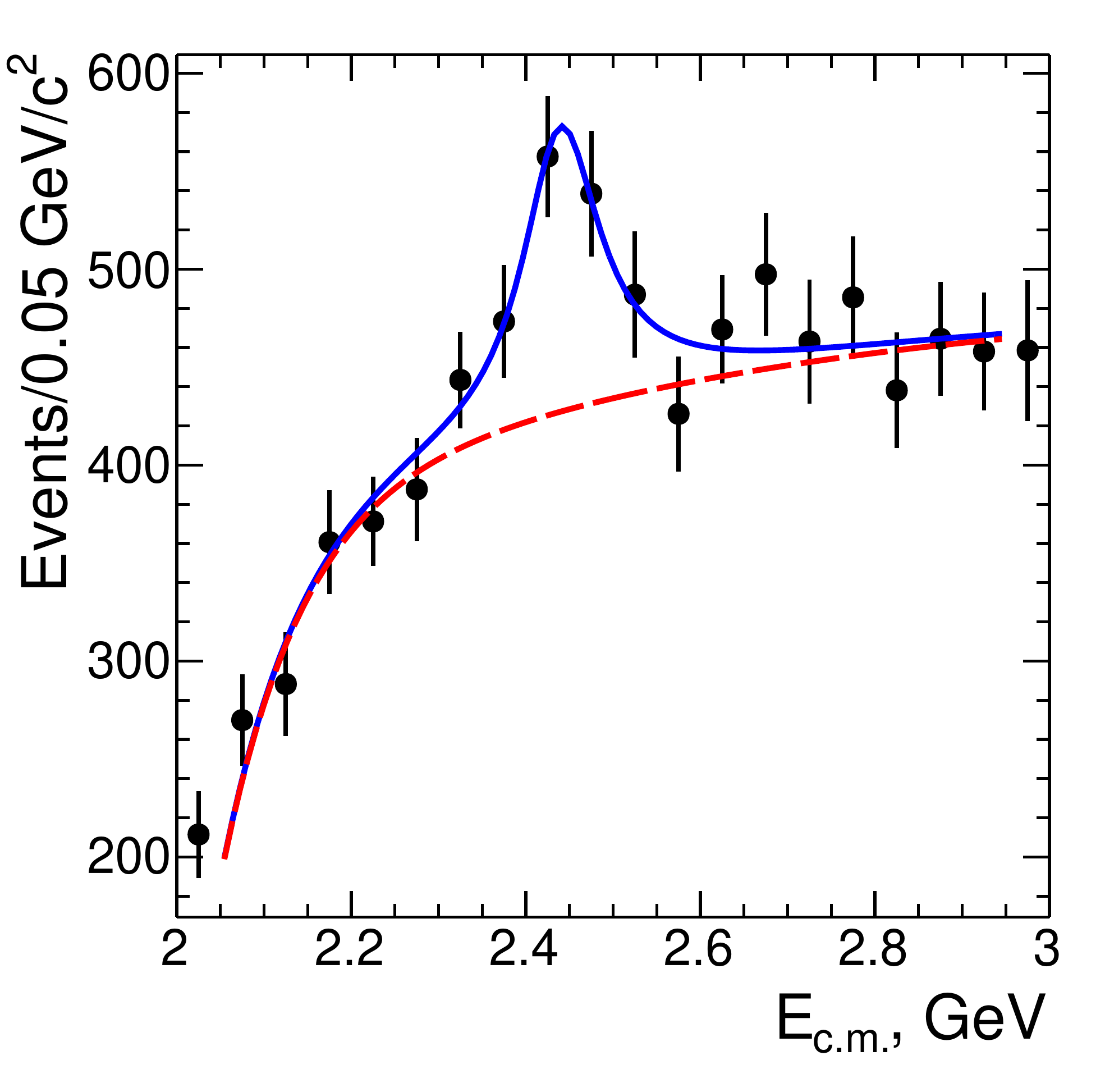}
\put(-25,90){\makebox(0,0)[lb]{\bf(b)}}
\vspace{-0.5cm}
\caption{
(a) The $m(\KS K3\pi)$ invariant mass distribution. The solid
curve shows the fit to the $X(2400)$ signal with a combinatorial background,
shown by the dashed curve.
(b) The sum of events from $\epem\to\KS\Kpi\pipi$, from the $\epem\to\KK\pipi\piz$,
and from the $\epem\to\pipi\pipi\ppz\piz$ reactions. The fit is the
same as for (a).
}
\label{nev_ksk3pi}
\end{center}
\end{figure}

\subsection{Structures at 2.4\gev}

In the cross section for the $\epem\to\KS
\Kpi\pipi$ reaction in Fig.~\ref{xs_bab}(c) some
structures are seen above 2\gev. We plot the number of signal events of
Fig.~\ref{nevents_data}(c) in bins of width 0.02\gevcc in the  hadronic
invariant mass and show them in Fig.~\ref{nev_ksk3pi}(a). An
indication of a bump is seen around 2.4\gevcc. We fit this region
with a BW function and a polynomial function for a non-resonant
background and obtain the following parameters:\\
 $~~~N = 108 \pm 50 ~{\rm events}$, \\
 $~~~m = 2.41 \pm 0.01 \gevcc$,\\
 $~~~\Gamma = 0.051 \pm 0.027\gev$.\\
 The significance of the signal is 2.9 standard deviations.
 Similar behavior with less statistical significance is seen in the
 $\epem\to\KS \Kpi\ppz$ and $\epem\to\KK\ppz\piz$ reactions of Fig.~\ref{xs_bab}(a,b).  We examine
 our other measurements of the cross sections and similar indications
 are seen in the $\epem\to\KK\pipi\piz$ reaction~\cite{isr5pi} and in
 the  $\epem\to\pipi\pipi\ppz\piz$ reaction~\cite{isr4pi3pi0}. We
 combine events from these two reactions with that from
 Fig.~\ref{nev_ksk3pi}(a) in 0.05\mev bins and perform a similar fit, shown in
 Fig.~\ref{nev_ksk3pi}(b). The signal has 3.5 standard deviations
 significance with the following parameters:\\
 $~~~N = 487 \pm 251 ~{\rm events}$, \\
 $~~~m = 2.44 \pm 0.02 \gevcc$,\\
 $~~~\Gamma = 0.107 \pm 0.049\gev$.\\
 This resonance structure was also seen  and discussed by
 ~\babar~\cite{isr2k2pi} in the $\epem\to\KK f_0(980)$
 (and not well seen in  $\epem\to\phi f_0(980)$)
 reaction, and  was studied by the
 Belle~\cite{Belle2400} experiment. Later, Shen and Yuan~\cite{shen} performed a fit to
 the  structure called X(2400)  using the combined data of the Belle and ~\babar~
 experiments. The mass and the width were determined to be $2436\pm26$\mevcc
and $121 \pm 35$\mev, respectively. However, its statistical
significance was less than 3$\sigma$, and the structure can be explained as a
threshold behaviour of the $\epem\to\phi f_0(1370)$ reaction.

\begin{figure}[tbh]
\begin{center}
\includegraphics[width=0.49\linewidth]{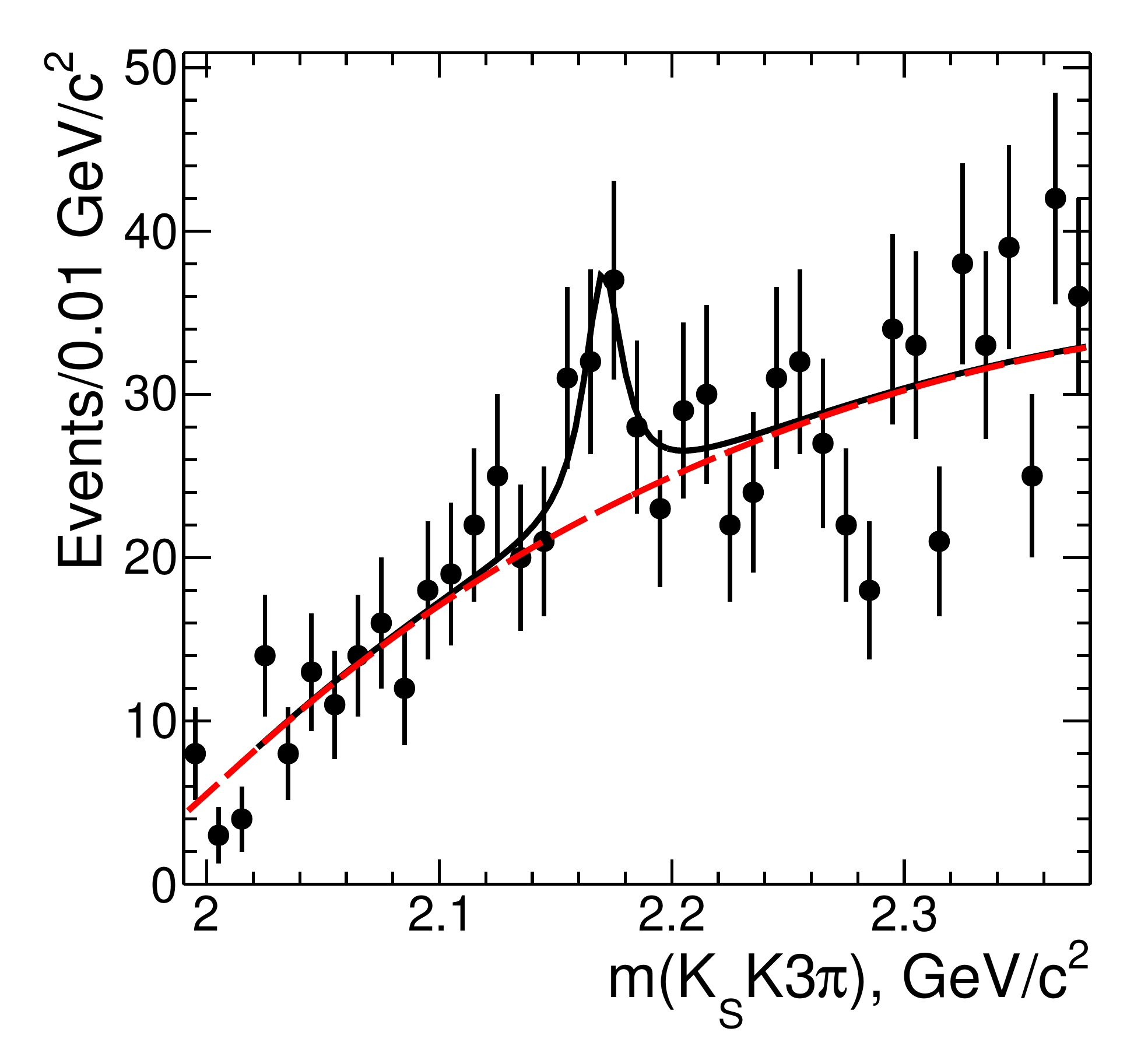}
\put(-90,90){\makebox(0,0)[lb]{\bf(a)}}
\includegraphics[width=0.49\linewidth]{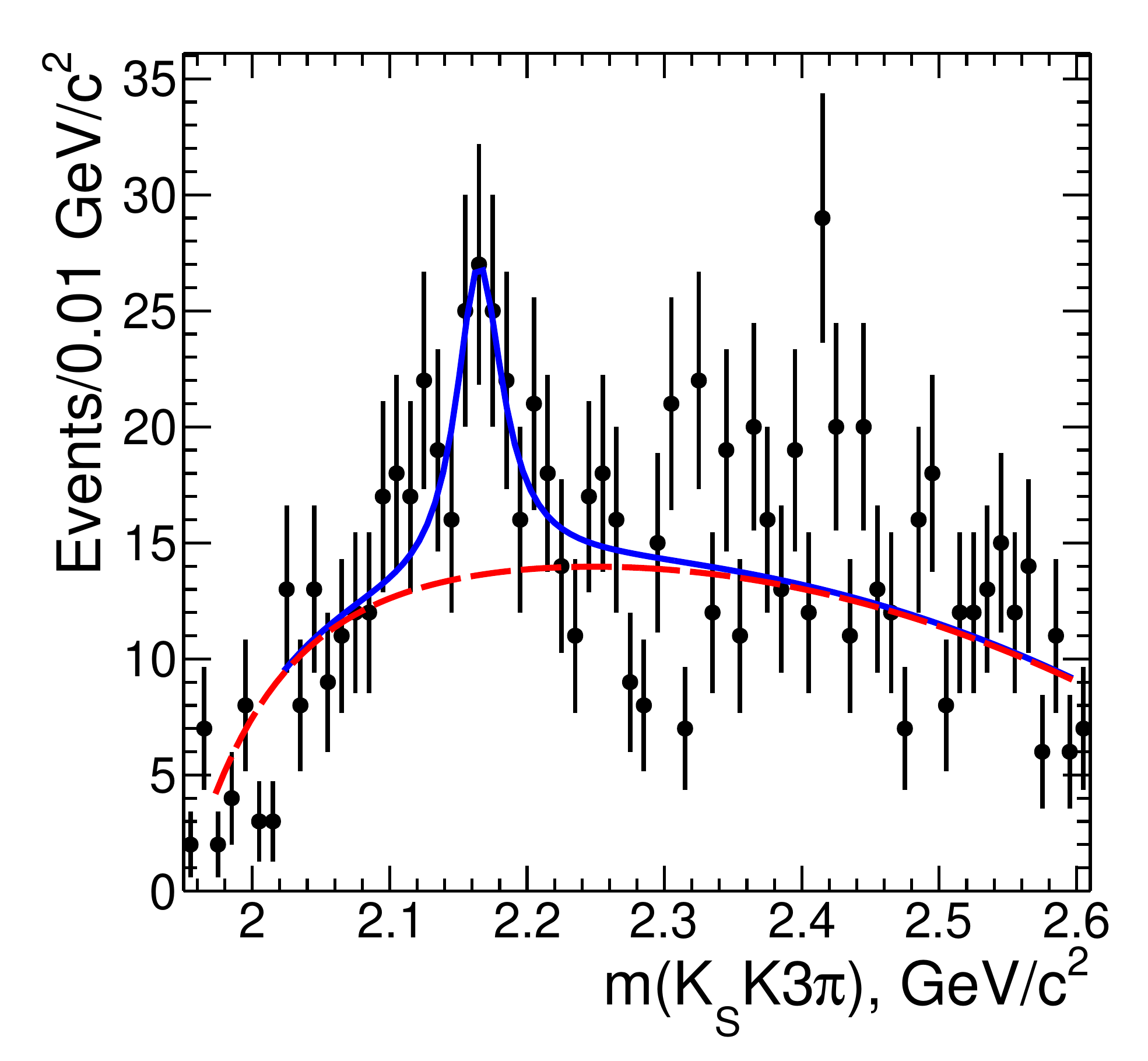}
\put(-28,90){\makebox(0,0)[lb]{\bf(b)}}
\vspace{-0.5cm}
\caption{
(a) The $m(\KS K3\pi)$ invariant mass distribution
around 2.17\gevcc.
The curves are fits to the $\phi(2170)$ signal with a combinatorial background.
(b) Same as (a) with the additional requirement
$m(\pipi)<0.7$\gevcc. The solid curve is a fit to the $\phi(2170)$ signal,
with the combinatorial background shown by the dashed curve.
}
\label{nev_2170}
\end{center}
\end{figure}

\begin{figure*}[tbh]
\includegraphics[width=0.32\linewidth]{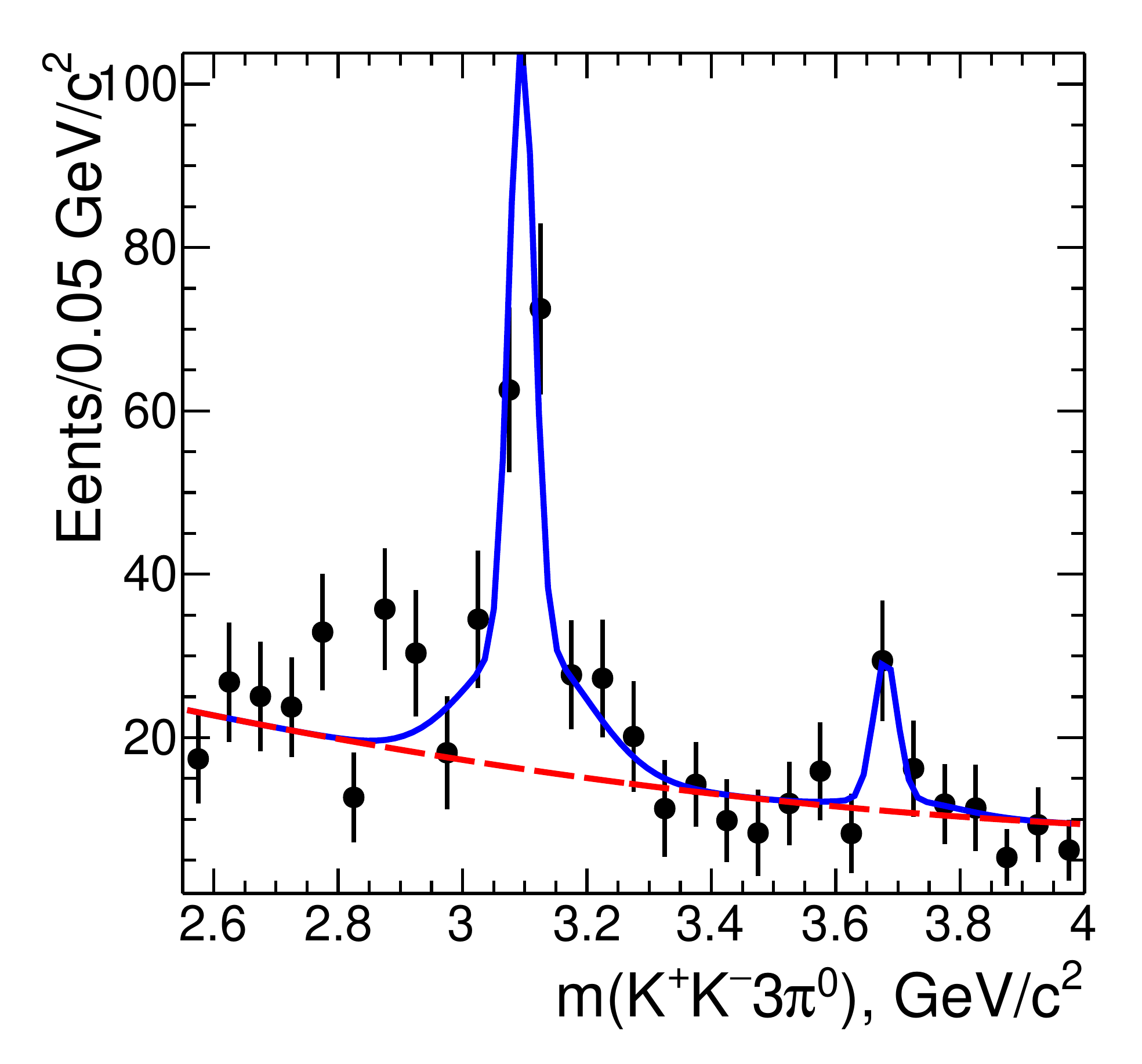}
\put(-40,90){\makebox(0,0)[lb]{\bf(a)}}
\includegraphics[width=0.32\linewidth]{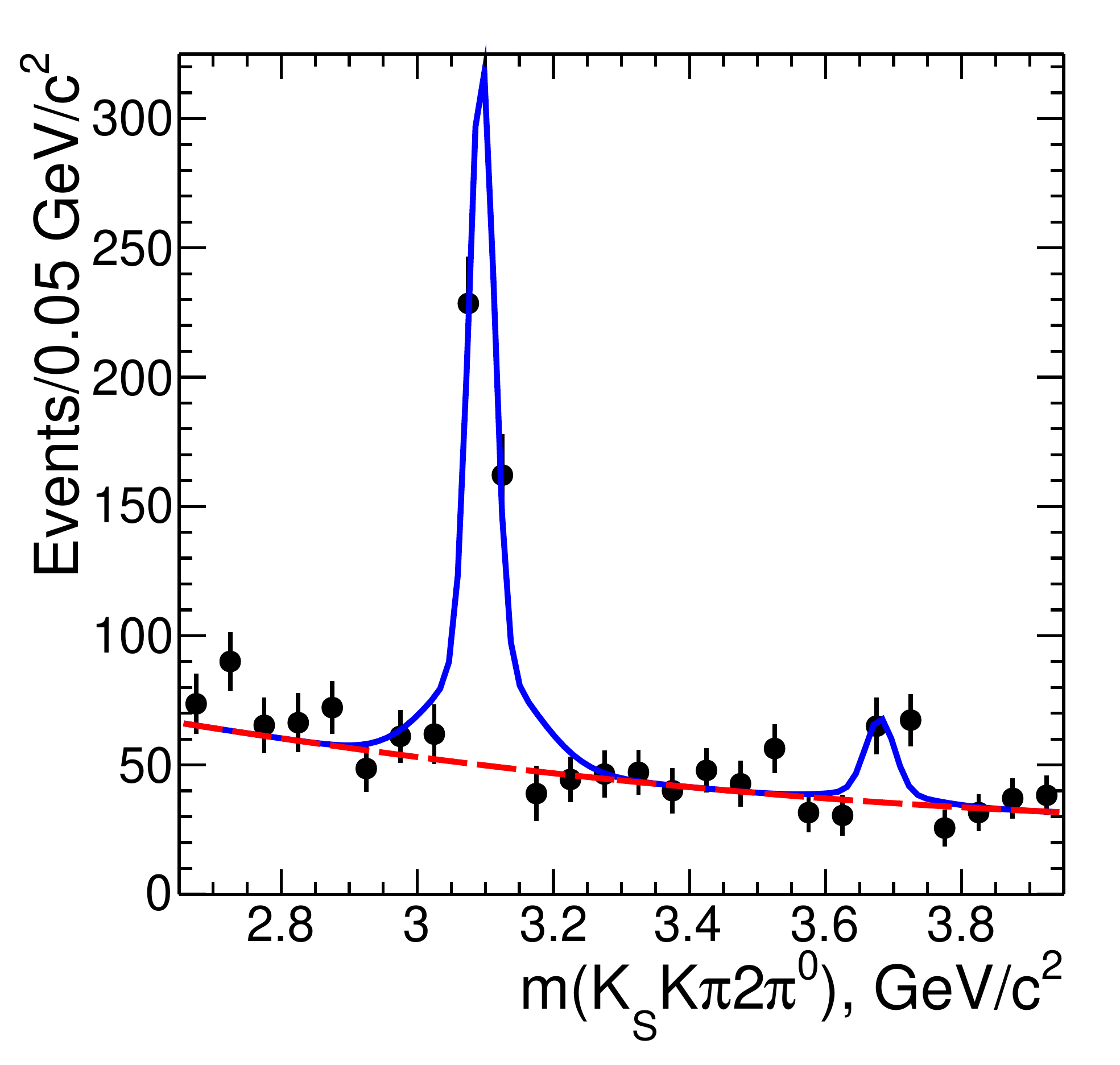}
\put(-40,90){\makebox(0,0)[lb]{\bf(b)}}
\includegraphics[width=0.32\linewidth]{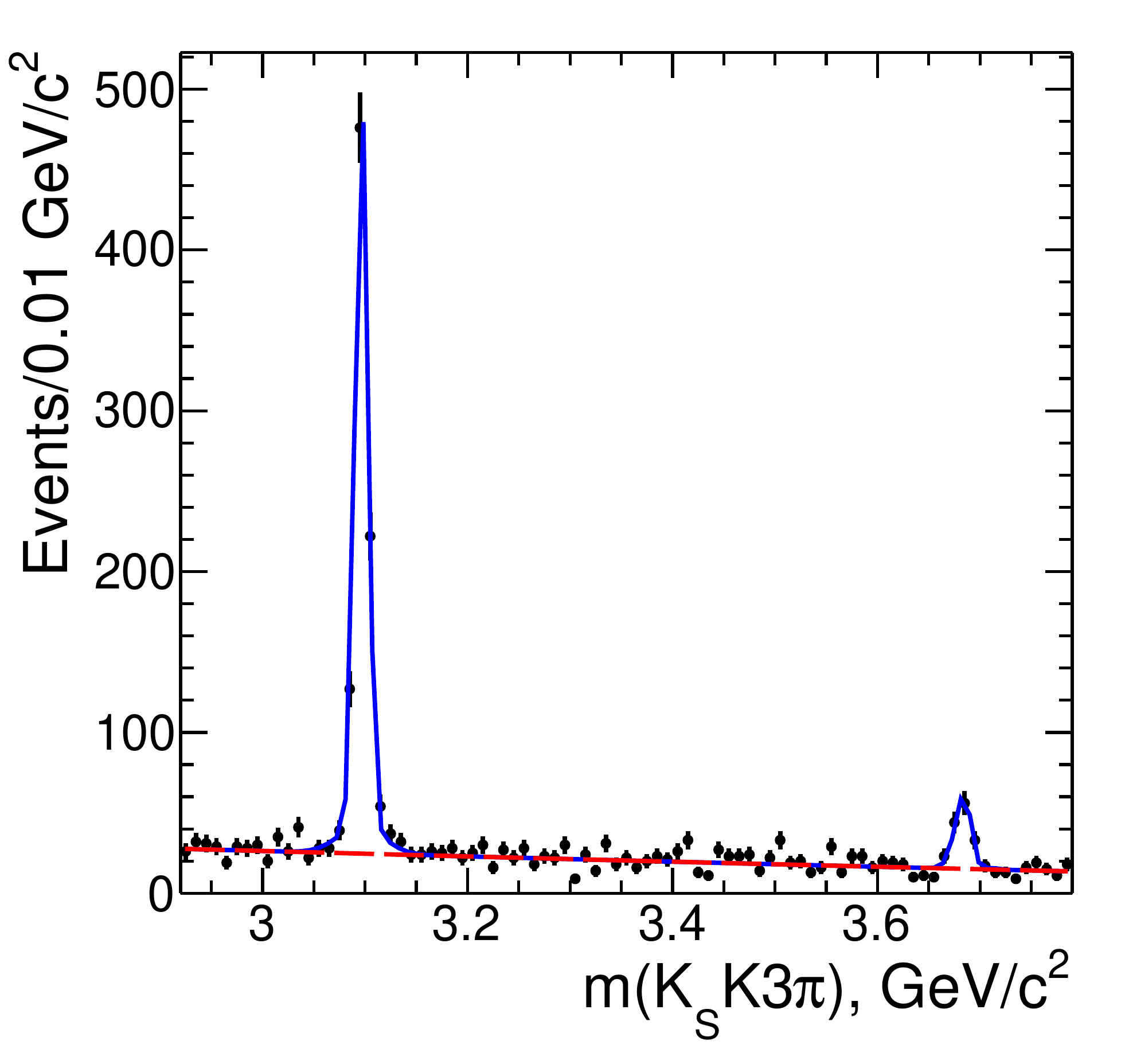}
\put(-40,90){\makebox(0,0)[lb]{\bf(c)}}
\vspace{-0.3cm}
\caption{
The $J/\psi$ invariant mass region for  (a) the $\KK\ppz\piz$, (b) 
the $\KS \Kpi\ppz$, and (c) the $\KS\Kpi\pipi$ events.
The curves show the fit
functions described in the text.
}
\label{jpsi}
\end{figure*}    

\subsection{Structure at 2.17\gev}

Figure~\ref{nev_ksk3pi}(a) also shows a few points in the region of the $\phi(2170)$~\cite{PDG}
resonance where the data lie above the fit.
We investigate this excess further.
  Figure~\ref{nev_2170}(a) shows the same plot with a 0.01\gevcc bin
 width for the $m(\KS K3\pi)$ invariant mass. A signal peak is seen, and a fit
 with a BW and a polynomial function yields a result with about 2.5$\sigma$
 significance.  We apply additional selection criteria trying to increase possible
 signal. Figure~\ref{nev_2170}(b) shows a similar plot with the additional requirement
 $m(\pipi)<0.7$\gevcc, which decreases the contribution from $\rho(770)$ in this region. 
 The signal is more prominent and the fit gives:\\
 $~~~N = 86 \pm 34 ~{\rm events}$, \\
 $~~~m = 2.164 \pm 0.006 \gevcc$,\\
 $~~~\Gamma = 0.041 \pm 0.020\gev$,\\
 with 3.9$\sigma$ significance. Additional selections that enlarge
 the  contribution from
 $K^{*0}$ or (and) $K^{*\pm}$ do not increase the signal. 
 The observed signal in the $\KS\Kpi\pipi$ final state could be one
 more decay channel for the $\phi(2170)$ resonance.

\section{\bf\boldmath The $J/\psi$ region}

Figure~\ref{jpsi} shows an expanded view of the $J/\psi$ mass region 
from Fig.~\ref{nevents_data} for the selected data sample. 
Signals from  $J/\psi$ and $\psi(2S)$ to $\KK\ppz\piz$ (a), $\KS
\Kpi\ppz$ (b), and $\KS \Kpi\pipi$ (c)  are  seen. 
The observed peak shapes are not purely Gaussian because of radiation
effects and resolution, and for the fit we  take shapes from the simulated signal distributions.
  The sum of two Gaussians  describes the shape well. The non-resonant
  background distribution is described by a second-order
  polynomial function in this region.
  We obtain $149\pm21$, $369\pm32$, and $815\pm31$  $J/\psi$ events
for the reactions shown in Fig.~\ref{jpsi}(a), (b), and (c), respectively.
The corresponding results for  $\psi(2S)$ events are
$23\pm19$, $44\pm15$, and $90\pm12$.
Using the results for the number of events, the detection
efficiency, and the ISR luminosity,
we determine the product of the decay rate to hadrons and the electronic width:
\begin{eqnarray}
  B_{J/\psi\to {\rm had}}\cdot\Gamma^{J/\psi}_{ee}
  = \frac{N(J/\psi\to {\rm had}) \cdot m_{J/\psi}^2}%
           {6\pi^2\cdot d{\cal L}/dE\cdot\epsilon^{\rm
  MC}\cdot\epsilon^{\rm corr}\cdot C}, \label{jpsieq}
\end{eqnarray}
where  $d{\cal L}/dE = 180~\invnb/\mev$ is the ISR luminosity 
at the $J/\psi$ mass $m_{J/\psi}$, $\epsilon^{\rm MC}$ is the detection
efficiency from simulation with the corrections $\epsilon^{\rm corr}$, discussed in 
Sec.~\ref{sec:Systematics},
and  $C = 3.894\times 10^{11}~\nb\mev^2$ is a
conversion constant ~\cite{PDG}. We estimate the systematic uncertainty for this
region to be 10\% according to Table~\ref{error_tab}.  

Using $\Gamma^{J/\psi}_{ee} =5.53\pm0.10~\kev$ ~\cite{PDG}, we obtain
$B_{J/\psi\to {\rm had}}$ for each inclusive final state. The measured
products, derived decay rates, and results of previous measurements
from the PDG~\cite{PDG} are listed in Table~\ref{jpsitab}. 

Using Eq.(\ref{jpsieq}) and the result $d{\cal L}/dE =
  228~\invnb/\mev$  at the $\psi(2S)$ mass, we obtain the
 products $B_{\psi(2S)\to {\rm had}}\cdot\Gamma^{\psi(2S)}_{ee}$ for each
 decay channel.
With $\Gamma^{\psi(2S)}_{ee} =2.33\pm0.04~\kev$ ~\cite{PDG}  we
find the corresponding $B_{\psi(2S)\to {\rm had}} $ and list them in
Table~\ref{jpsitab}.
These results represent the first measurements for these decay channels.

\begin{figure}[tbh]
\begin{center}
\includegraphics[width=0.49\linewidth,height=0.46\linewidth]{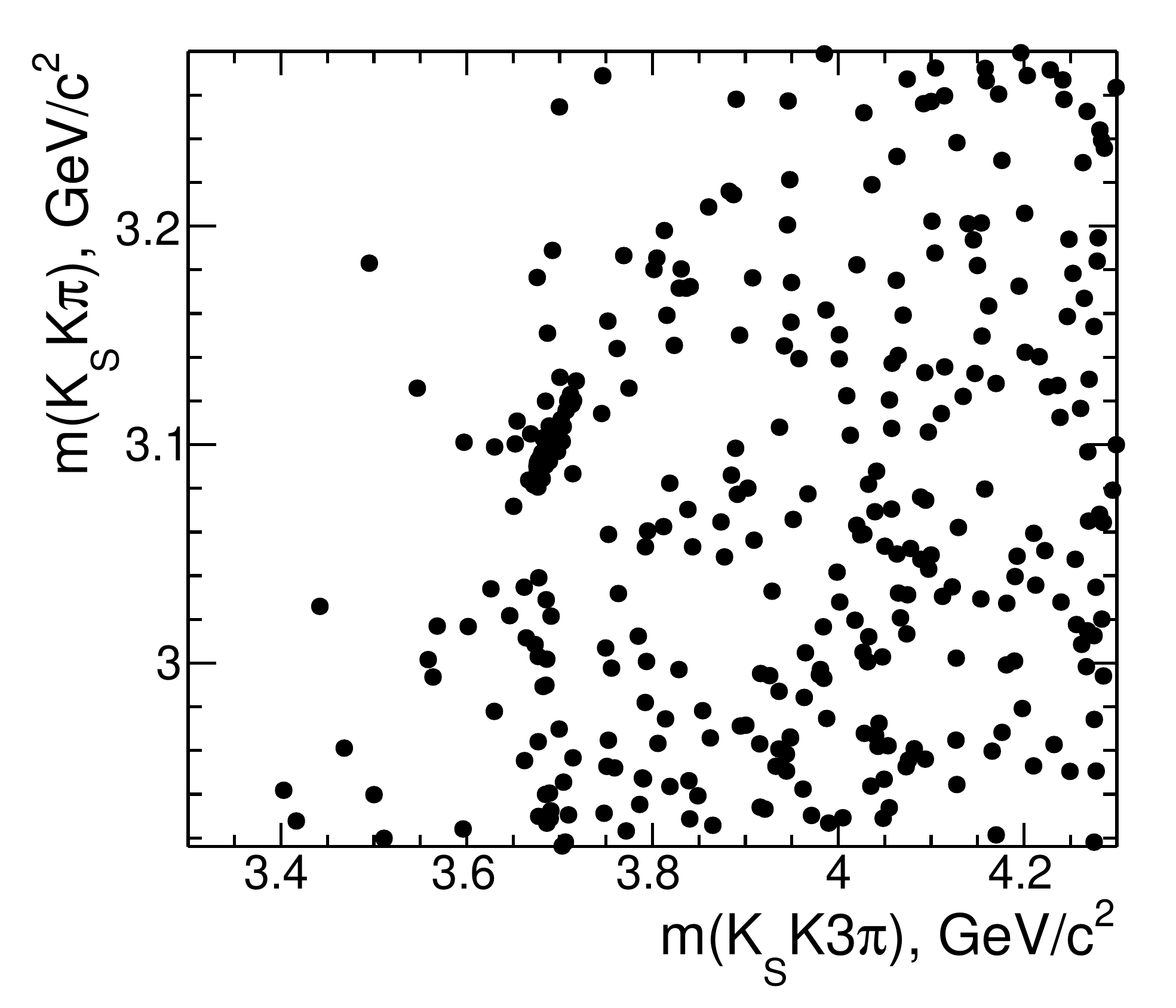}
\put(-85,90){\makebox(0,0)[lb]{\bf(a)}}
\includegraphics[width=0.49\linewidth]{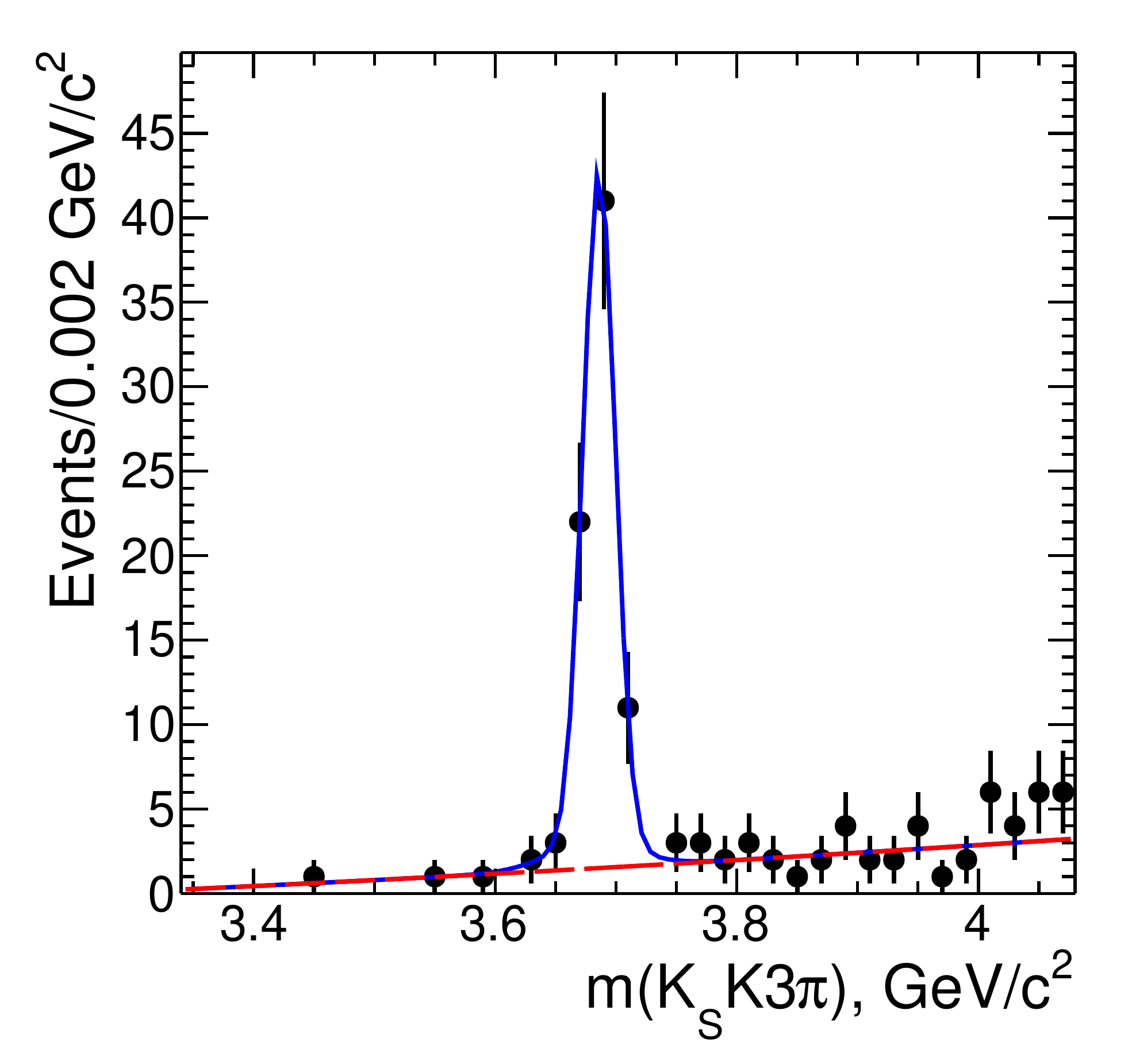}
\put(-25,90){\makebox(0,0)[lb]{\bf(b)}}
\vspace{-0.5cm}
\caption{
(a) The $m(\KS K\pi)$ invariant mass  vs $m(\KS
K3\pi)$ invariant mass for the $\KS\Kpi\pipi$ events around the $\psi(2S)$ signal.
(b) The $m(\KS K3\pi)$ invariant mass distribution
with the $J/\psi$ selection in the $\KS\Kpi$ invariant mass.
 The solid curve shows the fit to the $\psi(2S))$ signal, with a linear background
 shown by the dashed line.
}
\label{psitoJ}
\end{center}
\end{figure}

\begin{table*}[tbh]
\caption{
  Summary of the $J/\psi$ and $\psi(2S)$ branching fractions. Each value is quoted with its statistical and
    systematic uncertainties.
  }
\label{jpsitab}
% \begin{ruledtabular}
\begin{tabular}{r@{$\cdot$}l  r@{.}l@{$\pm$}l@{$\pm$}l 
                              r@{.}l@{$\pm$}l@{$\pm$}l
  r@{.}l@{$\pm$}l }
  \hline
\multicolumn{2}{c}{Measured} & \multicolumn{4}{c}{Measured}    &  
\multicolumn{7}{c}{$J/\psi$ or $\psi(2S)$ Branching Fraction  (10$^{-3}$)}\\
\multicolumn{2}{c}{Quantity} & \multicolumn{4}{c}{Value (\ev)} &
\multicolumn{4}{c}{Derived, this work}    & 
\multicolumn{3}{c}{PDG~\cite{PDG}} \\
\hline
$\Gamma^{J/\psi}_{ee}$  &  $B_{J/\psi \to \KK\ppz\piz}$  &
  8& 9 & 1.3 & 0.9  &   ~~~1&6 & 0.2 & 0.2  &   \multicolumn{3}{c}{no entry}  \\

$\Gamma^{J/\psi}_{ee}$  &  $B_{J/\psi  \to \eta\KK}
                           \cdot B_{\eta   \to \ppz\piz}$  &
  1&55 & 0.51 & 0.16  &   0&85 & 0.28 & 0.09  & \multicolumn{3}{c}{no entry}  \\
  
$\Gamma^{J/\psi}_{ee}$  &  $B_{J/\psi  \to \phi\eta}  
                        \cdot B_{\phi\to \KK} \cdot B_{\eta   \to \ppz\piz}$  &
   0&64& 0.26 & 0.06 &   0&72 & 0.29 & 0.07  & ~~~~0&74 & 0.08\\

$\Gamma^{J/\psi}_{ee}$  &  $B_{J/\psi  \to K^{*+}K^{*-}\piz}
                           \cdot B_{K^{*+}  \to K^+\piz}\cdot B_{K^{*-}  \to K^-\piz}$  &
  6&9 & 1.2 & 0.7  &   5&0 & 0.9 & 0.5  &    \multicolumn{3}{c}{no entry}   \\

$\Gamma^{\psi(2S)}_{ee}$  &  $B_{\psi(2S) \to\KK\ppz\piz} $  &
   1&54& 0.63& 0.15 &   0&66  & 0.27  & 0.07   & \multicolumn{3}{c}{no entry} \\
  
  $\Gamma^{\psi(2S)}_{ee}$  &  $B_{\psi(2S) \to J/\psi\ppz}
                              \cdot B_{J/\psi \to \KK\piz}$  &
   1&31& 0.35& 0.13 &   3&1  & 0.8  & 0.3   &  ~~~~2&88 & 0.13\\
  
 $\Gamma^{\psi(2S)}_{ee}$  &  $B_{\psi(2S) \to\eta\KK} \cdot B_{\eta   \to \ppz\piz}$  &
   ~~~~$<$0 & \multicolumn{3}{l}{2 at 90\% C.L.} & ~~~~ $<$0 &
                                                     \multicolumn{3}{l}{25
                                                     at 90\% C.L.} &      \multicolumn{3}{c}{no entry}\\ 

$\Gamma^{\psi(2S)}_{ee}$  &  $B_{\psi(2S)  \to K^{*+}K^{*-}\piz}
                           \cdot B_{K^{*+}  \to K^+\piz}\cdot B_{K^{*-}  \to K^-\piz}$  &
  0&94 & 0.45 & 0.10  &   1&6 & 0.8 & 0.2  &    \multicolumn{3}{c}{no entry}   \\

\hline
$\Gamma^{J/\psi}_{ee}$  &  $B_{J/\psi \to \KS\Kpi\ppz}$  &
  29& 3 & 2.6 & 2.9  &   ~~~5&3 & 0.5 & 0.5  &   \multicolumn{3}{c}{no entry}  \\

  $\Gamma^{J/\psi}_{ee}$  &  $B_{J/\psi \to K^{*\pm} K^{\mp}\ppz}
                                                          \cdot B_{K^{*\pm}  \to K^0\pi^{\pm}}\cdot B_{K^0  \to \KS}$     &
  2& 89 & 0.52 & 0.28  &   ~~~2&0 & 0.4 & 0.2  &   \multicolumn{3}{c}{no entry}  \\

 $\Gamma^{J/\psi}_{ee}$  &  $B_{J/\psi \to K^0 K^{*0} \ppz}
                                                          \cdot B_{K^{*0}  \to \Kpi}\cdot B_{K^0  \to \KS}$     &
  3& 73 & 0.53 & 0.37  &   ~~~2&7 & 0.4 & 0.3  &   \multicolumn{3}{c}{no entry}  \\

$\Gamma^{J/\psi}_{ee}$  &  $B_{J/\psi \to \KS K^{\pm}\rho^{\mp}\piz}$  &
  16& 0 & 4.1 & 1.6  &   ~~~2&9 & 0.7 & 0.3  &   \multicolumn{3}{c}{no entry}  \\
  
$\Gamma^{\psi(2S)}_{ee}$  &  $B_{\psi(2S) \to\KS\Kpi\ppz} $  &
   4&0& 1.4& 0.4 &   1&7  & 0.6  & 0.2  & \multicolumn{3}{c}{no entry} \\

$\Gamma^{\psi(2S)}_{ee}$  &  $B_{\psi(2S) \to J/\psi\ppz}
                              \cdot B_{J/\psi \to \KS\Kpi}$  &
   2&36& 0.59& 0.24 &   5&5  & 1.4  & 0.6   &  ~~~~5&6 & 0.5\\

 $\Gamma^{\psi(2S)}_{ee}$  &  $B_{\psi(2S) \to K^{*\pm} K^{\mp}\ppz}
                                                         \cdot B_{K^{*\pm}  \to K^0\pi^{\pm}}\cdot B_{K^0  \to \KS}$     &
  0& 54 & 0.22 & 0.05  &   ~~~0&92 & 0.37 & 0.09  &   \multicolumn{3}{c}{no entry}  \\

 $\Gamma^{\psi(2S)}_{ee}$  &  $B_{\psi(2S) \to K^0 K^{*0} \ppz}
                                                          \cdot B_{K^{*0}  \to \Kpi}\cdot B_{K^0  \to \KS}$     &
  0& 47 & 0.19 & 0.05  &   ~~~0&81 & 0.32 & 0.08  &   \multicolumn{3}{c}{no entry}  \\
  
$\Gamma^{\psi(2S)}_{ee}$  &  $B_{\psi(2S) \to\KS K^{\pm}\rho^{\mp}\piz} $  &
   ~~~~$<$1 & \multicolumn{3}{l}{6 at 90\% C.L.} & ~~~~ $<$0 &
                                                     \multicolumn{3}{l}{6
                                                     at 90\% C.L.} &      \multicolumn{3}{c}{no entry}\\ 
  \hline
$\Gamma^{J/\psi}_{ee}$  &  $B_{J/\psi \to \KS\Kpi\pipi}$  &
  34& 6 & 1.4 & 1.8  &   ~~~6&2 & 0.2 & 0.4  &   \multicolumn{3}{c}{no entry}  \\

$\Gamma^{J/\psi}_{ee}$  &  $B_{J/\psi \to K^{*\pm} K^{*0}\pi^{\mp}}
                      \cdot B_{K^{*\pm}  \to K^0\pi^{\pm}} \cdot B_{K^{*0}  \to \Kpi}\cdot B_{K^0  \to \KS}$     &
  5& 9 & 1.0 & 0.6  &   ~~~8&5 & 1.5 & 0.9  &   \multicolumn{3}{c}{no entry}  \\
  
 $\Gamma^{J/\psi}_{ee}$  &  $B_{J/\psi \to K^{*\pm} K^{\mp}\pipi}
                                                          \cdot B_{K^{*\pm}  \to K^0\pi^{\pm}}\cdot B_{K^0  \to \KS}$     &
  6& 2 & 2.1 & 0.6  &   ~~~4&4 & 1.5 & 0.4  &   \multicolumn{3}{c}{no entry}  \\

 $\Gamma^{J/\psi}_{ee}$  &  $B_{J/\psi \to K^0 K^{*0} \pipi}
                                                          \cdot B_{K^{*0}  \to \Kpi}\cdot B_{K^0  \to \KS}$     &
  6& 3 & 2.1 & 0.6  &   ~~~4&5 & 1.5 & 0.5  &   \multicolumn{3}{c}{no entry}  \\
  
$\Gamma^{J/\psi}_{ee}$  &  $B_{J/\psi \to \KS\Kpi\rho^0}$  &
  17& 3 & 2.1 & 1.7  &   ~~~3&1 & 0.4 & 0.3  &   \multicolumn{3}{c}{no entry}  \\
  
$\Gamma^{\psi(2S)}_{ee}$  &  $B_{\psi(2S) \to \KS\Kpi\pipi}$  &
  5& 1 & 0.7 & 0.4  &   ~~~2&2 & 0.3 & 0.2  &   \multicolumn{3}{c}{no entry}  \\
  
$\Gamma^{\psi(2S)}_{ee}$  &  $B_{\psi(2S) \to J/\psi\pipi}
                              \cdot B_{J/\psi \to \KS\Kpi}$  &
   4&14& 0.55& 0.29 &   5&1  & 0.7  & 0.1   &  ~~~~5&6 & 0.5\\

$\Gamma^{\psi(2S)}_{ee}$  &  $B_{\psi(2S) \to\KS \Kpi\rho^{0}} $  &
   ~~~~$<$1 & \multicolumn{3}{l}{6 at 90\% C.L.} & ~~~~ $<$0 &
                                                     \multicolumn{3}{l}{6
                                                     at 90\% C.L.} &      \multicolumn{3}{c}{no entry}\\ 
  
\hline
\end{tabular}
%\end{ruledtabular}
\end{table*}

The observed $\psi(2S)$ signals are partly due to the $\psi(2S)\to
J/\psi\pipi, J/\psi\ppz$ transitions. Indeed, if we plot $\KK\piz$ or $\KS
\Kpi$ invariant masses vs the full hadronic system mass, the
$J/\psi$ signal is seen. An example is shown in Fig.~\ref{psitoJ}(a)
where the $m(\KS K\pi)$ invariant mass is plotted vs the $m(\KS
K3\pi)$ invariant mass. The $J/\psi\to \KS
\Kpi$ decay signal is seen. We select this signal in the $\pm
50$\mevcc window around the $J/\psi$ mass, and plot the $m(\KS
K3\pi)$ invariant mass for the selected events, shown
in Fig.~\ref{psitoJ}(b) by dots. We fit the $\psi(2S)$ signal with the
sum of the Gaussian
 and linear functions (solid curve) and obtain $73\pm10$
events over the background (dashed curve) for the $\psi(2S)\to J/\psi\pipi, J/\psi\to\KS
\Kpi$ transition. A similar study gives $20\pm 5$ and
$26\pm6$ events for the  $\psi(2S)\to J/\psi\ppz, J/\psi\to\KK\piz$
and  $\psi(2S)\to J/\psi\ppz, J/\psi\to\KS\Kpi$ decay
channels, respectively. Using Eq.~(\ref{jpsieq}) we calculate the product of branching
fraction and electronic width for each decay chain and list the
results in Table~\ref{jpsitab}. Because the  $\psi(2S)\to
J/\psi\pipi, J/\psi\ppz$ transition rates are known with good accuracy, we
calculate the $J/\psi$ decay rates and compare them with the direct
measurements, presented in the fourth column of
Table~\ref{jpsitab}~\cite{PDG} and find  good agreement.

Because the $J/\psi$ and $\psi(2S)$ signals are narrow with relatively
small background we are able to determine exclusive decay rates that include
narrow $\eta$, $\phi$, $K^*$, $\rho$ intermediate resonances or correlated
production of them. Using event
selections for the intermediate structures described in
Secs.~\ref{inter_2k3pi0},\ref{inter_kskpi2pi0}, and~\ref{inter_ksk3pi} we 
extract  corresponding numbers of signal events and calculate the
product of branching fractions and electronic width. The obtained
values are listed in Table~\ref{jpsitab}. Using known values for the
electronic widths and known decay rates of narrow states we derived
the corresponding branching fractions for the $J/\psi$ and $\psi(2S)$
resonances, listed in the third column of Table~\ref{jpsitab}. Almost all
of them are measured for the first time.

\section{Summary}
\label{sec:Summary}
%\begin{itemize}
\noindent
The excellent photon-energy and charged-particle momentum 
resolutions, as well as  the particle
identification capabilities of the \babar\ detector, allow  the
reconstruction of the 
$\KK\ppz\piz$, $\KS \Kpi\ppz$, and $\KS\Kpi\pipi$
final states produced at  center-of-mass energies below 4.5\gev 
via ISR  in data collected at the $\Upsilon(4S)$  center-of-mass region.

The cross sections for the $\epem\to\KK\ppz\piz$,
the $\epem\to\KS\Kpi\ppz$, 
and the $\epem\to\KS\Kpi\pipi$ reactions
have been measured for the first time. The accuracy is about 10\%.
The cross sections for these channels can help to estimate
the contribution from the other $K\overline K 3\pi$ combinations and can
improve the reliability of the HVP calculation.

The selected multi-hadronic final states in the broad range of accessible
energies provide new information on hadron spectroscopy. The  
observed contribution from intermediate narrow $\eta$, $K^*$, and $\rho$
resonances  provide  additional information for the hadronic contribution
calculation of the muon $g_\mu-2$.  

The initial-state radiation events allow a study of $J/\psi$ and
$\psi(2S)$ production and a measurement of the corresponding products of
the decay branching fractions and $\epem$ width for most of
the studied channels, the majority of them for the first time.

%\end{linenumbers}

\section{Acknowledgments}
\label{sec:Acknowledgments}
% Specific acknowledgments for this paper; remove if not needed.
% Standard acknowledgments paragraph; must always be included.

We are grateful for the extraordinary contributions of our PEP-II colleagues in achieving the excellent luminosity and machine conditions 
that have made this work possible. The success of this project also relies critically on the expertise and dedication of the computing 
organizations that support BABAR. The collaborating institutions wish to thank SLAC for its support and the kind hospitality extended to 
them.

\clearpage
%\newpage

\begin{table*}
\caption{Summary of the $\epem\to\KK\ppz\piz$ 
cross section measurement. The uncertainties are statistical only.}
\label{2k3pi0_tab}
%\begin{ruledtabular}
%\hspace{-1.8cm}
\begin{tabular}{c c c c c c c c c c}
$E_{\rm c.m.}$, GeV & $\sigma$, nb  
& $E_{\rm c.m.}$, GeV & $\sigma$, nb 
& $E_{\rm c.m.}$, GeV & $\sigma$, nb 
& $E_{\rm c.m.}$, GeV & $\sigma$, nb  
& $E_{\rm c.m.}$, GeV & $\sigma$, nb  
\\
  \hline
1.525 & 0.03 $\pm$ 0.03 &2.125 & 0.14 $\pm$ 0.04 &2.725 & 0.08 $\pm$ 0.02 &3.325 & 0.03 $\pm$ 0.02 &3.925 & 0.02 $\pm$ 0.01 \\ 
1.575 & 0.16 $\pm$ 0.05 &2.175 & 0.09 $\pm$ 0.03 &2.775 & 0.10 $\pm$ 0.03 &3.375 & 0.04 $\pm$ 0.01 &3.975 & 0.01 $\pm$ 0.01 \\ 
1.625 & 0.36 $\pm$ 0.07 &2.225 & 0.03 $\pm$ 0.03 &2.825 & 0.04 $\pm$ 0.02 &3.425 & 0.02 $\pm$ 0.02 &4.025 & 0.03 $\pm$ 0.01 \\ 
1.675 & 0.45 $\pm$ 0.07 &2.275 & 0.09 $\pm$ 0.03 &2.875 & 0.11 $\pm$ 0.03 &3.475 & 0.02 $\pm$ 0.02 &4.075 & 0.02 $\pm$ 0.01 \\ 
1.725 & 0.38 $\pm$ 0.06 &2.325 & 0.05 $\pm$ 0.03 &2.925 & 0.09 $\pm$ 0.02 &3.525 & 0.03 $\pm$ 0.02 &4.125 & 0.03 $\pm$ 0.01 \\ 
1.775 & 0.29 $\pm$ 0.05 &2.375 & 0.08 $\pm$ 0.03 &2.975 & 0.05 $\pm$ 0.02 &3.575 & 0.04 $\pm$ 0.02 &4.175 & 0.02 $\pm$ 0.01 \\ 
1.825 & 0.21 $\pm$ 0.05 &2.425 & 0.06 $\pm$ 0.02 &3.025 & 0.10 $\pm$ 0.03 &3.625 & 0.02 $\pm$ 0.01 &4.225 & 0.01 $\pm$ 0.01 \\ 
1.875 & 0.21 $\pm$ 0.05 &2.475 & 0.12 $\pm$ 0.03 &3.075 & 0.18 $\pm$ 0.03 &3.675 & 0.07 $\pm$ 0.02 &4.275 & 0.01 $\pm$ 0.01 \\ 
1.925 & 0.17 $\pm$ 0.03 &2.525 & 0.07 $\pm$ 0.02 &3.125 & 0.21 $\pm$ 0.03 &3.725 & 0.04 $\pm$ 0.02 &4.325 & 0.02 $\pm$ 0.01 \\ 
1.975 & 0.09 $\pm$ 0.03 &2.575 & 0.06 $\pm$ 0.02 &3.175 & 0.08 $\pm$ 0.02 &3.775 & 0.03 $\pm$ 0.01 &4.375 & 0.03 $\pm$ 0.01 \\ 
2.025 & 0.09 $\pm$ 0.03 &2.625 & 0.09 $\pm$ 0.03 &3.225 & 0.08 $\pm$ 0.02 &3.825 & 0.02 $\pm$ 0.01 &4.425 & 0.01 $\pm$ 0.01 \\ 
2.075 & 0.21 $\pm$ 0.04 &2.675 & 0.09 $\pm$ 0.03 &3.275 & 0.05 $\pm$ 0.02 &3.875 & 0.01 $\pm$ 0.01 &4.475 & 0.01 $\pm$ 0.01 \\ 
\hline
\end{tabular}
%\end{ruledtabular}
\end{table*}

\begin{table*}
\caption{Summary of the $\epem\to\KS K^{\pm}\pi^{\mp}\ppz$ 
cross section measurement. The uncertainties are statistical only.}
\label{kskpi2pi0_tab}
%\begin{ruledtabular}
%\hspace{-1.8cm}
\begin{tabular}{c c c c c c c c c c}
$E_{\rm c.m.}$, GeV & $\sigma$, nb  
& $E_{\rm c.m.}$, GeV & $\sigma$, nb 
& $E_{\rm c.m.}$, GeV & $\sigma$, nb 
& $E_{\rm c.m.}$, GeV & $\sigma$, nb  
& $E_{\rm c.m.}$, GeV & $\sigma$, nb  
\\
  \hline
1.875 & 0.00 $\pm$ 0.01 &2.425 & 0.34 $\pm$ 0.06 &2.975 & 0.25 $\pm$ 0.05 &3.525 & 0.19 $\pm$ 0.04 &4.075 & 0.09 $\pm$ 0.02 \\ 
1.925 & 0.01 $\pm$ 0.03 &2.475 & 0.42 $\pm$ 0.07 &3.025 & 0.25 $\pm$ 0.05 &3.575 & 0.10 $\pm$ 0.03 &4.125 & 0.04 $\pm$ 0.02 \\ 
1.975 & 0.01 $\pm$ 0.03 &2.525 & 0.40 $\pm$ 0.06 &3.075 & 0.95 $\pm$ 0.08 &3.625 & 0.10 $\pm$ 0.03 &4.175 & 0.05 $\pm$ 0.02 \\ 
2.025 & 0.05 $\pm$ 0.03 &2.575 & 0.30 $\pm$ 0.05 &3.125 & 0.67 $\pm$ 0.07 &3.675 & 0.20 $\pm$ 0.04 &4.225 & 0.07 $\pm$ 0.02 \\ 
2.075 & 0.14 $\pm$ 0.05 &2.625 & 0.29 $\pm$ 0.06 &3.175 & 0.14 $\pm$ 0.05 &3.725 & 0.19 $\pm$ 0.04 &4.275 & 0.03 $\pm$ 0.02 \\ 
2.125 & 0.15 $\pm$ 0.05 &2.675 & 0.36 $\pm$ 0.06 &3.225 & 0.14 $\pm$ 0.04 &3.775 & 0.06 $\pm$ 0.03 &4.325 & 0.01 $\pm$ 0.02 \\ 
2.175 & 0.30 $\pm$ 0.06 &2.725 & 0.41 $\pm$ 0.06 &3.275 & 0.18 $\pm$ 0.04 &3.825 & 0.09 $\pm$ 0.03 &4.375 & 0.06 $\pm$ 0.02 \\ 
2.225 & 0.31 $\pm$ 0.07 &2.775 & 0.27 $\pm$ 0.05 &3.325 & 0.15 $\pm$ 0.04 &3.875 & 0.09 $\pm$ 0.03 &4.425 & 0.05 $\pm$ 0.02 \\ 
2.275 & 0.24 $\pm$ 0.05 &2.825 & 0.28 $\pm$ 0.06 &3.375 & 0.13 $\pm$ 0.04 &3.925 & 0.10 $\pm$ 0.03 &4.475 & 0.04 $\pm$ 0.02 \\ 
2.325 & 0.28 $\pm$ 0.06 &2.875 & 0.29 $\pm$ 0.05 &3.425 & 0.16 $\pm$ 0.03 &3.975 & 0.08 $\pm$ 0.03 & & \\ 
2.375 & 0.34 $\pm$ 0.06 &2.925 & 0.19 $\pm$ 0.04 &3.475 & 0.13 $\pm$ 0.04 &4.025 & 0.07 $\pm$ 0.03 & &  \\ 
\hline
\end{tabular}
%\end{ruledtabular}
\end{table*}

\begin{table*}
\caption{Summary of the $\epem\to\KS K^{\pm}\pi^{\mp}\pipi$ 
cross section measurement. The uncertainties are statistical only.}
\label{ksk3pi_tab}
%\begin{ruledtabular}
%\hspace{-1.8cm}
\begin{tabular}{c c c c c c c c c c}
$E_{\rm c.m.}$, GeV & $\sigma$, nb  
& $E_{\rm c.m.}$, GeV & $\sigma$, nb 
& $E_{\rm c.m.}$, GeV & $\sigma$, nb 
& $E_{\rm c.m.}$, GeV & $\sigma$, nb  
& $E_{\rm c.m.}$, GeV & $\sigma$, nb  
\\
  \hline
1.775 & 0.01 $\pm$ 0.01 &2.325 & 0.38 $\pm$ 0.04 &2.875 & 0.23 $\pm$ 0.03 &3.425 & 0.18 $\pm$ 0.02 &3.975 & 0.09 $\pm$ 0.02 \\ 
1.825 & 0.01 $\pm$ 0.01 &2.375 & 0.38 $\pm$ 0.04 &2.925 & 0.28 $\pm$ 0.03 &3.475 & 0.18 $\pm$ 0.02 &4.025 & 0.07 $\pm$ 0.02 \\ 
1.875 & 0.02 $\pm$ 0.01 &2.425 & 0.48 $\pm$ 0.04 &2.975 & 0.21 $\pm$ 0.03 &3.525 & 0.15 $\pm$ 0.02 &4.075 & 0.12 $\pm$ 0.02 \\ 
1.925 & 0.02 $\pm$ 0.01 &2.475 & 0.34 $\pm$ 0.04 &3.025 & 0.21 $\pm$ 0.03 &3.575 & 0.14 $\pm$ 0.02 &4.125 & 0.08 $\pm$ 0.02 \\ 
1.975 & 0.07 $\pm$ 0.02 &2.525 & 0.37 $\pm$ 0.03 &3.075 & 1.44 $\pm$ 0.06 &3.625 & 0.11 $\pm$ 0.02 &4.175 & 0.07 $\pm$ 0.01 \\ 
2.025 & 0.10 $\pm$ 0.02 &2.575 & 0.30 $\pm$ 0.03 &3.125 & 0.69 $\pm$ 0.05 &3.675 & 0.30 $\pm$ 0.03 &4.225 & 0.07 $\pm$ 0.01 \\ 
2.075 & 0.16 $\pm$ 0.03 &2.625 & 0.32 $\pm$ 0.04 &3.175 & 0.21 $\pm$ 0.03 &3.725 & 0.08 $\pm$ 0.02 &4.275 & 0.09 $\pm$ 0.02 \\ 
2.125 & 0.28 $\pm$ 0.03 &2.675 & 0.33 $\pm$ 0.03 &3.225 & 0.17 $\pm$ 0.03 &3.775 & 0.11 $\pm$ 0.02 &4.325 & 0.05 $\pm$ 0.02 \\ 
2.175 & 0.40 $\pm$ 0.04 &2.725 & 0.33 $\pm$ 0.03 &3.275 & 0.20 $\pm$ 0.03 &3.825 & 0.12 $\pm$ 0.02 &4.375 & 0.06 $\pm$ 0.02 \\ 
2.225 & 0.31 $\pm$ 0.04 &2.775 & 0.31 $\pm$ 0.03 &3.325 & 0.15 $\pm$ 0.02 &3.875 & 0.10 $\pm$ 0.02 &4.425 & 0.06 $\pm$ 0.01 \\ 
2.275 & 0.30 $\pm$ 0.03 &2.825 & 0.22 $\pm$ 0.03 &3.375 & 0.15 $\pm$ 0.02 &3.925 & 0.09 $\pm$ 0.02 &4.475 & 0.03 $\pm$ 0.01 \\ 
\hline
\end{tabular}
%\end{ruledtabular}
\end{table*}

\begin{table*}
\caption{Summary of the $\epem\to\KK\eta$ 
cross section measurement. The uncertainties are statistical only.}
\label{2k3pi0_2keta_tab}
%\begin{ruledtabular}
%\hspace{-1.8cm}
\begin{tabular}{c c c c c c c c c c}
$E_{\rm c.m.}$, GeV & $\sigma$, nb  
& $E_{\rm c.m.}$, GeV & $\sigma$, nb 
& $E_{\rm c.m.}$, GeV & $\sigma$, nb 
& $E_{\rm c.m.}$, GeV & $\sigma$, nb  
& $E_{\rm c.m.}$, GeV & $\sigma$, nb  
\\
  \hline
1.525 & 0.00 $\pm$ 0.04 &1.925 & 0.24 $\pm$ 0.09 &2.325 & 0.01 $\pm$ 0.02 &2.725 & 0.03 $\pm$ 0.03 &3.125 & 0.06 $\pm$ 0.03 \\ 
1.575 & 0.09 $\pm$ 0.12 &1.975 & 0.12 $\pm$ 0.07 &2.375 & 0.03 $\pm$ 0.04 &2.775 & 0.03 $\pm$ 0.02 &3.175 & 0.00 $\pm$ 0.01 \\ 
1.625 & 1.17 $\pm$ 0.23 &2.025 & 0.21 $\pm$ 0.08 &2.425 & 0.02 $\pm$ 0.04 &2.825 & 0.04 $\pm$ 0.03 &3.225 & 0.02 $\pm$ 0.02 \\ 
1.675 & 1.47 $\pm$ 0.25 &2.075 & 0.22 $\pm$ 0.08 &2.475 & 0.06 $\pm$ 0.05 &2.875 & 0.00 $\pm$ 0.01 &3.275 & 0.00 $\pm$ 0.01 \\ 
1.725 & 1.39 $\pm$ 0.21 &2.125 & 0.21 $\pm$ 0.08 &2.525 & 0.07 $\pm$ 0.04 &2.925 & 0.01 $\pm$ 0.01 &3.325 & 0.01 $\pm$ 0.02 \\ 
1.775 & 0.78 $\pm$ 0.18 &2.175 & 0.17 $\pm$ 0.06 &2.575 & 0.08 $\pm$ 0.04 &2.975 & 0.00 $\pm$ 0.02 &3.375 & 0.01 $\pm$ 0.02 \\ 
1.825 & 0.55 $\pm$ 0.14 &2.225 & 0.04 $\pm$ 0.04 &2.625 & 0.04 $\pm$ 0.04 &3.025 & 0.03 $\pm$ 0.03 &3.425 & 0.01 $\pm$ 0.01 \\ 
1.875 & 0.56 $\pm$ 0.14 &2.275 & 0.15 $\pm$ 0.05 &2.675 & 0.06 $\pm$ 0.03 &3.075 & 0.15 $\pm$ 0.05 &3.475 & 0.00 $\pm$ 0.01 \\ 
\hline
\end{tabular}
%\end{ruledtabular}
\end{table*}

\begin{table*}
\caption{Summary of the $\epem\to\phi(1020)\eta$ 
cross section measurement. The uncertainties are statistical only.}
\label{2k3pi0_phieta_tab}
%\begin{ruledtabular}
%\hspace{-1.8cm}
\begin{tabular}{c c c c c c c c c c}
$E_{\rm c.m.}$, GeV & $\sigma$, nb  
& $E_{\rm c.m.}$, GeV & $\sigma$, nb 
& $E_{\rm c.m.}$, GeV & $\sigma$, nb 
& $E_{\rm c.m.}$, GeV & $\sigma$, nb  
& $E_{\rm c.m.}$, GeV & $\sigma$, nb  
\\
  \hline
1.525 & 0.00 $\pm$ 0.09 &1.875 & 1.12 $\pm$ 0.23 &2.225 & 0.11 $\pm$ 0.07 &2.575 & 0.11 $\pm$ 0.05 &2.925 & 0.01 $\pm$ 0.03 \\ 
1.575 & 0.27 $\pm$ 0.25 &1.925 & 0.42 $\pm$ 0.14 &2.275 & 0.25 $\pm$ 0.08 &2.625 & 0.05 $\pm$ 0.04 &2.975 & 0.02 $\pm$ 0.03 \\ 
1.625 & 2.24 $\pm$ 0.44 &1.975 & 0.22 $\pm$ 0.11 &2.325 & 0.02 $\pm$ 0.03 &2.675 & 0.11 $\pm$ 0.05 &3.025 & 0.04 $\pm$ 0.03 \\ 
1.675 & 2.87 $\pm$ 0.47 &2.025 & 0.33 $\pm$ 0.11 &2.375 & 0.10 $\pm$ 0.05 &2.725 & 0.06 $\pm$ 0.04 &3.075 & 0.17 $\pm$ 0.06 \\ 
1.725 & 2.36 $\pm$ 0.40 &2.075 & 0.37 $\pm$ 0.11 &2.425 & 0.12 $\pm$ 0.06 &2.775 & 0.04 $\pm$ 0.03 &3.125 & 0.03 $\pm$ 0.03 \\ 
1.775 & 1.54 $\pm$ 0.30 &2.125 & 0.32 $\pm$ 0.11 &2.475 & 0.14 $\pm$ 0.06 &2.825 & 0.02 $\pm$ 0.02 &3.175 & 0.00 $\pm$ 0.01 \\ 
1.825 & 0.82 $\pm$ 0.22 &2.175 & 0.26 $\pm$ 0.09 &2.525 & 0.07 $\pm$ 0.05 &2.875 & 0.00 $\pm$ 0.01 &3.225 & 0.04 $\pm$ 0.03 \\
\hline
\end{tabular}
%\end{ruledtabular}
\end{table*}

\begin{table*}
\caption{Summary of the $\epem\to K^{*+}K^{*-}\piz$ 
cross section measurement. The uncertainties are statistical only.}
\label{2k3pi0_kstar_tab}
%\begin{ruledtabular}
%\hspace{-1.8cm}
\begin{tabular}{c c c c c c c c c c}
$E_{\rm c.m.}$, GeV & $\sigma$, nb  
& $E_{\rm c.m.}$, GeV & $\sigma$, nb 
& $E_{\rm c.m.}$, GeV & $\sigma$, nb 
& $E_{\rm c.m.}$, GeV & $\sigma$, nb  
& $E_{\rm c.m.}$, GeV & $\sigma$, nb  
\\
  \hline
1.875 & 0.00 $\pm$ 0.05 &2.425 & 0.24 $\pm$ 0.10 &2.975 & 0.29 $\pm$ 0.08 &3.525 & 0.20 $\pm$ 0.05 &4.075 & 0.07 $\pm$ 0.03 \\ 
1.925 & 0.03 $\pm$ 0.05 &2.475 & 0.39 $\pm$ 0.11 &3.025 & 0.17 $\pm$ 0.09 &3.575 & 0.13 $\pm$ 0.05 &4.125 & 0.06 $\pm$ 0.03 \\ 
1.975 & 0.06 $\pm$ 0.05 &2.525 & 0.32 $\pm$ 0.10 &3.075 & 0.90 $\pm$ 0.13 &3.625 & 0.10 $\pm$ 0.05 &4.175 & 0.06 $\pm$ 0.03 \\ 
2.025 & 0.16 $\pm$ 0.07 &2.575 & 0.39 $\pm$ 0.10 &3.125 & 0.37 $\pm$ 0.10 &3.675 & 0.22 $\pm$ 0.06 &4.225 & 0.08 $\pm$ 0.03 \\ 
2.075 & 0.07 $\pm$ 0.12 &2.625 & 0.38 $\pm$ 0.10 &3.175 & 0.24 $\pm$ 0.07 &3.725 & 0.17 $\pm$ 0.05 &4.275 & 0.08 $\pm$ 0.03 \\ 
2.125 & 0.41 $\pm$ 0.19 &2.675 & 0.20 $\pm$ 0.10 &3.225 & 0.20 $\pm$ 0.07 &3.775 & 0.06 $\pm$ 0.04 &4.325 & 0.01 $\pm$ 0.02 \\ 
2.175 & 0.29 $\pm$ 0.09 &2.725 & 0.22 $\pm$ 0.09 &3.275 & 0.19 $\pm$ 0.06 &3.825 & 0.12 $\pm$ 0.04 &4.375 & 0.05 $\pm$ 0.03 \\ 
2.225 & 0.16 $\pm$ 0.09 &2.775 & 0.24 $\pm$ 0.09 &3.325 & 0.10 $\pm$ 0.05 &3.875 & 0.08 $\pm$ 0.03 &4.425 & 0.04 $\pm$ 0.03 \\ 
2.275 & 0.31 $\pm$ 0.15 &2.825 & 0.16 $\pm$ 0.07 &3.375 & 0.06 $\pm$ 0.05 &3.925 & 0.13 $\pm$ 0.04 &4.475 & 0.05 $\pm$ 0.03 \\ 
2.325 & 0.17 $\pm$ 0.09 &2.875 & 0.26 $\pm$ 0.08 &3.425 & 0.07 $\pm$ 0.05 &3.975 & 0.02 $\pm$ 0.03 & &  \\ 
2.375 & 0.45 $\pm$ 0.18 &2.925 & 0.33 $\pm$ 0.10 &3.475 & 0.10 $\pm$ 0.05 &4.025 & 0.04 $\pm$ 0.03 & &  \\ 
\hline
\end{tabular}
%\end{ruledtabular}
\end{table*}

\begin{table*}
\caption{Summary of the $\epem\to\KS K^{*}(892)^0\ppz$ 
cross section measurement. The uncertainties are statistical only.}
\label{kskpi2pi0_kstarn_tab}
%\begin{ruledtabular}
%\hspace{-1.8cm}
\begin{tabular}{c c c c c c c c c c}
$E_{\rm c.m.}$, GeV & $\sigma$, nb  
& $E_{\rm c.m.}$, GeV & $\sigma$, nb 
& $E_{\rm c.m.}$, GeV & $\sigma$, nb 
& $E_{\rm c.m.}$, GeV & $\sigma$, nb  
& $E_{\rm c.m.}$, GeV & $\sigma$, nb  
\\
  \hline
1.875 & 0.01 $\pm$ 0.04 &2.425 & 0.25 $\pm$ 0.09 &2.975 & 0.13 $\pm$ 0.07 &3.525 & 0.04 $\pm$ 0.03 &4.075 & 0.05 $\pm$ 0.03 \\ 
1.925 & 0.02 $\pm$ 0.04 &2.475 & 0.17 $\pm$ 0.09 &3.025 & 0.06 $\pm$ 0.05 &3.575 & 0.01 $\pm$ 0.02 &4.125 & 0.02 $\pm$ 0.02 \\ 
1.975 & 0.01 $\pm$ 0.04 &2.525 & 0.31 $\pm$ 0.10 &3.075 & 0.38 $\pm$ 0.12 &3.625 & 0.04 $\pm$ 0.03 &4.175 & 0.02 $\pm$ 0.02 \\ 
2.025 & 0.01 $\pm$ 0.04 &2.575 & 0.18 $\pm$ 0.06 &3.125 & 0.36 $\pm$ 0.09 &3.675 & 0.11 $\pm$ 0.05 &4.225 & 0.02 $\pm$ 0.02 \\ 
2.075 & 0.11 $\pm$ 0.07 &2.625 & 0.06 $\pm$ 0.05 &3.175 & 0.05 $\pm$ 0.04 &3.725 & 0.01 $\pm$ 0.02 &4.275 & 0.01 $\pm$ 0.01 \\ 
2.125 & 0.06 $\pm$ 0.05 &2.675 & 0.09 $\pm$ 0.06 &3.225 & 0.09 $\pm$ 0.05 &3.775 & 0.01 $\pm$ 0.02 &4.325 & 0.02 $\pm$ 0.02 \\ 
2.175 & 0.21 $\pm$ 0.10 &2.725 & 0.13 $\pm$ 0.07 &3.275 & 0.04 $\pm$ 0.04 &3.825 & 0.07 $\pm$ 0.03 &4.375 & 0.02 $\pm$ 0.02 \\ 
2.225 & 0.44 $\pm$ 0.11 &2.775 & 0.17 $\pm$ 0.08 &3.325 & 0.09 $\pm$ 0.05 &3.875 & 0.04 $\pm$ 0.03 &4.425 & 0.01 $\pm$ 0.02 \\ 
2.275 & 0.32 $\pm$ 0.09 &2.825 & 0.10 $\pm$ 0.06 &3.375 & 0.06 $\pm$ 0.04 &3.925 & 0.03 $\pm$ 0.02 &4.475 & 0.01 $\pm$ 0.01 \\ 
2.325 & 0.23 $\pm$ 0.09 &2.875 & 0.12 $\pm$ 0.07 &3.425 & 0.03 $\pm$ 0.03 &3.975 & 0.02 $\pm$ 0.02 & & \\ 
2.375 & 0.28 $\pm$ 0.10 &2.925 & 0.15 $\pm$ 0.07 &3.475 & 0.08 $\pm$ 0.04 &4.025 & 0.00 $\pm$ 0.02 & & \\ 
\hline
\end{tabular}
%\end{ruledtabular}
\end{table*}

\begin{table*}
\caption{Summary of the $\epem\to K^{*}(892)^{\pm} K^{\mp}\ppz$ 
cross section measurement. The uncertainties are statistical only.}
\label{kskpi2pi0_kstarc_tab}
%\begin{ruledtabular}
%\hspace{-1.8cm}
\begin{tabular}{c c c c c c c c c c}
$E_{\rm c.m.}$, GeV & $\sigma$, nb  
& $E_{\rm c.m.}$, GeV & $\sigma$, nb 
& $E_{\rm c.m.}$, GeV & $\sigma$, nb 
& $E_{\rm c.m.}$, GeV & $\sigma$, nb  
& $E_{\rm c.m.}$, GeV & $\sigma$, nb  
\\
  \hline
1.925 & 0.00 $\pm$ 0.04 &2.475 & 0.13 $\pm$ 0.10 &3.025 & 0.16 $\pm$ 0.08 &3.575 & 0.06 $\pm$ 0.04 &4.125 & 0.06 $\pm$ 0.03 \\ 
1.975 & 0.02 $\pm$ 0.03 &2.525 & 0.14 $\pm$ 0.10 &3.075 & 0.37 $\pm$ 0.11 &3.625 & 0.03 $\pm$ 0.03 &4.175 & 0.02 $\pm$ 0.02 \\ 
2.025 & 0.09 $\pm$ 0.06 &2.575 & 0.17 $\pm$ 0.08 &3.125 & 0.25 $\pm$ 0.08 &3.675 & 0.12 $\pm$ 0.04 &4.225 & 0.06 $\pm$ 0.03 \\ 
2.075 & 0.07 $\pm$ 0.04 &2.625 & 0.14 $\pm$ 0.07 &3.175 & 0.19 $\pm$ 0.06 &3.725 & 0.08 $\pm$ 0.05 &4.275 & 0.03 $\pm$ 0.02 \\ 
2.125 & 0.13 $\pm$ 0.07 &2.675 & 0.14 $\pm$ 0.08 &3.225 & 0.08 $\pm$ 0.05 &3.775 & 0.03 $\pm$ 0.03 &4.325 & 0.00 $\pm$ 0.02 \\ 
2.175 & 0.20 $\pm$ 0.10 &2.725 & 0.22 $\pm$ 0.10 &3.275 & 0.04 $\pm$ 0.04 &3.825 & 0.06 $\pm$ 0.02 &4.375 & 0.07 $\pm$ 0.02 \\ 
2.225 & 0.28 $\pm$ 0.10 &2.775 & 0.17 $\pm$ 0.08 &3.325 & 0.13 $\pm$ 0.05 &3.875 & 0.05 $\pm$ 0.03 &4.425 & 0.03 $\pm$ 0.02 \\ 
2.275 & 0.17 $\pm$ 0.07 &2.825 & 0.26 $\pm$ 0.08 &3.375 & 0.09 $\pm$ 0.04 &3.925 & 0.00 $\pm$ 0.02 &4.475 & 0.03 $\pm$ 0.02 \\ 
2.325 & 0.18 $\pm$ 0.09 &2.875 & 0.12 $\pm$ 0.07 &3.425 & 0.03 $\pm$ 0.03 &3.975 & 0.03 $\pm$ 0.03 & & \\ 
2.375 & 0.13 $\pm$ 0.10 &2.925 & 0.23 $\pm$ 0.07 &3.475 & 0.03 $\pm$ 0.03 &4.025 & 0.00 $\pm$ 0.02 & &  \\ 
2.425 & 0.19 $\pm$ 0.09 &2.975 & 0.10 $\pm$ 0.06 &3.525 & 0.10 $\pm$ 0.05 &4.075 & 0.02 $\pm$ 0.02 & &  \\ 
\hline
\end{tabular}
%\end{ruledtabular}
\end{table*}

\begin{table*}
\caption{Summary of the $\epem\to\KS K^{\pm}\rho^{\mp}\piz$ 
cross section measurement. The uncertainties are statistical only.}
\label{kskpi2pi0_rho_tab}
%\begin{ruledtabular}
%\hspace{-1.8cm}
\begin{tabular}{c c c c c c c c c c}
$E_{\rm c.m.}$, GeV & $\sigma$, nb  
& $E_{\rm c.m.}$, GeV & $\sigma$, nb 
& $E_{\rm c.m.}$, GeV & $\sigma$, nb 
& $E_{\rm c.m.}$, GeV & $\sigma$, nb  
& $E_{\rm c.m.}$, GeV & $\sigma$, nb  
\\
  \hline
2.225 & 0.00 $\pm$ 0.01 &2.675 & 0.33 $\pm$ 0.08 &3.125 & 0.40 $\pm$ 0.09 &3.575 & 0.08 $\pm$ 0.04 &4.025 & 0.06 $\pm$ 0.03 \\ 
2.275 & 0.03 $\pm$ 0.06 &2.725 & 0.22 $\pm$ 0.07 &3.175 & 0.16 $\pm$ 0.06 &3.625 & 0.10 $\pm$ 0.04 &4.075 & 0.06 $\pm$ 0.03 \\ 
2.325 & 0.13 $\pm$ 0.07 &2.775 & 0.16 $\pm$ 0.07 &3.225 & 0.10 $\pm$ 0.05 &3.675 & 0.06 $\pm$ 0.05 &4.125 & 0.04 $\pm$ 0.03 \\ 
2.375 & 0.06 $\pm$ 0.07 &2.825 & 0.12 $\pm$ 0.06 &3.275 & 0.22 $\pm$ 0.06 &3.725 & 0.17 $\pm$ 0.05 &4.175 & 0.05 $\pm$ 0.03 \\ 
2.425 & 0.22 $\pm$ 0.08 &2.875 & 0.16 $\pm$ 0.07 &3.325 & 0.17 $\pm$ 0.05 &3.775 & 0.06 $\pm$ 0.03 &4.225 & 0.01 $\pm$ 0.03 \\ 
2.475 & 0.18 $\pm$ 0.08 &2.925 & 0.20 $\pm$ 0.07 &3.375 & 0.06 $\pm$ 0.04 &3.825 & 0.07 $\pm$ 0.04 &4.275 & 0.02 $\pm$ 0.02 \\ 
2.525 & 0.27 $\pm$ 0.08 &2.975 & 0.23 $\pm$ 0.07 &3.425 & 0.16 $\pm$ 0.05 &3.875 & 0.07 $\pm$ 0.03 &4.325 & 0.04 $\pm$ 0.02 \\ 
2.575 & 0.15 $\pm$ 0.07 &3.025 & 0.15 $\pm$ 0.07 &3.475 & 0.11 $\pm$ 0.04 &3.925 & 0.05 $\pm$ 0.03 &4.375 & 0.07 $\pm$ 0.03 \\ 
2.625 & 0.26 $\pm$ 0.07 &3.075 & 0.51 $\pm$ 0.10 &3.525 & 0.11 $\pm$ 0.05 &3.975 & 0.03 $\pm$ 0.03 &4.425 & 0.03 $\pm$ 0.02 \\
\hline
\end{tabular}
%\end{ruledtabular}
\end{table*}

\begin{table*}
\caption{Summary of the $\epem\to f_1(1285)\pipi$ 
cross section measurement. The uncertainties are statistical only.}
\label{ksk3pi_f1_tab}
%\begin{ruledtabular}
%\hspace{-1.8cm}
\begin{tabular}{c c c c c c c c c c}
$E_{\rm c.m.}$, GeV & $\sigma$, nb  
& $E_{\rm c.m.}$, GeV & $\sigma$, nb 
& $E_{\rm c.m.}$, GeV & $\sigma$, nb 
& $E_{\rm c.m.}$, GeV & $\sigma$, nb  
& $E_{\rm c.m.}$, GeV & $\sigma$, nb  
\\
  \hline
1.650 & 0.02 $\pm$ 0.09 &2.250 & 0.48 $\pm$ 0.27 &2.850 & 0.02 $\pm$ 0.09 &3.450 & 0.21 $\pm$ 0.09 &4.050 & 0.09 $\pm$ 0.06 \\ 
1.750 & 0.30 $\pm$ 0.16 &2.350 & 0.20 $\pm$ 0.19 &2.950 & 0.20 $\pm$ 0.12 &3.550 & 0.05 $\pm$ 0.06 &4.150 & 0.04 $\pm$ 0.05 \\ 
1.850 & 0.28 $\pm$ 0.18 &2.450 & 0.49 $\pm$ 0.25 &3.050 & 0.03 $\pm$ 0.06 &3.650 & 0.08 $\pm$ 0.09 &4.250 & 0.02 $\pm$ 0.05 \\ 
1.950 & 0.90 $\pm$ 0.26 &2.550 & 0.11 $\pm$ 0.16 &3.150 & 0.15 $\pm$ 0.09 &3.750 & 0.12 $\pm$ 0.12 &4.350 & 0.00 $\pm$ 0.05 \\ 
2.050 & 0.84 $\pm$ 0.28 &2.650 & 0.12 $\pm$ 0.15 &3.250 & 0.06 $\pm$ 0.08 &3.850 & 0.00 $\pm$ 0.09 &4.450 & 0.00 $\pm$ 0.05 \\ 
2.150 & 0.97 $\pm$ 0.32 &2.750 & 0.06 $\pm$ 0.12 &3.350 & 0.02 $\pm$ 0.05 &3.950 & 0.15 $\pm$ 0.09 &4.550 & 0.00 $\pm$ 0.00 \\ 
\hline
\end{tabular}
%\end{ruledtabular}
\end{table*}

\end{document}